\documentclass[prx,twocolumn,superscriptaddress,citeautoscript,showpacs,amsart,longbibliography]{revtex4-2}

\usepackage{blindtext}
\usepackage{graphicx}
\usepackage{bm}
\usepackage{epsfig}
\usepackage{amssymb}
\usepackage{amsfonts}
\usepackage{braket}
\usepackage{color}
\usepackage{epstopdf}
\usepackage{hyperref}
\usepackage{float}
\restylefloat{table}
\usepackage{bibentry}
\usepackage{color}
\usepackage{multirow}
\usepackage[caption=false]{subfig}

\usepackage{amsfonts}
\usepackage{amsthm}
\usepackage{graphicx}
\usepackage{multirow}
\usepackage{color}
\usepackage{bbold}
\usepackage{bm}
\usepackage{times}
\usepackage{amsmath,bm,amsfonts}
\usepackage{dcolumn}
\usepackage{graphicx}
\usepackage{latexsym}

\newcommand{\bpm}{\begin{pmatrix}}
\newcommand{\epm}{\end{pmatrix}}
\newcommand{\ba}{\begin{eqnarray}}
\newcommand{\ea}{\end{eqnarray}}
\newcommand{\bd}{\begin{displaymath}}
\renewcommand{\v}[1]{{\bf #1}}

\graphicspath{{figures/}}

\begin{document}
\title{Sign-tunable anomalous Hall effect induced by two-dimensional symmetry-protected nodal structures in ferromagnetic perovskite oxide thin films}

\author{Byungmin Sohn}
\thanks {These authors contributed equally to this work.}
\affiliation{Department of Physics and Astronomy, Seoul National University, Seoul 08826, Korea}
\affiliation{Center for Correlated Electron Systems, Institute for Basic Science, Seoul 08826, Korea}
\author{Eunwoo Lee}
\thanks {These authors contributed equally to this work.}
\affiliation{Department of Physics and Astronomy, Seoul National University, Seoul 08826, Korea}
\affiliation{Center for Correlated Electron Systems, Institute for Basic Science, Seoul 08826, Korea}
\affiliation{Center for Theoretical Physics (CTP), Seoul National University, Seoul 08826, Korea}
\author{Se Young Park}
\affiliation{Department of Physics, Soongsil University, Seoul, 06978 Korea}
\affiliation{Integrative Institute of Basic Sciences, Soongsil University, Seoul, 06978 Korea}
\author{Wonshik Kyung}
\affiliation{Department of Physics and Astronomy, Seoul National University, Seoul 08826, Korea}
\affiliation{Center for Correlated Electron Systems, Institute for Basic Science, Seoul 08826, Korea}
\author{Jinwoong Hwang}
\affiliation{Advanced Light Source, Lawrence Berkeley National Laboratory, Berkeley, CA 94720, USA}
\author{Jonathan D. Denlinger}
\affiliation{Advanced Light Source, Lawrence Berkeley National Laboratory, Berkeley, CA 94720, USA}
\author{Minsoo Kim}
\affiliation{Department of Physics and Astronomy, Seoul National University, Seoul 08826, Korea}
\affiliation{Center for Correlated Electron Systems, Institute for Basic Science, Seoul 08826, Korea}
\author{Donghan Kim}
\affiliation{Department of Physics and Astronomy, Seoul National University, Seoul 08826, Korea}
\affiliation{Center for Correlated Electron Systems, Institute for Basic Science, Seoul 08826, Korea}
\author{Bongju Kim}
\affiliation{Department of Physics and Astronomy, Seoul National University, Seoul 08826, Korea}
\affiliation{Center for Correlated Electron Systems, Institute for Basic Science, Seoul 08826, Korea}
\author{Hanyoung Ryu}
\affiliation{Department of Physics and Astronomy, Seoul National University, Seoul 08826, Korea}
\affiliation{Center for Correlated Electron Systems, Institute for Basic Science, Seoul 08826, Korea}
\author{Soonsang Huh}
\affiliation{Department of Physics and Astronomy, Seoul National University, Seoul 08826, Korea}
\affiliation{Center for Correlated Electron Systems, Institute for Basic Science, Seoul 08826, Korea}
\author{Ji Seop Oh}
\affiliation{Department of Physics and Astronomy, Seoul National University, Seoul 08826, Korea}
\affiliation{Center for Correlated Electron Systems, Institute for Basic Science, Seoul 08826, Korea}
\author{Jong Keun Jung}
\affiliation{Department of Physics and Astronomy, Seoul National University, Seoul 08826, Korea}
\affiliation{Center for Correlated Electron Systems, Institute for Basic Science, Seoul 08826, Korea}
\author{Dongjin Oh}
\affiliation{Department of Physics and Astronomy, Seoul National University, Seoul 08826, Korea}
\affiliation{Center for Correlated Electron Systems, Institute for Basic Science, Seoul 08826, Korea}
\author{Younsik Kim}
\affiliation{Department of Physics and Astronomy, Seoul National University, Seoul 08826, Korea}
\affiliation{Center for Correlated Electron Systems, Institute for Basic Science, Seoul 08826, Korea}
\author{Moonsup Han}
\affiliation{Department of Physics, University of Seoul, Seoul, 02504, Korea}
\author{Tae Won Noh}
\affiliation{Department of Physics and Astronomy, Seoul National University, Seoul 08826, Korea}
\affiliation{Center for Correlated Electron Systems, Institute for Basic Science, Seoul 08826, Korea}
\author{Bohm-Jung Yang}
\email[Electronic address:$~~$]{bjyang@snu.ac.kr}
\affiliation{Department of Physics and Astronomy, Seoul National University, Seoul 08826, Korea}
\affiliation{Center for Correlated Electron Systems, Institute for Basic Science, Seoul 08826, Korea}
\affiliation{Center for Theoretical Physics (CTP), Seoul National University, Seoul 08826, Korea}
\author{Changyoung Kim}
\email[Electronic address:$~~$]{changyoung@snu.ac.kr}
\affiliation{Department of Physics and Astronomy, Seoul National University, Seoul 08826, Korea}
\affiliation{Center for Correlated Electron Systems, Institute for Basic Science, Seoul 08826, Korea}

\date{\today}

\begin{abstract}
	
Magnetism and spin-orbit coupling (SOC) are two quintessential ingredients underlying novel topological transport phenomena in itinerant ferromagnets. When spin-polarized bands support nodal points/lines with band degeneracy that can be lifted by SOC, the nodal structures become a source of Berry curvature; this leads to a large anomalous Hall effect (AHE). Contrary to three-dimensional systems that naturally host nodal points/lines, two-dimensional (2D) systems can possess stable nodal structures only when proper crystalline symmetry exists. Here we show that 2D spin-polarized band structures of perovskite oxides generally support symmetry-protected nodal lines and points that govern both the sign and the magnitude of the AHE. To demonstrate this, we performed angle-resolved photoemission studies of ultrathin films of SrRuO$_3$, a representative metallic ferromagnet with SOC. We show that the sign-changing AHE upon variation in the film thickness, magnetization, and chemical potential can be well explained by theoretical models. Our study is the first to directly characterize the topological band structure of 2D spin-polarized bands and the corresponding AHE, which could facilitate new switchable devices based on ferromagnetic ultrathin films.

\end{abstract}
\maketitle

\section*{1. Introduction}

The interplay between magnetism and spin-orbit coupling (SOC) underlies the various novel topological transport phenomena seen in metallic ferromagnets~\cite{Burkor14,Ye18,Groenendijk2020}. In three-dimensional (3D) ferromagnets, spin-polarized bands often accompany nodal points or nodal lines (NLs)~\cite{Chang16,Chang18,Kim18}. The lifting of the band degeneracy at a nodal structure due to SOC induces enhanced Berry curvature around it, which leads to various topological transport phenomena such as the anomalous Hall effect (AHE)~\cite{Nagaosa10,Zeng06,Wang18}. For instance, in the SrRuO$_3$ (SRO) bulk, theoretical predictions have indicated that nodal structures with an SOC-induced gap play a role as magnetic monopoles in momentum space, which leads to a non-monotonous change in the AHE~\cite{Fang03}.

The 3D SRO bulk has been considered as a canonical system, in which the topological band structure induces a large AHE~\cite{Fang03, Chen13}. However, the existence of magnetic monopoles in SRO has not yet been verified experimentally. Part of the reason for this is the lack of a suitable single crystal, as well as the difficulty in preparing clean, cleaved surfaces for angle-resolved photoemission spectroscopy (ARPES) study due to its 3D structure. Meanwhile, it has recently been shown that the NLs of 3D spin-polarized bands can also induce a large AHE, as seen in a layered Fe$_3$GeTe$_2$ ferromagnet~\cite{Kim18}. The AHE driven by nodal points (so-called Weyl points) has also been reported in various 3D magnetic metals~\cite{Vazifeh13,Zyuzin16}.

\begin{figure}[htbp]
	\includegraphics[width=0.475\textwidth]{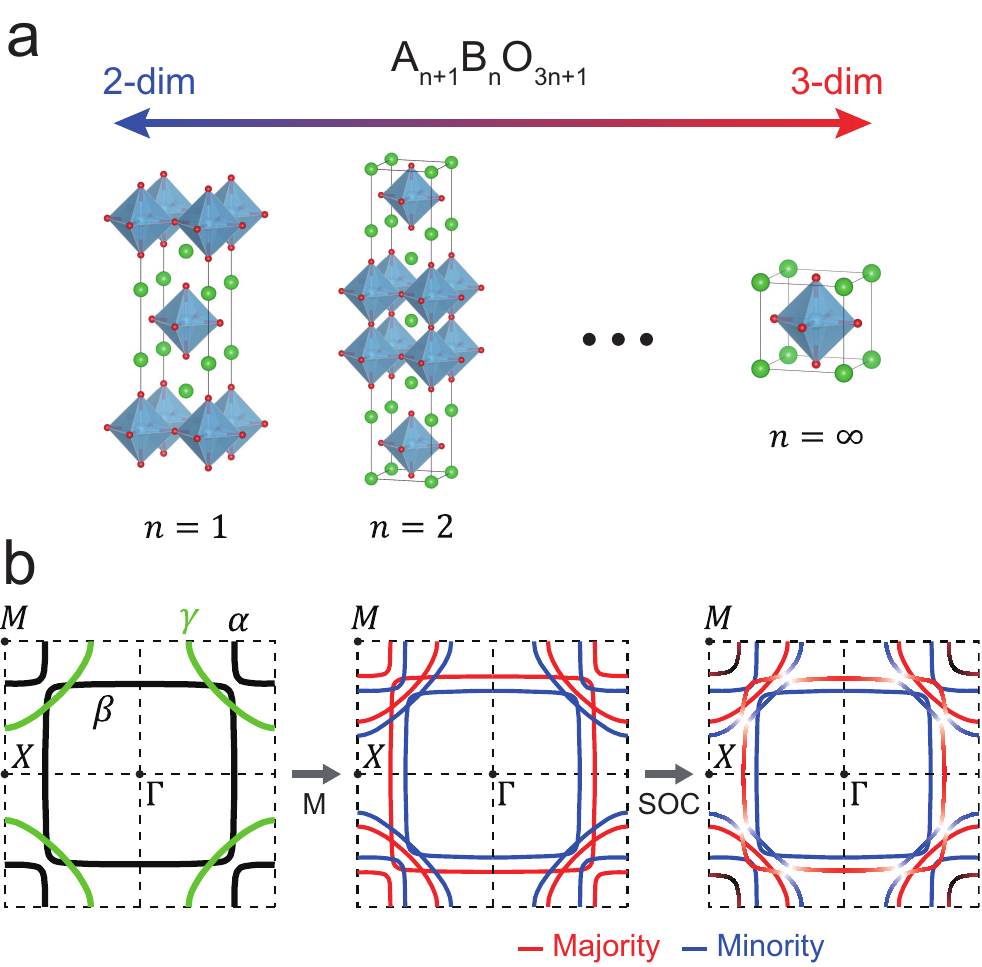}
	\caption{{\bf Fermi surface of two-dimensional (2D) ferromagnetic perovskites.} 
		({\bf a}) Structures of layered perovskite oxides with the chemical formula A$_{n+1}$B$_{n}$O$_{3n+1}$ where $n=1,2,...$ denotes a natural number. 
		({\bf b}) Schematic Fermi surfaces (FSs) of SrRuO$_3$ (SRO) in the 2D limit. When magnetization (M) and spin-orbit coupling (SOC) are absent, all bands are spin degenerate. When M is finite, spin degeneracy is lifted, such that majority (red) and minority (blue) bands form. On initiating SOC, the FSs become hybridized, inducing a finite Berry curvature.}
	\label{fig:1}
\end{figure}
The relationship between the topological band structure and corresponding transport phenomena remains largely unexplored in two-dimensional (2D) metallic ferromagnets~\cite{Groenendijk2020}. The nodal structures in 2D ferromagnets are more fragile and unstable compared with 3D structures. Only when proper symmetry conditions are satisfied, can nodal points or NLs manifest as symmetry-protected band degeneracy in 2D bands~\cite{Young15,Niu17}.
Moreover, the increased correlations expected in the reduced dimensionality could alter both the quasi-particle energy spectrum and intrinsic AHE such that careful comparison between theoretical and experimental results is indispensable to verify the topological band structure in 2D magnets~\cite{PhysRevB.90.165143, PhysRevB.83.205101}.

In this work, we demonstrate the relation between the symmetry-protected nodal structures of 2D spin-polarized bands and the AHE. For this purpose, we performed ARPES measurements on SRO ultrathin films, revealing their band structures for the first time. Based on tight-binding models, first-principles calculations, and symmetry analysis, we propose that the spin-polarized bands derived from $t_{2g}$ orbitals in layered perovskite oxides generally support i) nodal points with quadratic band crossing (QBC) protected by four-fold rotation symmetry and ii) NLs arising from the crossing between majority and minority spin bands. When SOC is included, these nodal points and lines are gapped and generate large Berry curvature in the surrounding area. Because the energies of nodal points and lines are different, when the Berry curvatures that arise from them have the opposite signs, the magnitude and sign of the AHE can generally be varied by changing the Fermi level. We demonstrate that the AHE of SRO thin films exhibits sign reversal, depending on the film thickness, temperature, magnetization, and chemical potential, due to the symmetry-protected nodal structures of the 2D spin-polarized bands.

\section*{2. Results}

Figure 1(a) shows the general lattice structures of layered Ruddlesden-Popper perovskite oxides, with the chemical formula A$_{n+1}$B$_n$O$_{3n+1}$, where $n$ denotes a natural number. As $n$ increases, the material takes on a more 3D character, interpolating between the 2D limit with $n=1$ (A$_{2}$BO$_{4}$) and the 3D limit with $n=\infty$ (ABO$_{3}$). Among the perovskite materials, SRO is a representative example of 3D ferromagnetic metals, with an AHE that has been attributed to the magnetic monopoles in momentum space of the topological band structure~\cite{Fang03}.

As the 2D Sr$_2$RuO$_4$ is non-magnetic~\cite{Neumeier94}, it is not an appropriate system for our study. Therefore, as an alternative, we studied the quasi-2D limit of SRO by growing ultrathin films of SRO with a thickness of 4~unit cells (u.c.). Due to its quasi-2D nature, the Fermi surfaces (FSs) of SRO ultrathin films are expected to be somewhat similar to those of Sr$_2$RuO$_4$~\cite{Puchkov98}. We confirm this through our ARPES measurements as will be shown later. For Sr$_2$RuO$_4$, it is well known that FSs consist of three bands, $\alpha,\beta,$ and $\gamma$, where $\alpha$ and $\beta$ FSs are composed of $d_{xz,yz}$ orbitals while $\gamma$ FS arises from the $d_{xy}$ orbital as schematically shown in Fig. 1(b)~\cite{Puchkov98, Lu96, Damascelli00}. When ferromagnetism develops, spin-degenerate bands split into majority and minority bands, which results in six bands derived from $t_{2g}$ orbitals appearing at the Fermi level. With SOC, the majority and minority bands are hybridized at the points where they are crossed. This provides the general idea on how Berry curvature is generated in ferromagnetic 2D perovskites.

\begin{figure*}[htbp]
\includegraphics[width=1\textwidth]{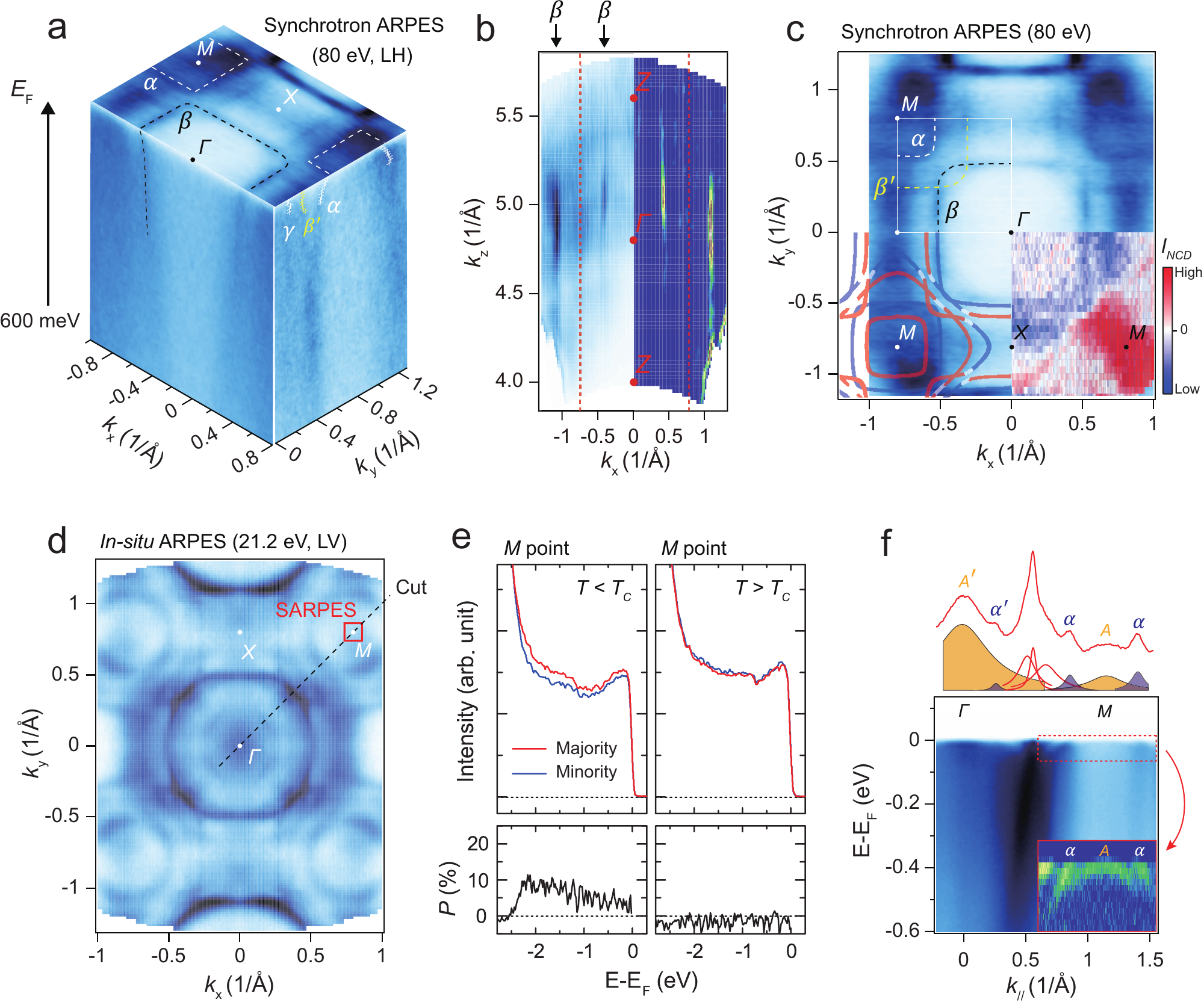}
\caption{{\bf Angle-resolved photoemission spectroscopy (ARPES) data of a 4 unit cell (u.c.) SRO thin film.}
({\bf a}) Intensity plot of ARPES data obtained with linear horizontal (LH) polarized synchrotron light with a photon energy of 80~eV. An integration window of $E_F\pm 8$~meV is used. $\alpha$, a folded $\beta$ ($\beta^\prime$) and $\gamma$ bands are marked in the $E$-$k_y$ cut. The folded $\beta$ band is induced due to the $\sqrt{2} \times \sqrt{2}$ octahedron rotation in the tetragonal structure of SRO. 
({\bf b}) FS map in the ($k_x$, $k_z$) plane obtained with an integration window of $E_F\pm 20$~meV (left) and its 2D curvature (right). The red dotted lines indicate the zone boundary.
({\bf c}) FS map of the data in ({\bf a}). The white, black, and yellow dotted lines indicate the $\alpha$, $\beta$, and $\beta^\prime$ bands, respectively. The thick lines from the tight-binding model fit are superimposed on the bottom-left part of the map. Shown on the right-bottom is normalized circular-dichroism (CD) FS with an integration window of $E_F\pm 27$~meV. 
({\bf d}) FS map obtained {\it in-situ} with linear vertical (LV) polarized He $I\alpha$ light (21.2~eV). An integration range of $E_F\pm 10$~meV was used. 
({\bf e}) Spin-resolved ARPES (SARPES) data of the valence band at the $M$ point. Measurements were done at 10~K ($T<T_C$) and 125~K ($T> T_C$). Lower panels show the spin polarization. 
({\bf f}) A $\Gamma$-$M$ high-symmetry cut (lower) and a momentum distribution curve obtained by integrating $E_F\pm10$~meV (upper). Shown with the MDC is its fit result with Lorentzian peaks. The blue (orange) peaks near $M$ come from the $\alpha$ ($A$) band. Replica peaks ($A^\prime$ and $\alpha^\prime$) appear near the $\Gamma$ point due to the band folding. (inset) 2D-curvature band dispersion near the $M$ point.
}
\label{fig:2}
\end{figure*}


To experimentally verify the band structure of a 2D ferromagnetic perovskite, we performed ARPES measurements on SRO ultrathin films. Figure 2(a) shows a FS map as well as the energy-momentum ($E$-$k$) spectra of a 4~u.c. SRO thin film measured with 80~eV light with linear horizontal (LH) polarization. $\alpha$ and $\beta$ FSs are clearly visible and are similar to the corresponding FSs reported for Sr$_2$RuO$_4$~\cite{Puchkov98, Lu96, Damascelli00, Mackenzie96, Mackenzie98}. Due to the rotational distortion of RuO$_6$ octahedra in SRO ultrathin films grown on SrTiO$_3$ (001) substrates~\cite{Chang11, Sohn18}, the FS from the folded-$\beta$ band ($\beta^\prime$) is also observed.

Whether ultrathin films indeed possess 2D electronic structures requires an examination of the $k_z$ (i.e., photon energy) dependence. The measured $k_z$ dispersion is shown in Fig. 2(b). Here $k_z$ is calculated based on an inner potential of 14~eV obtained from experiments on SRO single crystals~\cite{Oh20}. Unlike 3D materials exhibiting $k_z$ dispersion, the $\beta$ band does not show any $k_z$ dispersion as expected. This clearly illustrates that SRO ultrathin films have quasi-2D band structures, similar to Sr$_2$RuO$_4$~\cite{Damascelli00, Mackenzie98} and Sr$_3$Ru$_2$O$_7$~\cite{Hase97, Singh01}. 

In Fig. 2(c), we reproduce the measured FSs. In addition to the $\alpha$ and $\beta$ pockets shown by the dotted lines, there are more detailed structures. Superimposed on the map as thick lines are results obtained from an effective 2D tight-binding model fit (see Methods for details). It is seen that the effective 2D model can explain the major features of the experimental dispersions. We also performed circular dichroism (CD) ARPES to examine the Berry curvature and the result is shown in the bottom-right corner of the figure (See SM for analysis of the CD-ARPES data). Moving from $\Gamma$ to $M$, the intensity of CD data varies from zero to negative, and then to positive. A clear sign-changing behavior is observed as a function of the electron momentum. Detailed discussions on the tight-binding fit and CD-ARPES results will be discussed later with Fig.~5.

Since ARPES measurement is sensitive to the surface condition, clearer band features can be resolved via {\it in-situ} measurements. Figure 2(d) shows an FS map of the 4~u.c. SRO thin film {\it in-situ} measured with linear vertical (LV) polarized He-$I\alpha$ light (21.2~eV). Spin polarization of the bands was also obtained by performing spin-resolved ARPES (SARPES) measurements below and above the Curie temperature, $T_C\sim 110~K$. Figure 2(e) shows the spin-resolved energy distribution curves (EDCs) and spin polarizations at the $M$ point at 10~K (below $T_C$) and 125~K (above $T_C$). The EDCs below $T_C$ show a considerable difference between the majority and minority spins, whereas no difference was observed above $T_C$. The difference is also presented in the form of spin polarization, $P=(I_{\uparrow}-I_{\downarrow})/(I_{\uparrow}+I_{\downarrow})$, in the lower panels. Spin-resolved EDCs along the high-symmetry lines ($\Gamma$-$M$ and $\Gamma$-$X$) can be found in Supplementary Materials (SM). These observations are consistent with itinerant ferromagnetism~\cite{Chang09, Jeong13}.

The high resolution FS map in Fig. 2(d) shows more detailed features in comparison to the result from {\it ex-situ} films in Fig. 2(a). An important feature visible in the high resolution data is heavy bands observed at the $\Gamma$ and $M$ points. To investigate those heavy bands, $\Gamma$-$M$ cut of the data in Fig. 2(d) and its momentum distribution curve (MDC) at the Fermi level are shown in Fig. 2(f). To see the dispersion more clearly, we plot 2D curvature~\cite{Zhang11} of the data near the $M$ point in the inset. Two hole-like bands are observed at the M point; one from the $\alpha$ band and the other, unknown and labeled as $A$, located within the $\alpha$ pocket. The latter does not exist in the case of Sr$_2$RuO$_4$. We fitted the MDC with Lorentziaion peaks as shown in the upper panel of Fig. 2(f). Near the $M$ point, three peaks are observed: two from the $\alpha$ band and one from the $A$ band. We also find that the heavy bands near the $\Gamma$ point are $\alpha$ and $A$ replica bands due to the $\sqrt{2}\times\sqrt{2}$ rotational distortion of RuO$_6$ octahedra~\cite{Chang11, Sohn18} (See SM for low-energy electron diffraction (LEED) patterns).

\begin{figure*}[htbp]
\includegraphics[width=0.8235\textwidth]{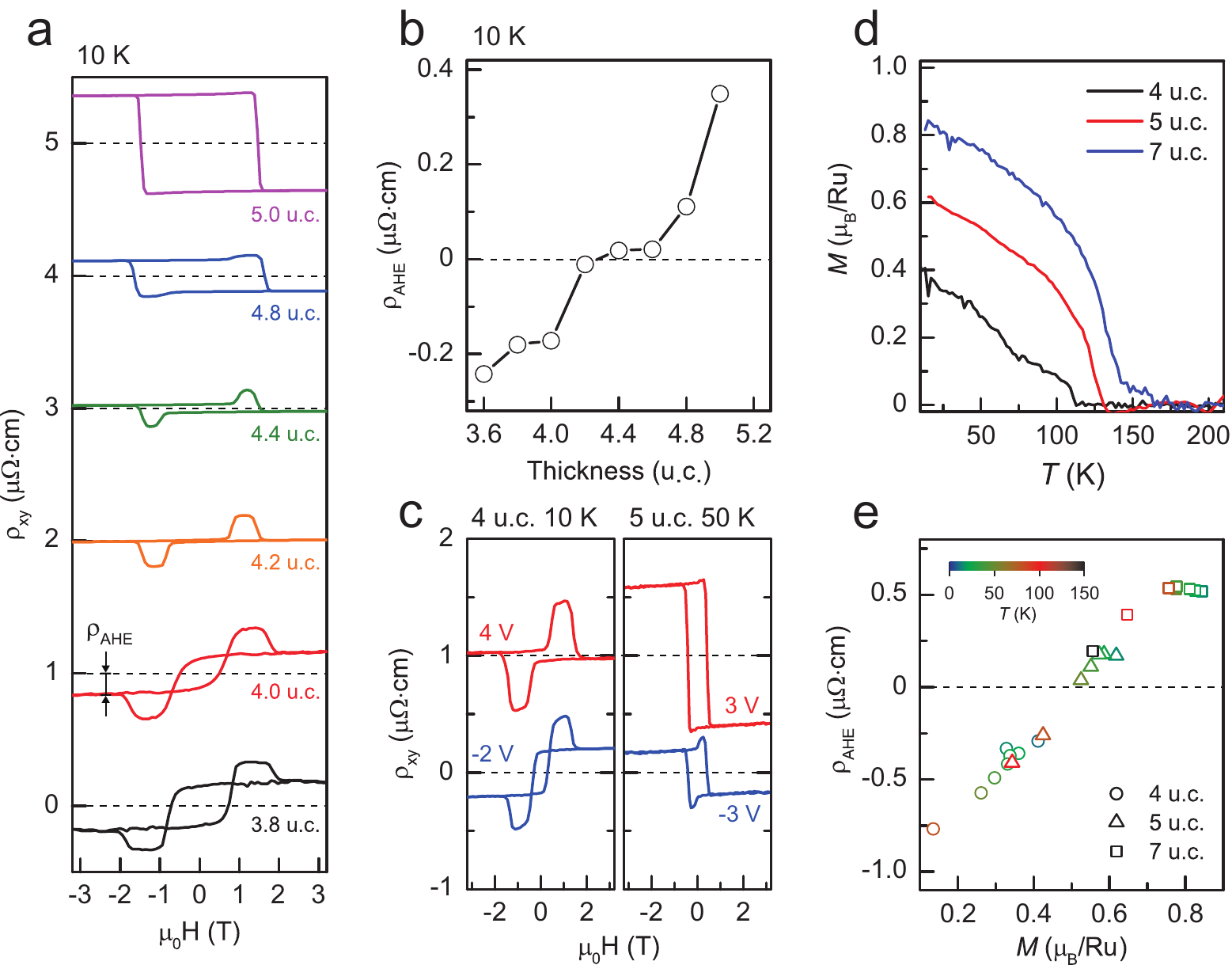}
\caption{{\bf Non-monotonous anomalous Hall effect (AHE) in SRO ultrathin films.} ({\bf a}) Thickness-dependent $\rho_{xy}$ of SRO ultrathin films at 10~K. $\rho_{\rm AHE}$ is defined from the saturated Hall resistivity $\rho_{xy}$. ({\bf b}) $\rho_{\rm AHE}$ $v.s.$ thickness curve for SRO ultrathin films at 10~K. The sign of $\rho_{\rm AHE}$ changes as the film thickness varies. ({\bf c}) $\rho_{xy}$ of 4 and 5~u.c. SRO thin films with ionic liquid gating at 10 and 50~K, respectively. The sign and magnitude of $\rho_{\rm AHE}$ change with the bias voltage. ({\bf d}) Out-of-plane magnetization of 4, 5, and 7~u.c. SRO thin films. ({\bf e}) $\rho_{\rm AHE}$ $v.s.$ $M$ plot for 4, 5 and 7 u.c. SRO thin films at various temperatures. The sign of $\rho_{\rm AHE}$ changes near $0.5$~$\mu_B$/Ru.}
\label{fig:3}
\end{figure*}

\begin{figure*}[htbp]
\includegraphics[width=0.765\textwidth]{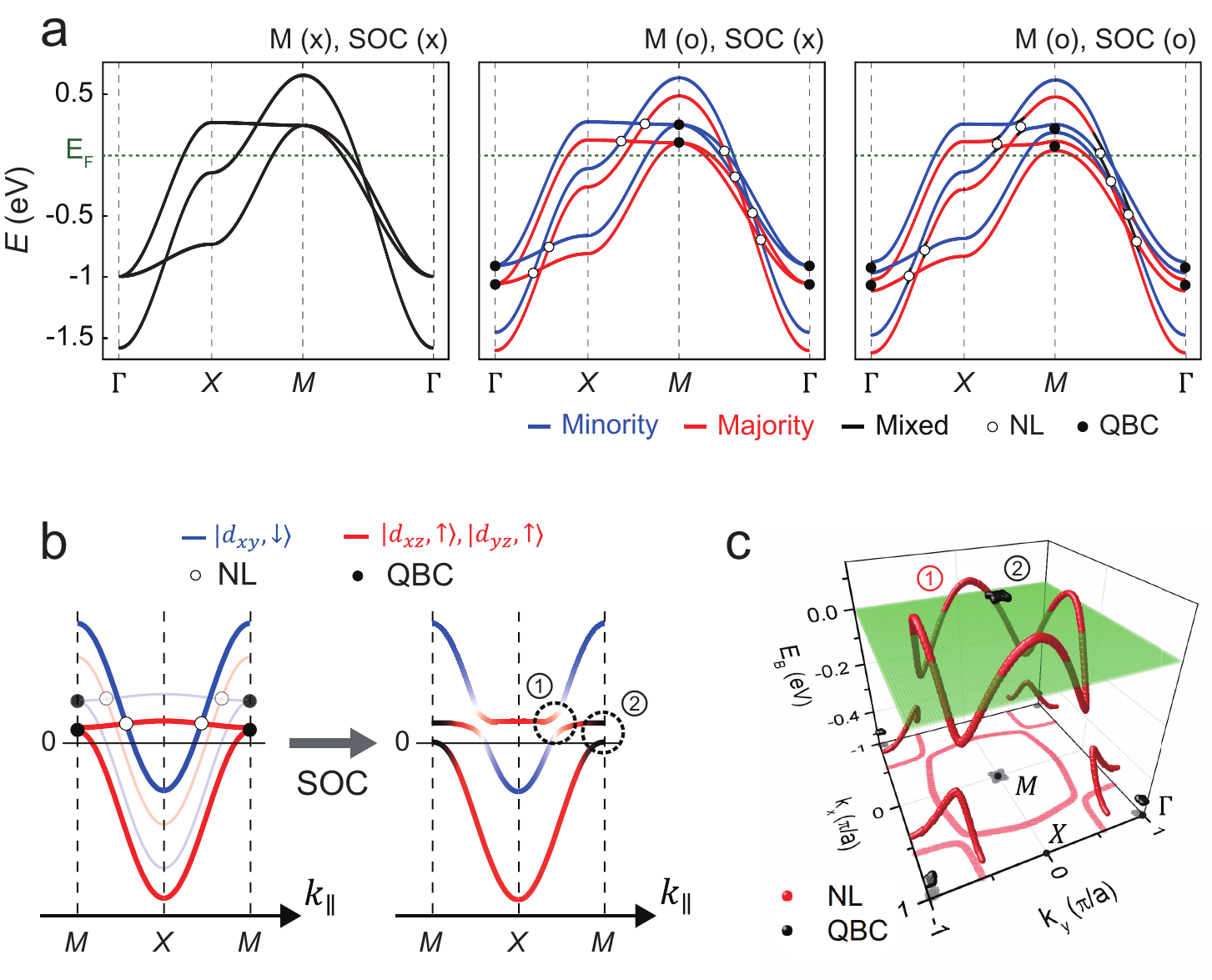}
\caption{{\bf Mechanism for the sign-tunable AHE induced by nodal lines (NLs) and points in a 2D ferromagnetic perovskite.}
({\bf a}) Band structures of SRO calculated from an effective six-band model relevant to 1~u.c. SRO. Without SOC, quadratic band crossings (QBCs) appear at the $\Gamma$ and $M$ points, and nodal lines (NLs) are formed when majority and minority bands cross.
({\bf b}) A calculated band structure along the $M$-$X$-$M$ line. For easier understanding, we focus on the two majority bands derived from $d_{xz,yz}$ orbitals (red) and the one minority band from the $d_{xy}$ orbital (blue). QBC occurs at the $M$ point and an NL is formed when the majority and minority bands cross. SOC can lift the degeneracy of the nodal structures, as denoted by circled numbers in the right panel. In ({\bf a}) and ({\bf b}), a white (black) circle denotes a source of Berry curvature from the NL (QBC). 
({\bf c}) Configuration of nodal structures in the momentum space. Red lines (black dots) denote the NLs (QBCs).}
\label{fig:4}
\end{figure*}

\begin{figure*}[htbp]
	\includegraphics[width=1\textwidth]{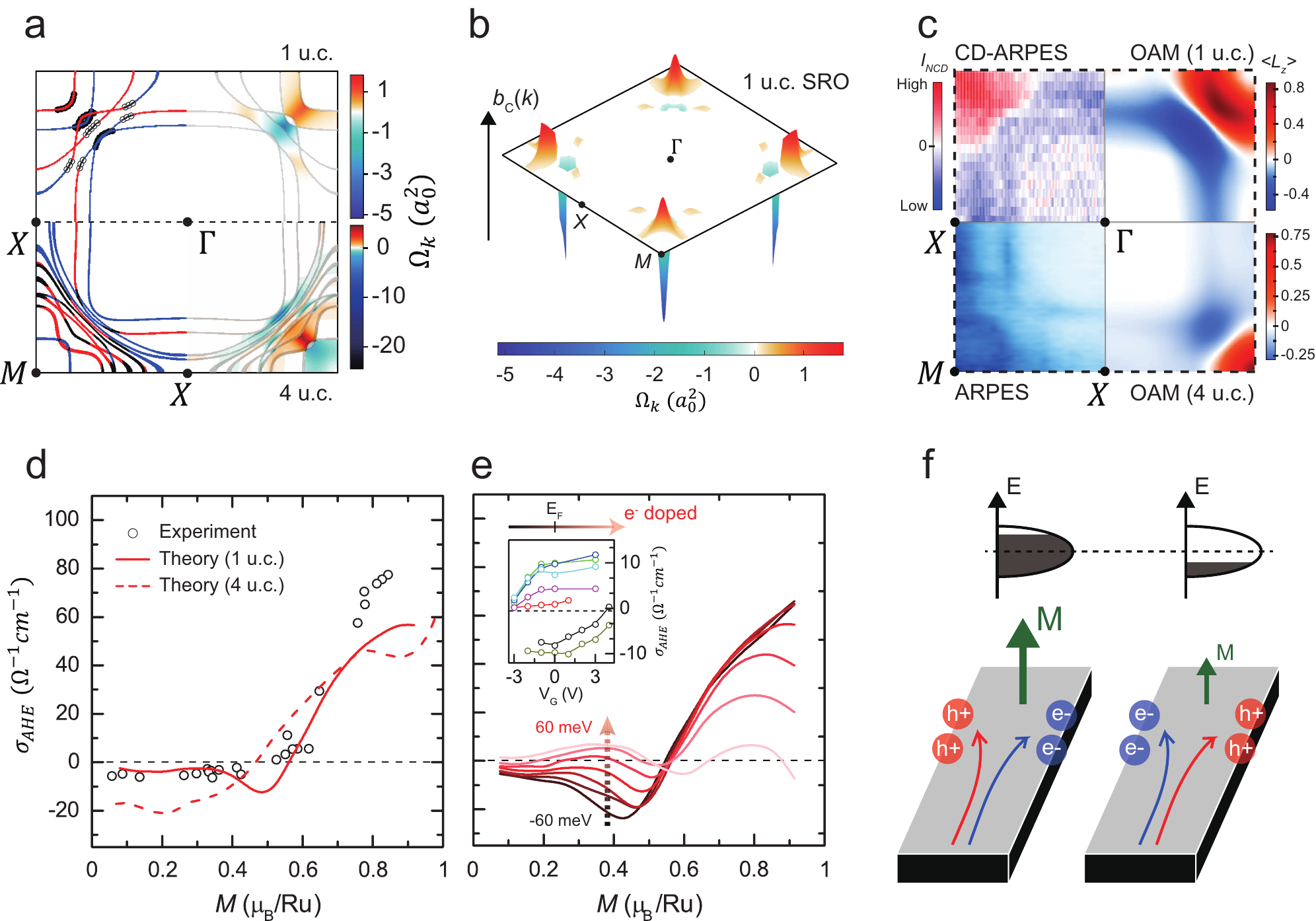}
	\caption{{\bf Berry curvature hot spots from nodal structures and switchable AHE of the SRO ultrathin film.}
		({\bf a}) Calculated band structure and Berry curvature of SRO at the Fermi level. Here, SOC is turned on. While the majority (blue) and minority (red) bands are spin-polarized, the mixed bands (black lines) have comparable contributions from up and down spins. The upper (lower) half of the figure is derived from the 1~u.c. (4~u.c.) tight-binding model. The left (right) half of the figure indicates the band structures (Berry curvature distribution) at the Fermi level.
		({\bf b}) Calculated Berry curvature distribution of 1~u.c. SRO in the momentum space.
		({\bf c}) (top-left) CD- and (bottom-left) normal ARPES FSs. Orbital angular momentum (OAM) at the Fermi level from tight-binding model of 1~u.c. (top-right) and 4~u.c. (bottom-right) SRO.
		({\bf d}) Measured (circles) and calculated (solid (dashed) line for the 1 (4)~u.c. model) magnetization dependence of anomalous Hall conductivity (AHC). When the magnetization is small (large), the AHC is negative (positive). 
		({\bf e}) Chemical potential dependence of AHC. The AHC increases with the chemical potential as the magnetization is about $0.4$~$\mu_B$/Ru. (inset) Ionic liquid gating measurements on 4 and 5~u.c. SRO thin films. AHC increases with the gate voltage.
		({\bf f}) A schematic showing the sign-tunable AHE. The sign of anomalous Hall resistivity can change as the magnetization or Fermi energy is varied due to Berry curvature hot spots near the Fermi level induced by nodal structures.}
	\label{fig:5}
\end{figure*}

With the full experimental electronic structure identified, we turn our attention to the transport and magnetic properties of SRO ultrathin films presented in Fig. 3. Figure 3(a) shows the thickness-dependent Hall resistivity measured at 10 K for $3.8$~-~$5.0$ u.c. SRO ultrathin films under an out-of-plane magnetic field. Here, the thickness of the thin film was estimated from the corresponding {\it in-situ} reflection high-energy electron diffraction (RHEED) intensity plot (see SM for details). In general, the ordinary Hall effect (OHE), AHE, and the hump-like features which are observed near the coercive field can contribute to the Hall resistivity $\rho_{xy}$, which is thus given by $\rho_{xy}=\rho_{\rm OHE} + \rho_{\rm AHE} + \rho_{\rm hump}$~\cite{Matsuno16, Sohn18, Sohn20}. The OHE term, which is proportional to an applied magnetic field, was subtracted from all Hall resistivity data presented in this paper. As $\rho_{\rm hump}$ is non-zero only near the coercive field, $\rho_{\rm AHE}$ can be determined from the saturated $\rho_{xy}$ in the high field limit.

The thickness dependence of $\rho_{\rm AHE}$ is shown in Fig. 3(b). Here, $\rho_{\rm AHE}$ decreased as the thickness became thinner and eventually took on a negative value with a sign change in between. We also performed ionic liquid gating of SRO ultrathin films to investigate how the change in the chemical potential affects $\rho_{xy}$. Figure 3(c) shows the results of ionic gating experiments for 4 and 5~u.c. SRO thin films measured at 10 and 50~K, respectively. In both cases, the magnitude of $\rho_{xy}$ changed significantly with the gate voltage. Figure 3(d) shows the out-of-plane magnetization of 4, 5, and 7~u.c. SRO thin films. Combining the thickness and temperature-dependent Hall effect results with the magnetization data, we can derive $\rho_{\rm AHE}$ $vs.$ $M$ data, as plotted in Fig. 3(e) (see SM for details). It should be noted that the data points fall on a single line, implying that $M$ may be the key parameter for $\rho_{\rm AHE}$. The sign reversal of $\rho_{\rm AHE}$ can also be clearly observed as the magnetization approaches $0.5$~$\mu_B$/Ru.

It is worth noting that the sign-switching behavior of $\rho_{\rm AHE}$ is highly unusual~\cite{Fang03}. To understand the origin of the unusual AHE, we conducted a tight-binding model analysis combined with first-principles calculations. Figure 4(a) shows the band structure in the effective six-band tight-binding model (relevant to 1 u.c. SRO), the parameters of which were adjusted to describe the FS in Fig. 2(c) with a magnetization $M$~=~$0.33$~$\mu_B$/Ru. With magnetization and SOC, several bands are hybridized near the points at which they are crossed. We find that two different origins of symmetry-protected nodal structures, which we call as NL (white circle) and QBC (black circle), are formed at the crossing points. The NLs and QBCs of spin-polarized bands, respectively, become the sources of the large Berry curvature when they are gapped due to SOC. We note that the lowest lying QBC band at the M point is the $A$ band that we observed in Fig. 2(f).
	
To explain the characteristics of NL and QBC in details, a band structure along the $M$-$X$-$M$ line is presented in Fig. 4(b). For simplicity, we focus on three out of the six bands, in which two majority bands are derived from $d_{xz,yz}$ orbitals (red) and one minority band originates from the $d_{xy}$ orbital (blue). The properties of the system as a whole can be understood simply by doubling the number of bands.

When the crossing between bands with opposite spin-polarization exists, it generally leads to 2D NLs in spin-polarized systems~\cite{jin2020two,zhou2021time,jin2020ferromagnetic}. In the absence of SOC, the majority and minority bands are not hybridized, so that NLs form when they intersect~\cite{jin2020ferromagnetic}. Meanwhile, due to the tetragonal crystalline symmetry of thin films, QBCs can appear at the $\Gamma$ and $M$ points with $C_{4v}$ point group symmetry made up of bands with the same spin characte~\cite{sun2009topological, chong2008effective}. These NLs and QBCs are generic nodal structures that exist in the spin-polarized bands of layered 2D perovskite structures (See Methods for details). Figure 4(c) shows the configuration of NLs and QBCs in the momentum space formed by three out of the six bands in Fig. 4(b). When SOC is turned on, these nodal structures are gapped and Berry curvature is thus generated, which leads to the AHE (see Methods for details).

Multiple Berry curvature sources appearing near the Fermi level can induce fluctuating Berry curvature distribution with alternating signs.
The upper half of Fig. 5(a) shows that QBCs formed by minority bands induce negative Berry curvature while the others generate positive Berry curvature. For better visualization, we plot the Berry curvature distribution of 1~u.c. SRO. in Fig. 5(b). The resulting Berry curvature distribution is similar to a previously reported one~\cite{Groenendijk2020}.

Since the sign-alternating Berry curvature distribution is generated by the topological band crossings, it would be desirable to experimentally show the Berry curvature behavior in the momentum space. During the past decade, the connection between Berry curvature and orbital angular momentum (OAM) has been firmly established~\cite{Xiao07, Go18}. In addition, measurement of OAM (thus Berry curvature) with CD ARPES has been well studied~\cite{Liu11, Park12_2, Cho18, 2020momentumspace,Park12, Schuler20}. In fact, CD-ARPES has been used to study chiral structures of pseudo-spins of topological bands~\cite{Park12,Schuler20,Chen13,Wang13,Wang11}. Therefore, the CD-ARPES in Fig. 2(c) may be compared with calculated OAM of our 1~u.c. tight-binding model. The upper half of Fig. 5(c) shows a clear match between the CD-ARPES data and OAM distribution. Positive CD intensity and OAM are observed near the $M$ point but they turn negative away from the $M$ point. Eventually, both experimental and theoretical results turn to zero close to the $\Gamma$ point. Interestingly, we found that both of the calculated OAM and Berry curvature flip their signs along the $\Gamma$-$M$ high-symmetry line due to the intrinsic nature of QBC (See Methods for the relationship between OAM and Berry curvature).

To demonstrate that the QBCs and NLs are present, and also the associated sign-switching of the AHE signal is a generic property of SRO ultrathin films independent of film thickness, we also constructed a tight-binding model for 4~u.c. SRO and compared it with the effective six-band model. The hopping parameters for the 4~u.c. tight-binding Hamiltonian were extracted from first-principles calculations and adjusted further by fitting the experimentally observed FSs. For realistic calculations, the atomic positions of 4~u.c. SRO were obtained from the experiment using coherent Bragg rod analysis (COBRA)~\cite{Sohn18} (See Methods for the detailed description of the model). The lower half of Fig. 5(a) shows the FS and Berry curvature derived from the 4~u.c. model, where large Berry curvatures are generated by gapped QBCs and NLs. It is interesting to note that the 1~u.c. and 4~u.c. models predict similar FS and Berry curvature distributions. Both models also show consistent OAM distributions as shown in the right half in Fig. 5(c). 


Due to multiple sources of enhanced Berry curvature generated by QBCs and NLs as shown in Figs. 5(a) and (b), the sign and magnitude of the AHE can vary depending on the energies of the nodal structures relative to the Fermi level (see Methods for details). We show that the sign and magnitude of the AHE can be controlled by different parameters, such as the magnetization and chemical potential. The calculated anomalous Hall conductivity (AHC), $\sigma_{\rm AHE} = \rho_{xy}/(\rho_{xy}^2+\rho_{xx}^2)$ using two tight-binding models for 1 u.c. and 4~u.c. SRO as a function of magnetization, well matches the experimental results (Fig. 5(d)). It is noteworthy that the sign of the AHC reverses when the amplitude of the magnetization varies. Note that the sign-switching AHC appears in both 1~u.c. and 4~u.c. models.

The chemical potential dependence of AHC was also computed using the tight-binding model (Fig. 5(e)). According to the calculation with $M$~$\sim$~$0.4$~$\mu_B$/Ru, AHC increases with the chemical potential. This tendency is in accordance with the ionic liquid gating measurements on 4 and 5~u.c. SRO ultrathin films (Fig. 5(e), inset). Given that both films have a magnetization of $M$~$\sim$~$0.4$~$\mu_B$/Ru, AHC tends to increase as the gate voltage increases. We belive that these results can support that the sign changing AHE comes from competition between Berry curvature sources from multiple topological features as magnetization and chemical potential change. 
Especially, we found that as the chemical potential increases, the positive Berry curvatures from NL and majority band QBC dominate the negative Berry curvature from minority band QBC, which eventually leads to the sign-reversal of AHC.
Interestingly, AHC is not proportional to the net magnetization and the sign of AHC can change with a small variation of magnetization and chemical potential as schematically shown in Fig. 5(f).
	

For last decades, nonmonotonous AHE of SRO has been studied not only in the ultrathin limit (i.e. 2D limit)~\cite{Groenendijk2020, Matsuno16, Sohn18, Schultz09} but also in the 3D limit like single crystals and thick films~\cite{Fang03, mathieu2004scaling}. Interestingly, the sign-changing AHE v.s. magnetization in the 2D SRO shows a similar tendency to that of 3D SROs.
Such a similarity of AHE between 2D and 3D SROs is quite surprising because 3D topological band structure is generally not adiabatically connected to the band structure of 2D layers. As the interlayer coupling generates several band inversion processes, which strongly modify the nodal structures, the topological nodal structures and related Berry curvature distribution of 2D and 3D are not smoothly connected in general. Especially, QBCs and NLs of 2D SRO cannot be understood from a simple 2D confinement of magnetic monopoles in 3D SRO. Generic evolution pattern of the band structure between 2D and 3D limits is further examined using tight-binding model shown in SM.

Because the topological nodal structures (QBCs and NLs) are protected only by symmetries, similar nodal structures can appear in any 2D systems sharing the same space group symmetry as SRO. To confirm this idea, we also performed first-principles calculations on monolayer SrCoO$_3$, another ferromagnetic perovskite oxide, in which the sign-switching AHE was recently observed in thin films~\cite{Zhang19}. Interestingly, the NLs and QBCs were also located close to the Fermi energy and generated Berry curvature with opposite signs, as in SRO (see SM for details). 

Let us briefly mention the stability of NLs and QBCs under rotation or tilting of oxygen octahedra. NLs are stable against rotation or tiling as they do not affect spins when SOC is absent. QBCs are stable against rotation distortion that preserves $C_4$ rotation. Although the tilting breaks $C_4$ symmetry, the topological properties of QBCs should still remain intact as long as the SOC-induced gap is larger than the gap caused by tiling. We believe that a similar mechanism may be applicable to understanding mysterious AHE in other ferromagnetic perovskite oxides and emergent interfacial ferromagnetism.

\section*{3. Conclusion}
In conclusion, by combining ARPES, transport measurements, and theoretical analysis, we demonstrated the topological band structure of ferromagnetic SRO thin films. In particular, the band structures of SRO film in the ultrathin 2D limit were observed and defined for the first time. Through theoretical analysis, we also showed that the spin-polarized bands of 2D ferromagnets generally possess nodal points and lines that become the source of enhanced Berry curvature. Comparing the measured band structure with the tight-binding model, we identified the Berry curvature hot spots originating from multi-nodal structures, which led to the unconventional AHE. Competing contributions from different Berry curvature hot spots induce a sign-changing AHE, which can be controlled by varying the film thickness, temperature, magnetization, and chemical potential. We believe that our findings will open up new avenues for investigating novel transport phenomena driven by symmetry-protected nodal structures of 2D magnetic systems, and facilitate the development of magnetic devices based on the engineering of magnetic topological band structures.

\section*{4. Methods}

\subsection{SrRuO$_3$ thin film fabrication}
SRO ultrathin films were grown on TiO$_2$-terminated SrTiO$_3$ (STO) single crystal substrates using pulsed laser deposition (PLD). TiO$_2$-terminated STO substrates from Shinkosha were used for SRO thin film growth. To dissolve the Sr compounds that can form on the surface of STO substrates, the STO substrates were prepared by deionized (DI) water etching~\cite{Koster98}. The DI water-treated STO substrates were pre-annealed $in$-$situ$ at $1070\,^{\circ}{\rm C}$ for 30 min with an oxygen partial pressure (PO$_2$) of $5\times$10$^{-6}$ Torr. We deposited epitaxial SRO thin film in an oxygen partial pressure of PO$_2$=100~mTorr; the growth temperature of the STO substrate was $700\,^{\circ}{\rm C}$. A KrF excimer laser (wavelength: 248 nm) irradiated a stoichiometric SRO target with a fluence of 1-2~J/cm$^{2}$ and repetition rate of 2~Hz. RHEED was used to monitor the growth dynamics. To clean the surface of SRO thin films, we post-annealed them at $550\,^{\circ}{\rm C}$ for 10~min.

\subsection{Transport and magnetic measurements}
For the Hall effect measurement of SRO thin films, we prepared a $60$-nm-thick Au electrode on top of the SRO thin films with a Hall bar geometry using an electron beam evaporator. Electric transport measurement was carried out using a physical property measurement system (Quantum Design Inc.). The magnetic characterization was performed using superconducting quantum interface device (SQUID) magnetometry with out-of-plane geometry. Given that the easy axis of the SRO ultrathin film is perpendicular to the thin film on STO (001) substrate, we measured the out-of-plane magnetization via SQUID magnetometry~\cite{klein96, Schultz09}. For ionic liquid gating, we used diethylmethyl(2-methoxyethyl)ammonium bis(trifluoromethylsulfonyl)imide (DEME-TFSI) as an electrolyte. We applied a gate voltage at 260~K for 30~min to form an electric double layer.

\subsection{Angle-resolved photoemission spectroscopy}
{\it In-situ} ARPES measurements were performed at 10~K using the home lab system equipped with a Scienta DA30 analyzer and a discharge lamp from Fermi instrument. He I$\alpha$ ($hv=21.2$~eV) light was mostly used. For the photon energy-dependent and CD studies, ARPES measurements were performed at the beam line 4.0.3 end station of the Advanced Light Source equipped with a Scienta R8000. Photon energy dependent ARPES measurements were performed at 10~K with photon energies ranging from 50 to 120~eV. CD-ARPES measurements were performed at 20~K with left- and right-circularly polarized (LCP and RCP, respectively) 80~eV light. The light incident angle was $45 \pm 10^{\circ}$.

Spin polarization was measured with a spin-resolved ARPES system in our laboratory. The system was equipped with a SPECS PHOIBOS 225 analyzer and a very low energy electron diffraction (VLEED) spin detector. For the spin detector, an oxidized iron film deposited on W(100) was used as the scattering target. He I$\alpha$ ($hv=21.2$~eV) light was used as the light source. The energy resolution was set to $\sim$ 60~meV with a pass energy of 10~eV. We used a Sherman function value of 0.29~$\pm$~0.01 to obtain the spin polarization.

\subsection{First-principles calculations}
We performed first-principles density functional theory (DFT) calculations with the generalized gradient approximation using the Vienna {\it ab-initio} simulation package (VASP) \cite{Kresse96,Kresse99}. Perdew-Becke-Erzenhof parametrization \cite{Perdew96} for the exchange-correlation functional and the projector augmented wave method \cite{Blochl94} were used, with an energy cutoff of 500 eV and a $k$-point sampling on a $8\times 8\times 1$ grid. The electronic structures of 4 u.c. SRO film were obtained by analyzing a slab of the SRO/STO heterostructure consisting of 4 layer of SRO and 4.5 layers of STO (including an additional SrO at the surface), and a vacuum of 19~\AA. We used the atomic positions obtained from the experimental using COBRA~\cite{Sohn18}, having a $\sqrt{2}\times\sqrt{2}$ in-plane u.c. to include octahedral rotation. The ferromagnetic ground state was obtained with an average Ru magnetic moment of 1.2 $\mu_{B}$. The tight-binding parameters for Ru-$d$-derived bands of the SRO/STO heterostructure were obtained using the Wannier90 package~\cite{Pizzi20}. For the GGA+$U$ calculations presented in the supplementary materials, we used the rotationally invariant form of the on-site Coulomb interaction \cite{Liechtenstein95} with $U$~=~3.23~eV and $J$~=~0.74~eV from constrained random-phase approximation calculation~\cite{vaugier2012hubbard}.

The electronic structure of 2D SrCoO$_{3}$ was calculated using a slab geometry of 1.5 layers of SrCoO$_{3}$ (including an additional SrO layer at the surface) with a vacuum of 10 {\AA} in which the experimental atomic positions of bulk SrCoO$_{3}$ were used \cite{Bezdicka93}. The space group of the atomic structure including the vacuum was $P4/mmm$ (124). The energy cutoff of 600 eV and $k$-point sampling on a $10\times 10\times 1$ grid were used. SOC was included. The ferromagnetic ground state was obtained with the Co magnetic moment of 1.6~$\mu_{B}$, in which the out-of-plane magnetization direction was preferred with magnetic anisotropy energy of 1 meV. The Berry curvature was calculated using Wannier90 package~\cite{Pizzi20} based on the tight-binding Hamiltonian constructed from the Wannier functions of Co-$d$ and O-$p$ derived bands.

\subsection{Effective tight-binding model for monolayer SRO}
In the main text, we assumed that electronic structure of 4 u.c. SRO can be well described by effective two-dimensional models, which are supported by $k_z$-dependent ARPES measurement. Following the Slater-Koster method, tight-binding Hamiltonian of ferromagnetic monolayer SrRuO$_3$ is constructed as follows.
\begin{align}
	H=\sum_\textbf{k}[(\epsilon_{\textbf{k}\sigma}^a )\delta_{ab} \delta_{\sigma\sigma'}+f_{\textbf{k}}^{ab} \delta_{\sigma\sigma'}+i\lambda\epsilon^{abc} \tau_{\sigma\sigma'}^c] d_{\textbf{k}a\sigma}^{\dagger} d_{\textbf{k}b\sigma},
\end{align}
where
\begin{align}
\epsilon_{k\sigma}^{1=yz}&=-2t_1\cos{k_y}-2t_2\cos{k_x}-4t_3\cos{k_x}\cos{k_y}-m\tau_{\sigma\sigma}^z, \nonumber\\
\epsilon_{k\sigma}^{2=xz}&=-2t_1\cos{k_x}-2t_2\cos{k_y}-4t_3\cos{k_x}\cos{k_y}-m\tau_{\sigma\sigma}^z, \nonumber\\
\epsilon_{k\sigma}^{3=xy}&=-2t_1(\cos{k_x}+\cos{k_y})-4t_4\cos{k_x}\cos{k_y}-m\tau_{\sigma\sigma}^z, \nonumber\\
f_{\textbf{k}}^{12}&=-4f\sin{k_x}\sin{k_y}=f_{\textbf{k}}^{21},
\end{align}
where $t_1$ and $t_2$ are the amplitudes for nearest neighbor interactions, and $t_3$, $t_4$, and $f$ are the amplitudes of next nearest neighbor interactions. $\lambda$ denotes the amplitude of SOC and $m$ is the amplitude of Zeeman interaction. We fit tight-binding parameters with ARPES data to describe the anomalous Hall conductivity quantitatively (see Table I).
\begin{center}
\begin{table}
\begin{tabular}{|c | c | c | c | c | c | c| c |c|}
\hline
$E_F$  & $t_1$   & $t_2$   & $t_3$    & $t_4$ & $m$   & $f$ & $\lambda$ & $M$  \\
\hline
0.3 & 0.28 & 0.03 & 0.018 & 0.04 & 0.08   & 0.015  & $0.045$ & $0.33~\mu_B/Ru$ \\
\hline
\end{tabular}
\caption{Parameters for tight-binding model Hamiltonian describing 1 u.c. SRO. Here the units are eV except for magnetization.}
\label{table1}
\end{table}
\end{center}

With the Hamiltonian, we can directly calculate Berry curvature of each energy band $\Omega_{n}(\textbf{k})$ by applying the following formula:
\begin{align}
\Omega_{n}(\textbf{k})=i\sum_{n\neq n'}\frac{\langle n|\partial H/\partial k_x|n'\rangle\langle n'|\partial H/\partial k_y|n\rangle-(x\leftrightarrow y)}{(\epsilon_{n}(\textbf{k})-\epsilon_{n'}(\textbf{k}))^{2}},
\end{align}
where $\epsilon_{n}$ is the energy of the $n$-th band represented by $|n\rangle$.
The anomalous Hall conductivity $\sigma_{xy}$ is given by integrating the Berry curvature over the Brillouin zone below the Fermi energy:
\begin{align}
    \sigma_{xy}=\frac{c}{2\pi}\sum_{n=1}^{N}\int dk_xdk_y \Omega_{n}({\bf k})\theta(\epsilon_{n}(\textbf{k})-E_f),
\end{align}
where $\theta(x)$ is step function that is $0 (1)$ when $x\geq0 ~(x<0)$ Also, $c=(R_{\rm H}\times l)^{-1}$, where $R_{\rm H}$ is Hall resistance and $l$ is the thickness of SRO thin film.

\subsection{Tight-binding model for 4 u.c. SRO}
We construct 4~u.c. SRO tight-binding model where the parameters are extracted from the DFT calculation and adjusted further by fitting the experimentally observed FSs.
The DFT calculation is performed based on the structure of 4~u.c. SRO ultrathin films, where the atomic structure information is obtained by COBRA that determines the layer-by-layer atomic positions of SRO ultrathin films. In the tight-binding Hamiltonian, only $t_{2g}$ orbitals are considered since they accounts for the most of the density of states near the Fermi level. Also, the tiny lattice rotation is neglected and the tetragonal structure is assumed.
The Hamiltonian $\hat{\mathcal{H}}=\psi^{\dagger}H\psi$ is described by the matrix $H$ and the basis $\psi^{\dagger}$, where
\begin{align}
H&=
\begin{bmatrix}
E_1 & t_{12} & 0& 0 \\
t_{12}^{\dagger}&  E_2 & t_{23}& 0 \\
0 & t_{32}^{\dagger} & E_3 & t_{34} \\
0 & 0 & t_{34}^{\dagger}& E_4
\end{bmatrix}, \\
\psi^{\dagger}&=(d_{1,\textbf{k}a\sigma}^{\dagger},d_{2,\textbf{k}a\sigma}^{\dagger},d_{3,\textbf{k}a\sigma}^{\dagger},d_{4,\textbf{k}a\sigma}^{\dagger}).
\end{align}
Here $E_n$ describes the intralayer part of the Hamiltonian $E_n d_{n,\textbf{k}a\sigma}^{\dagger} d_{n,\textbf{k}b\sigma}$ where $n$ denotes the index for the SRO layer. $t_{nm}$ and $t_{mn}$ describe the hopping interaction between $n$-th layer and $m$-th layer. Let us note that $t_{mn}=t_{nm}^{\dagger}$, due to the Hermiticity of the Hamiltonian.

The intralayer Hamiltonian $\hat{\mathcal{H}}_{n, {\rm intra}}=E_n d_{n,\textbf{k}a\sigma}^{\dagger} d_{n,\textbf{k}b\sigma}$ is written as
\begin{align}
&\hat{\mathcal{H}}_{n, {\rm intra}}\nonumber\\
&=\sum_\textbf{k}[(\epsilon_{n,\textbf{k}\sigma}^a )\delta_{ab} \delta_{\sigma\sigma'}+f_{\textbf{k}}^{ab} \delta_{\sigma\sigma'}+i\lambda\epsilon^{abc} \tau_{\sigma\sigma'}^c] d_{n,\textbf{k}a\sigma}^{\dagger} d_{n,\textbf{k}b\sigma},
\end{align}
where
\begin{align}
\epsilon_{n, k\sigma}^{1=yz}&=-2t_{1,n}\cos{k_y}-2t_{2,n}\cos{k_x}-m\tau_{\sigma\sigma}^z, \nonumber\\
\epsilon_{n, k\sigma}^{2=xz}&=-2t_{1,n}\cos{k_x}-2t_{2,n}\cos{k_y}-m\tau_{\sigma\sigma}^z, \nonumber\\
\epsilon_{n, k\sigma}^{3=xy}&=U-2t_{3,n}(\cos{k_x}+\cos{k_y})-4t_{4, n}\cos{k_x}\cos{k_y}\nonumber\\
&-m\tau_{\sigma\sigma}^z, \nonumber\\
f_{\textbf{k}}^{12}&=-4f\sin{k_x}\sin{k_y}=f_{\textbf{k}}^{21},
\end{align}
where $t_{1,n}, t_{2,n},$ and $t_{3,n}$ are the amplitudes of the nearest neighbor interactions, and $t_{4,n}$ and $f$ are the amplitudes of the next nearest neighbor interactions. $U$ is the on-site energy difference between $d_{xy}$ and $d_{xz, yz}$ orbitals, $\lambda$ denotes the amplitude of SOC, and $m$ is the magnitude of Zeeman interaction.

On the other hand, the interlayer hopping interaction is written as
\begin{align}
&t_{nm}d_{n,\textbf{k}a\sigma}^{\dagger} d_{m,\textbf{k}b\sigma}\nonumber\\
&=\sum_\textbf{k}[(\epsilon_{n,\textbf{k}\sigma}^a )\delta_{ab} \delta_{\sigma\sigma'}+f_{n,\textbf{k}}^{ab} \delta_{\sigma\sigma'}] d_{n,\textbf{k}a\sigma}^{\dagger} d_{m,\textbf{k}b\sigma},
\end{align}

where
\begin{align}
\epsilon_{n,k\sigma}^{1=yz}&=-u_{1,n}-u_{2,n}\cos{k_y}, \nonumber\\
\epsilon_{n,k\sigma}^{2=xz}&=-u_{1,n}-u_{2,n}\cos{k_x}, \nonumber\\
\epsilon_{n,k\sigma}^{3=xy}&=-t_{2,n}, \nonumber\\
f_{n, \textbf{k}}^{13}&=f_{n, \textbf{k}}^{31}=2if\sin{k_x} \nonumber\\
f_{n, \textbf{k}}^{23}&=f_{n, \textbf{k}}^{32}=2if\sin{k_y}.
\end{align}

Here we assume $n>m$ without losing generality. $u_{1,n}$ is the amplitude of the nearest neighbor interaction and $u_{2,n}$ is the amplitude of the next nearest neighbor interaction.
The parameters of the effective tight-binding Hamiltonian are summarized in Table II.
\begin{center}
\begin{table}
\begin{tabular}{|c |c |c | c | c | c | c | c | c| c |c |}
\hline
 & $t_{1,n}$ & $t_{2,n}$ & $t_{3,n}$ & $t_{4,n}$ & $U$ & $u_{1,n}$  & $u_{2,n}$ &$f$ & $\lambda$ & $E_F$    \\
\hline
$n=1$ &$0.367$ & $-0.03$ & $0.35$ & $0.12$ &      &$0.32$   & $0.15$   & 		  &            & \\
$n=2$ &$0.348$ & $-0.03$ & $0.35$ & $0.12$ & 0.2 &$0.253$ & $0.15$   & $-0.03$ & $0.06$ & $0.75$\\
$n=3$ &$0.290$ & $-0.03$ & $0.30$ & $0.1$   &      &$0.213$ & $0.15$   &             &            &  \\
$n=4$ &$0.250$ & $-0.03$ & $0.25$ & $0.01$ &      &            &             &              &            & \\
\hline
\end{tabular}
\caption{Parameters for 4~u.c. effective tight-binding model Hamiltonian. Here the units are eV.}
\label{table2}
\end{table}
\end{center}

\subsection{Quadratic Band Crossing (QBC)}
Negative AHC under small magnetization in SRO can be explained by the $\textbf{k}\cdot\textbf{p}$ Hamiltonian expanded near QBCs.
Neglecting rotation and tilting, a perovskite thin film has a square lattice structure. Without SOC, space group generators are four-fold rotation about the z-axis ($C_{4z}$), and three mirrors $M_x$, $M_y$, and $M_z$, where $M_{\alpha}$ changes the sign of the $\alpha$ coordinate. The $\textbf{k}\cdot\textbf{p}$ Hamiltonian valid near $\Gamma$ or $M$ points is written as
\begin{align}
H_{\text{QBC}}=E_0({\textbf{k}})+\beta(k_x^2-k_y^2)\tau_x+2\beta k_xk_y\tau_y,
\end{align}
where $(1,0)^t=|d_{xz} \rangle+i|d_{yz} \rangle \equiv |+1\rangle$ and $(0,1)^t=|d_{xz}\rangle-i|d_{yz}\rangle\equiv|-1\rangle$ are the basis near the QBC. We now discuss the influence of SOC. The on-site spin orbit coupling term is expressed as $\lambda L\cdot S=\lambda(L_x S_x+L_y S_y+L_z S_z)$. Since $\langle+1,\sigma|L_zS_z |+1,\sigma\rangle=-\langle-1,\sigma|L_zS_z |-1,\sigma\rangle=\lambda\sigma_z$, the effective Hamiltonian with SOC is written as follows
\begin{align}
H&=H_{\text{QBC}}+H_{\text{SOC}}\nonumber\\
&=E_0({\textbf{k}})+\beta(k_x^2-k_y^2)\tau_x+2\beta k_xk_y\tau_y+\lambda\sigma_z\tau_z,
\end{align}
where $\sigma_z$ is $+1(-1)$ for majority (minority) bands. Thus the gap is opened and the relevant Berry curvature is given by $\Omega(\textbf{k})=\frac{\lambda\sigma_z}{\beta}\frac{2k^2}{(k^4+\lambda^2/\beta^2)^{3/2}}$, where the total Berry curvature $\int\Omega(\textbf{k})d\textbf{k}$ is quantized to $2\pi$. Unlike Berry curvature induced by linear band crossing which is maximum at $|\textbf{k}|=0$, Berry curvature induced by QBC is zero at $|\textbf{k}|=0$ and maximum at $|\textbf{k}|=(2\lambda^2/\beta^2)^{1/4}$.

In SRO, although QBC formed by majority bands is closer to the Fermi level than that of minority bands, the former induces smaller Berry curvature. Since the bands at M point are all hole-like, Fermi level is between the majority bands for $k_1<|\textbf{k}|<k_2$ and minority bands for $k_1+\Delta_1<|\textbf{k}|<k_2+\Delta_2$. Then Berry curvature that contributes to anomalous Hall effect is $\int_{k_1}^{k_2}\Omega_{maj}(\textbf{k})d\textbf{k}+\int_{k_1+\Delta_1}^{k_2+\Delta_2}\Omega_{min}(\textbf{k})d\textbf{k}=\int_{k_1}^{k_1+\Delta_1}\frac{\lambda}{\beta}\frac{2k^2}{(k^4+\lambda^2/\beta^2)^{3/2}}-\int_{k_2}^{k_2+\Delta_2}\frac{\lambda}{\beta}\frac{2k^2}{(k^4+\lambda^2/\beta^2)^{3/2}}$.
For small magnetization, Hall conductance can be negative, whereas for large magnetization, Berry curvature induced by minority bands becomes negligible and Hall conductance becomes positive.

\subsection{Nodal line (NL)}
Without SOC, the crossing between spin up and spin down bands form a nodal line. The corresponding Hamiltonian is written as
\begin{align}
H_{\text{NL}}=
\begin{pmatrix}
   E_{\uparrow}(\textbf{k}) & 0 \\
   0 & E_{\downarrow}(\textbf{k})
\end{pmatrix}
=E_0(\textbf{k})+\Delta E(\textbf{k})\sigma_z,
\end{align}
where $(1,0)^t=|\psi_1(\textbf{k}),\uparrow\rangle,(1,0)^t=|\psi_2 (\textbf{k}),\downarrow\rangle.$ Due to the SOC, spin U(1) symmetry is broken and the gap is opened. SOC term is expressed as $\lambda(L_x S_x+L_y S_y+L_z S_z )=\frac{\lambda}{2} (L_+ S_-+L_- S_+ )+\lambda L_z S_z$,
\begin{align}
&\langle\psi_2 (\textbf{k}),\downarrow| L_+ S_- |\psi_1 (\textbf{k}),\uparrow\rangle=\langle \psi_2 (\textbf{k})| L_+ |\psi_1 (\textbf{k})\rangle \nonumber \\
&=\langle\psi_1 (\textbf{k}),\uparrow|L_- S_+ |\psi_2 (\textbf{k}),\downarrow\rangle^*=\langle \psi_1 (\textbf{k})|L_- |\psi_2 (\textbf{k})\rangle^*\nonumber \\
&=\frac{2}{\lambda} [\alpha(\textbf{k})+i\beta(\textbf{k})]. \\
&\therefore H_{\text{SOC}}=\alpha(\textbf{k}) \sigma_x+\beta(\textbf{k}) \sigma_y
\end{align}
Thus, the total Hamiltonian $H=H_{\text{NL}}+H_{\text{SOC}}=E_0 (\textbf{k})+\Delta E(\textbf{k}) \sigma_z+\alpha(\textbf{k})\sigma_x+\beta(\textbf{k})\sigma_y$. If Fermi energy lies between two energy bands, the Berry curvature is given by
\begin{align}
\Omega(\textbf{k})=\frac{i \langle-|\partial H/(\partial k_x )|+\rangle\langle+|\partial H/(\partial k_y )|-\rangle+ c.c.}{(\epsilon_--\epsilon_+ )^2},
\end{align}
 where $|+\rangle(|-\rangle)$ is higher (lower) energy state. Similar to the calculation for QBC, the total Berry curvature is quantized to $\pm 2\pi$.

Let us note, however, that some nodal lines are intact under SOC. This is due to $M_z:z\rightarrow-z$ symmetry, which is local in both real and momentum space in two dimensions: $M_z |d_{xy}\rangle=|d_{xy}\rangle, M_z |d_{xz,yz}\rangle=-|d_{xz,yz}\rangle$, represented as $M_z=$$\begin{pmatrix}
   -1 & 0 & 0\\
   0 & -1 & 0\\
   0 & 0 &1
\end{pmatrix}$ $\otimes i\sigma_z$. Two bands with different mirror eigenvalue do not hybridize.

\subsection{Relationship between OAM and BC}

Here, we explain that orbital angular momentum (OAM) and Berry curvature show similar distributions.
For simplicity, let us consider a two-band $k\cdot p$ Hamiltonian where the gap is opened by the spin orbit coupling.

The Hamiltonian is written as
\begin{align}
H=E_0({\bf k})+\beta(k_x^2-k_y^2)\tau_x+2\beta k_xk_y\tau_y+\lambda\sigma_z\tau_z,
\end{align}
where $\sigma_z=\pm1$, which represents the direction of spin polarization.
The energy spectrum is given as $E=E_0\pm\sqrt{\lambda^2+\beta^2k^4}$ where the gap due to the spin orbit coupling is proportional to $\lambda$.

The Berry curvature $\Omega({\bf k})$ of the occupied band is
\begin{align}
\Omega({\bf k})=2\lambda \frac{\beta^2k^2}{(\lambda^2+\beta^2k^4)^{3/2}}\sigma_z.
\end{align}

On the other hand, the $z$-component of the OAM of the occupied band $\langle L_z\rangle=\langle occ|\tau_z|occ\rangle$ is
\begin{align}
L_z({\bf k})=2\lambda \frac{\lambda+\sqrt{\lambda^2+\beta^2k^4}}{(\lambda+\sqrt{\lambda^2+\beta^2k^4})^2+\beta^2k^4}\sigma_z.
\end{align}
The Berry curvature and OAM show the similar tendency since both of them are induced by the spin-orbit coupling that opens the gap at QBC. In particular, we note that the Berry curvature and OAM have the same sign determined by $\lambda$.

\acknowledgments
We gratefully acknowledge discussions with J .R. Kim.
This work is supported by IBS-R009-D1 and IBS-R009-G2 through the IBS Center for Correlated Electron Systems.B.-J.Y. was supported by the Institute for Basic Science in Korea (Grant No. IBS-R009-D1),
the National Research Foundation of Korea (NRF) grant funded by the Korea government (MSIT) (No.2021R1A2C4002773),
and the U.S. Army Research Office and Asian Office of Aerospace Research \& Development (AOARD) under Grant number W911NF-18-1-0137.The Advanced Light Source is supported by the Office of Basic Energy Sciences of the U.S. DOE under Contract No. DE-AC02-05CH11231. S.Y.P was supported by the National Research Foundation of Korea (NRF) grant funded by the Korea government (MSIT) (No. 2020R1F1A1076742).


\begin{thebibliography}{61}%
	\makeatletter
	\providecommand \@ifxundefined [1]{%
		\@ifx{#1\undefined}
	}%
	\providecommand \@ifnum [1]{%
		\ifnum #1\expandafter \@firstoftwo
		\else \expandafter \@secondoftwo
		\fi
	}%
	\providecommand \@ifx [1]{%
		\ifx #1\expandafter \@firstoftwo
		\else \expandafter \@secondoftwo
		\fi
	}%
	\providecommand \natexlab [1]{#1}%
	\providecommand \enquote  [1]{``#1''}%
	\providecommand \bibnamefont  [1]{#1}%
	\providecommand \bibfnamefont [1]{#1}%
	\providecommand \citenamefont [1]{#1}%
	\providecommand \href@noop [0]{\@secondoftwo}%
	\providecommand \href [0]{\begingroup \@sanitize@url \@href}%
	\providecommand \@href[1]{\@@startlink{#1}\@@href}%
	\providecommand \@@href[1]{\endgroup#1\@@endlink}%
	\providecommand \@sanitize@url [0]{\catcode `\\12\catcode `\$12\catcode
		`\&12\catcode `\#12\catcode `\^12\catcode `\_12\catcode `\%12\relax}%
	\providecommand \@@startlink[1]{}%
	\providecommand \@@endlink[0]{}%
	\providecommand \url  [0]{\begingroup\@sanitize@url \@url }%
	\providecommand \@url [1]{\endgroup\@href {#1}{\urlprefix }}%
	\providecommand \urlprefix  [0]{URL }%
	\providecommand \Eprint [0]{\href }%
	\providecommand \doibase [0]{https://doi.org/}%
	\providecommand \selectlanguage [0]{\@gobble}%
	\providecommand \bibinfo  [0]{\@secondoftwo}%
	\providecommand \bibfield  [0]{\@secondoftwo}%
	\providecommand \translation [1]{[#1]}%
	\providecommand \BibitemOpen [0]{}%
	\providecommand \bibitemStop [0]{}%
	\providecommand \bibitemNoStop [0]{.\EOS\space}%
	\providecommand \EOS [0]{\spacefactor3000\relax}%
	\providecommand \BibitemShut  [1]{\csname bibitem#1\endcsname}%
	\let\auto@bib@innerbib\@empty
	\bibitem [{\citenamefont {Burkov}(2014)}]{Burkor14}%
	\BibitemOpen
	\bibfield  {author} {\bibinfo {author} {\bibfnamefont {A.}~\bibnamefont
			{Burkov}},\ }\href@noop {} {\bibfield  {journal} {\bibinfo  {journal} {Phys.
				Rev. Lett.}\ }\textbf {\bibinfo {volume} {113}},\ \bibinfo {pages} {187202}
		(\bibinfo {year} {2014})}\BibitemShut {NoStop}%
	\bibitem [{\citenamefont {Ye}\ \emph {et~al.}(2018)\citenamefont {Ye},
		\citenamefont {Kang}, \citenamefont {Liu}, \citenamefont {Von~Cube},
		\citenamefont {Wicker}, \citenamefont {Suzuki}, \citenamefont {Jozwiak},
		\citenamefont {Bostwick}, \citenamefont {Rotenberg}, \citenamefont {Bell}
		\emph {et~al.}}]{Ye18}%
	\BibitemOpen
	\bibfield  {author} {\bibinfo {author} {\bibfnamefont {L.}~\bibnamefont
			{Ye}}, \bibinfo {author} {\bibfnamefont {M.}~\bibnamefont {Kang}}, \bibinfo
		{author} {\bibfnamefont {J.}~\bibnamefont {Liu}}, \bibinfo {author}
		{\bibfnamefont {F.}~\bibnamefont {Von~Cube}}, \bibinfo {author}
		{\bibfnamefont {C.~R.}\ \bibnamefont {Wicker}}, \bibinfo {author}
		{\bibfnamefont {T.}~\bibnamefont {Suzuki}}, \bibinfo {author} {\bibfnamefont
			{C.}~\bibnamefont {Jozwiak}}, \bibinfo {author} {\bibfnamefont
			{A.}~\bibnamefont {Bostwick}}, \bibinfo {author} {\bibfnamefont
			{E.}~\bibnamefont {Rotenberg}}, \bibinfo {author} {\bibfnamefont {D.~C.}\
			\bibnamefont {Bell}}, \emph {et~al.},\ }\href@noop {} {\bibfield  {journal}
		{\bibinfo  {journal} {Nature}\ }\textbf {\bibinfo {volume} {555}},\ \bibinfo
		{pages} {638} (\bibinfo {year} {2018})}\BibitemShut {NoStop}%
	\bibitem [{\citenamefont {Groenendijk}\ \emph {et~al.}(2020)\citenamefont
		{Groenendijk}, \citenamefont {Autieri}, \citenamefont {van Thiel},
		\citenamefont {Brzezicki}, \citenamefont {Hortensius}, \citenamefont
		{Afanasiev}, \citenamefont {Gauquelin}, \citenamefont {Barone}, \citenamefont
		{van~den Bos}, \citenamefont {van Aert}, \citenamefont {Verbeeck},
		\citenamefont {Filippetti}, \citenamefont {Picozzi}, \citenamefont {Cuoco},\
		and\ \citenamefont {Caviglia}}]{Groenendijk2020}%
	\BibitemOpen
	\bibfield  {author} {\bibinfo {author} {\bibfnamefont {D.~J.}\ \bibnamefont
			{Groenendijk}}, \bibinfo {author} {\bibfnamefont {C.}~\bibnamefont
			{Autieri}}, \bibinfo {author} {\bibfnamefont {T.~C.}\ \bibnamefont {van
				Thiel}}, \bibinfo {author} {\bibfnamefont {W.}~\bibnamefont {Brzezicki}},
		\bibinfo {author} {\bibfnamefont {J.~R.}\ \bibnamefont {Hortensius}},
		\bibinfo {author} {\bibfnamefont {D.}~\bibnamefont {Afanasiev}}, \bibinfo
		{author} {\bibfnamefont {N.}~\bibnamefont {Gauquelin}}, \bibinfo {author}
		{\bibfnamefont {P.}~\bibnamefont {Barone}}, \bibinfo {author} {\bibfnamefont
			{K.~H.~W.}\ \bibnamefont {van~den Bos}}, \bibinfo {author} {\bibfnamefont
			{S.}~\bibnamefont {van Aert}}, \bibinfo {author} {\bibfnamefont
			{J.}~\bibnamefont {Verbeeck}}, \bibinfo {author} {\bibfnamefont
			{A.}~\bibnamefont {Filippetti}}, \bibinfo {author} {\bibfnamefont
			{S.}~\bibnamefont {Picozzi}}, \bibinfo {author} {\bibfnamefont
			{M.}~\bibnamefont {Cuoco}},\ and\ \bibinfo {author} {\bibfnamefont {A.~D.}\
			\bibnamefont {Caviglia}},\ }\href
	{https://doi.org/10.1103/PhysRevResearch.2.023404} {\bibfield  {journal}
		{\bibinfo  {journal} {Phys. Rev. Res.}\ }\textbf {\bibinfo {volume} {2}},\
		\bibinfo {pages} {023404} (\bibinfo {year} {2020})}\BibitemShut {NoStop}%
	\bibitem [{\citenamefont {Chang}\ \emph {et~al.}(2016)\citenamefont {Chang},
		\citenamefont {Xu}, \citenamefont {Zheng}, \citenamefont {Singh},
		\citenamefont {Hsu}, \citenamefont {Bian}, \citenamefont {Alidoust},
		\citenamefont {Belopolski}, \citenamefont {Sanchez}, \citenamefont {Zhang}
		\emph {et~al.}}]{Chang16}%
	\BibitemOpen
	\bibfield  {author} {\bibinfo {author} {\bibfnamefont {G.}~\bibnamefont
			{Chang}}, \bibinfo {author} {\bibfnamefont {S.-Y.}\ \bibnamefont {Xu}},
		\bibinfo {author} {\bibfnamefont {H.}~\bibnamefont {Zheng}}, \bibinfo
		{author} {\bibfnamefont {B.}~\bibnamefont {Singh}}, \bibinfo {author}
		{\bibfnamefont {C.-H.}\ \bibnamefont {Hsu}}, \bibinfo {author} {\bibfnamefont
			{G.}~\bibnamefont {Bian}}, \bibinfo {author} {\bibfnamefont {N.}~\bibnamefont
			{Alidoust}}, \bibinfo {author} {\bibfnamefont {I.}~\bibnamefont
			{Belopolski}}, \bibinfo {author} {\bibfnamefont {D.~S.}\ \bibnamefont
			{Sanchez}}, \bibinfo {author} {\bibfnamefont {S.}~\bibnamefont {Zhang}},
		\emph {et~al.},\ }\href@noop {} {\bibfield  {journal} {\bibinfo  {journal}
			{Sci. Rep.}\ }\textbf {\bibinfo {volume} {6}},\ \bibinfo {pages} {38839}
		(\bibinfo {year} {2016})}\BibitemShut {NoStop}%
	\bibitem [{\citenamefont {Chang}\ \emph {et~al.}(2018)\citenamefont {Chang},
		\citenamefont {Singh}, \citenamefont {Xu}, \citenamefont {Bian},
		\citenamefont {Huang}, \citenamefont {Hsu}, \citenamefont {Belopolski},
		\citenamefont {Alidoust}, \citenamefont {Sanchez}, \citenamefont {Zheng}
		\emph {et~al.}}]{Chang18}%
	\BibitemOpen
	\bibfield  {author} {\bibinfo {author} {\bibfnamefont {G.}~\bibnamefont
			{Chang}}, \bibinfo {author} {\bibfnamefont {B.}~\bibnamefont {Singh}},
		\bibinfo {author} {\bibfnamefont {S.-Y.}\ \bibnamefont {Xu}}, \bibinfo
		{author} {\bibfnamefont {G.}~\bibnamefont {Bian}}, \bibinfo {author}
		{\bibfnamefont {S.-M.}\ \bibnamefont {Huang}}, \bibinfo {author}
		{\bibfnamefont {C.-H.}\ \bibnamefont {Hsu}}, \bibinfo {author} {\bibfnamefont
			{I.}~\bibnamefont {Belopolski}}, \bibinfo {author} {\bibfnamefont
			{N.}~\bibnamefont {Alidoust}}, \bibinfo {author} {\bibfnamefont {D.~S.}\
			\bibnamefont {Sanchez}}, \bibinfo {author} {\bibfnamefont {H.}~\bibnamefont
			{Zheng}}, \emph {et~al.},\ }\href@noop {} {\bibfield  {journal} {\bibinfo
			{journal} {Phys. Rev. B}\ }\textbf {\bibinfo {volume} {97}},\ \bibinfo
		{pages} {041104} (\bibinfo {year} {2018})}\BibitemShut {NoStop}%
	\bibitem [{\citenamefont {Kim}\ \emph {et~al.}(2018)\citenamefont {Kim},
		\citenamefont {Seo}, \citenamefont {Lee}, \citenamefont {Ko}, \citenamefont
		{Kim}, \citenamefont {Jang}, \citenamefont {Ok}, \citenamefont {Lee},
		\citenamefont {Jo}, \citenamefont {Kang} \emph {et~al.}}]{Kim18}%
	\BibitemOpen
	\bibfield  {author} {\bibinfo {author} {\bibfnamefont {K.}~\bibnamefont
			{Kim}}, \bibinfo {author} {\bibfnamefont {J.}~\bibnamefont {Seo}}, \bibinfo
		{author} {\bibfnamefont {E.}~\bibnamefont {Lee}}, \bibinfo {author}
		{\bibfnamefont {K.-T.}\ \bibnamefont {Ko}}, \bibinfo {author} {\bibfnamefont
			{B.}~\bibnamefont {Kim}}, \bibinfo {author} {\bibfnamefont {B.~G.}\
			\bibnamefont {Jang}}, \bibinfo {author} {\bibfnamefont {J.~M.}\ \bibnamefont
			{Ok}}, \bibinfo {author} {\bibfnamefont {J.}~\bibnamefont {Lee}}, \bibinfo
		{author} {\bibfnamefont {Y.~J.}\ \bibnamefont {Jo}}, \bibinfo {author}
		{\bibfnamefont {W.}~\bibnamefont {Kang}}, \emph {et~al.},\ }\href@noop {}
	{\bibfield  {journal} {\bibinfo  {journal} {Nat. Mater.}\ }\textbf {\bibinfo
			{volume} {17}},\ \bibinfo {pages} {794} (\bibinfo {year} {2018})}\BibitemShut
	{NoStop}%
	\bibitem [{\citenamefont {Nagaosa}\ \emph {et~al.}(2010)\citenamefont
		{Nagaosa}, \citenamefont {Sinova}, \citenamefont {Onoda}, \citenamefont
		{MacDonald},\ and\ \citenamefont {Ong}}]{Nagaosa10}%
	\BibitemOpen
	\bibfield  {author} {\bibinfo {author} {\bibfnamefont {N.}~\bibnamefont
			{Nagaosa}}, \bibinfo {author} {\bibfnamefont {J.}~\bibnamefont {Sinova}},
		\bibinfo {author} {\bibfnamefont {S.}~\bibnamefont {Onoda}}, \bibinfo
		{author} {\bibfnamefont {A.~H.}\ \bibnamefont {MacDonald}},\ and\ \bibinfo
		{author} {\bibfnamefont {N.~P.}\ \bibnamefont {Ong}},\ }\href@noop {}
	{\bibfield  {journal} {\bibinfo  {journal} {Rev. Mod. Phys.}\ }\textbf
		{\bibinfo {volume} {82}},\ \bibinfo {pages} {1539} (\bibinfo {year}
		{2010})}\BibitemShut {NoStop}%
	\bibitem [{\citenamefont {Zeng}\ \emph {et~al.}(2006)\citenamefont {Zeng},
		\citenamefont {Yao}, \citenamefont {Niu},\ and\ \citenamefont
		{Weitering}}]{Zeng06}%
	\BibitemOpen
	\bibfield  {author} {\bibinfo {author} {\bibfnamefont {C.}~\bibnamefont
			{Zeng}}, \bibinfo {author} {\bibfnamefont {Y.}~\bibnamefont {Yao}}, \bibinfo
		{author} {\bibfnamefont {Q.}~\bibnamefont {Niu}},\ and\ \bibinfo {author}
		{\bibfnamefont {H.~H.}\ \bibnamefont {Weitering}},\ }\href
	{https://doi.org/10.1103/PhysRevLett.96.037204} {\bibfield  {journal}
		{\bibinfo  {journal} {Phys. Rev. Lett.}\ }\textbf {\bibinfo {volume} {96}},\
		\bibinfo {pages} {037204} (\bibinfo {year} {2006})}\BibitemShut {NoStop}%
	\bibitem [{\citenamefont {Wang}\ \emph {et~al.}(2018)\citenamefont {Wang},
		\citenamefont {Xu}, \citenamefont {Lou}, \citenamefont {Liu}, \citenamefont
		{Li}, \citenamefont {Huang}, \citenamefont {Shen}, \citenamefont {Weng},
		\citenamefont {Wang},\ and\ \citenamefont {Lei}}]{Wang18}%
	\BibitemOpen
	\bibfield  {author} {\bibinfo {author} {\bibfnamefont {Q.}~\bibnamefont
			{Wang}}, \bibinfo {author} {\bibfnamefont {Y.}~\bibnamefont {Xu}}, \bibinfo
		{author} {\bibfnamefont {R.}~\bibnamefont {Lou}}, \bibinfo {author}
		{\bibfnamefont {Z.}~\bibnamefont {Liu}}, \bibinfo {author} {\bibfnamefont
			{M.}~\bibnamefont {Li}}, \bibinfo {author} {\bibfnamefont {Y.}~\bibnamefont
			{Huang}}, \bibinfo {author} {\bibfnamefont {D.}~\bibnamefont {Shen}},
		\bibinfo {author} {\bibfnamefont {H.}~\bibnamefont {Weng}}, \bibinfo {author}
		{\bibfnamefont {S.}~\bibnamefont {Wang}},\ and\ \bibinfo {author}
		{\bibfnamefont {H.}~\bibnamefont {Lei}},\ }\href@noop {} {\bibfield
		{journal} {\bibinfo  {journal} {Nat. commun.}\ }\textbf {\bibinfo {volume}
			{9}},\ \bibinfo {pages} {3681} (\bibinfo {year} {2018})}\BibitemShut
	{NoStop}%
	\bibitem [{\citenamefont {Fang}\ \emph {et~al.}(2003)\citenamefont {Fang},
		\citenamefont {Nagaosa}, \citenamefont {Takahashi}, \citenamefont {Asamitsu},
		\citenamefont {Mathieu}, \citenamefont {Ogasawara}, \citenamefont {Yamada},
		\citenamefont {Kawasaki}, \citenamefont {Tokura},\ and\ \citenamefont
		{Terakura}}]{Fang03}%
	\BibitemOpen
	\bibfield  {author} {\bibinfo {author} {\bibfnamefont {Z.}~\bibnamefont
			{Fang}}, \bibinfo {author} {\bibfnamefont {N.}~\bibnamefont {Nagaosa}},
		\bibinfo {author} {\bibfnamefont {K.~S.}\ \bibnamefont {Takahashi}}, \bibinfo
		{author} {\bibfnamefont {A.}~\bibnamefont {Asamitsu}}, \bibinfo {author}
		{\bibfnamefont {R.}~\bibnamefont {Mathieu}}, \bibinfo {author} {\bibfnamefont
			{T.}~\bibnamefont {Ogasawara}}, \bibinfo {author} {\bibfnamefont
			{H.}~\bibnamefont {Yamada}}, \bibinfo {author} {\bibfnamefont
			{M.}~\bibnamefont {Kawasaki}}, \bibinfo {author} {\bibfnamefont
			{Y.}~\bibnamefont {Tokura}},\ and\ \bibinfo {author} {\bibfnamefont
			{K.}~\bibnamefont {Terakura}},\ }\href@noop {} {\bibfield  {journal}
		{\bibinfo  {journal} {Science}\ }\textbf {\bibinfo {volume} {302}},\ \bibinfo
		{pages} {92} (\bibinfo {year} {2003})}\BibitemShut {NoStop}%
	\bibitem [{\citenamefont {Chen}\ \emph {et~al.}(2013)\citenamefont {Chen},
		\citenamefont {Bergman},\ and\ \citenamefont {Burkov}}]{Chen13}%
	\BibitemOpen
	\bibfield  {author} {\bibinfo {author} {\bibfnamefont {Y.}~\bibnamefont
			{Chen}}, \bibinfo {author} {\bibfnamefont {D.}~\bibnamefont {Bergman}},\ and\
		\bibinfo {author} {\bibfnamefont {A.}~\bibnamefont {Burkov}},\ }\href@noop {}
	{\bibfield  {journal} {\bibinfo  {journal} {Phys. Rev. B}\ }\textbf {\bibinfo
			{volume} {88}},\ \bibinfo {pages} {125110} (\bibinfo {year}
		{2013})}\BibitemShut {NoStop}%
	\bibitem [{\citenamefont {Vazifeh}\ and\ \citenamefont
		{Franz}(2013)}]{Vazifeh13}%
	\BibitemOpen
	\bibfield  {author} {\bibinfo {author} {\bibfnamefont {M.~M.}\ \bibnamefont
			{Vazifeh}}\ and\ \bibinfo {author} {\bibfnamefont {M.}~\bibnamefont
			{Franz}},\ }\href {https://doi.org/10.1103/PhysRevLett.111.027201} {\bibfield
		{journal} {\bibinfo  {journal} {Phys. Rev. Lett.}\ }\textbf {\bibinfo
			{volume} {111}},\ \bibinfo {pages} {027201} (\bibinfo {year}
		{2013})}\BibitemShut {NoStop}%
	\bibitem [{\citenamefont {Zyuzin}\ and\ \citenamefont
		{Tiwari}(2016)}]{Zyuzin16}%
	\BibitemOpen
	\bibfield  {author} {\bibinfo {author} {\bibfnamefont {A.~A.}\ \bibnamefont
			{Zyuzin}}\ and\ \bibinfo {author} {\bibfnamefont {R.~P.}\ \bibnamefont
			{Tiwari}},\ }\href@noop {} {\bibfield  {journal} {\bibinfo  {journal} {JETP
				letters}\ }\textbf {\bibinfo {volume} {103}},\ \bibinfo {pages} {717}
		(\bibinfo {year} {2016})}\BibitemShut {NoStop}%
	\bibitem [{\citenamefont {Young}\ and\ \citenamefont {Kane}(2015)}]{Young15}%
	\BibitemOpen
	\bibfield  {author} {\bibinfo {author} {\bibfnamefont {S.~M.}\ \bibnamefont
			{Young}}\ and\ \bibinfo {author} {\bibfnamefont {C.~L.}\ \bibnamefont
			{Kane}},\ }\href@noop {} {\bibfield  {journal} {\bibinfo  {journal} {Phys.
				Rev. Lett.}\ }\textbf {\bibinfo {volume} {115}},\ \bibinfo {pages} {126803}
		(\bibinfo {year} {2015})}\BibitemShut {NoStop}%
	\bibitem [{\citenamefont {Niu}\ \emph {et~al.}(2017)\citenamefont {Niu},
		\citenamefont {Buhl}, \citenamefont {Bihlmayer}, \citenamefont {Wortmann},
		\citenamefont {Dai}, \citenamefont {Bl{\"u}gel},\ and\ \citenamefont
		{Mokrousov}}]{Niu17}%
	\BibitemOpen
	\bibfield  {author} {\bibinfo {author} {\bibfnamefont {C.}~\bibnamefont
			{Niu}}, \bibinfo {author} {\bibfnamefont {P.~M.}\ \bibnamefont {Buhl}},
		\bibinfo {author} {\bibfnamefont {G.}~\bibnamefont {Bihlmayer}}, \bibinfo
		{author} {\bibfnamefont {D.}~\bibnamefont {Wortmann}}, \bibinfo {author}
		{\bibfnamefont {Y.}~\bibnamefont {Dai}}, \bibinfo {author} {\bibfnamefont
			{S.}~\bibnamefont {Bl{\"u}gel}},\ and\ \bibinfo {author} {\bibfnamefont
			{Y.}~\bibnamefont {Mokrousov}},\ }\href@noop {} {\bibfield  {journal}
		{\bibinfo  {journal} {Phys. Rev. B}\ }\textbf {\bibinfo {volume} {95}},\
		\bibinfo {pages} {235138} (\bibinfo {year} {2017})}\BibitemShut {NoStop}%
	\bibitem [{\citenamefont {Zhang}\ \emph {et~al.}(2014)\citenamefont {Zhang},
		\citenamefont {Huang}, \citenamefont {Haule},\ and\ \citenamefont
		{Vanderbilt}}]{PhysRevB.90.165143}%
	\BibitemOpen
	\bibfield  {author} {\bibinfo {author} {\bibfnamefont {H.}~\bibnamefont
			{Zhang}}, \bibinfo {author} {\bibfnamefont {H.}~\bibnamefont {Huang}},
		\bibinfo {author} {\bibfnamefont {K.}~\bibnamefont {Haule}},\ and\ \bibinfo
		{author} {\bibfnamefont {D.}~\bibnamefont {Vanderbilt}},\ }\href
	{https://doi.org/10.1103/PhysRevB.90.165143} {\bibfield  {journal} {\bibinfo
			{journal} {Phys. Rev. B}\ }\textbf {\bibinfo {volume} {90}},\ \bibinfo
		{pages} {165143} (\bibinfo {year} {2014})}\BibitemShut {NoStop}%
	\bibitem [{\citenamefont {Wan}\ \emph {et~al.}(2011)\citenamefont {Wan},
		\citenamefont {Turner}, \citenamefont {Vishwanath},\ and\ \citenamefont
		{Savrasov}}]{PhysRevB.83.205101}%
	\BibitemOpen
	\bibfield  {author} {\bibinfo {author} {\bibfnamefont {X.}~\bibnamefont
			{Wan}}, \bibinfo {author} {\bibfnamefont {A.~M.}\ \bibnamefont {Turner}},
		\bibinfo {author} {\bibfnamefont {A.}~\bibnamefont {Vishwanath}},\ and\
		\bibinfo {author} {\bibfnamefont {S.~Y.}\ \bibnamefont {Savrasov}},\ }\href
	{https://doi.org/10.1103/PhysRevB.83.205101} {\bibfield  {journal} {\bibinfo
			{journal} {Phys. Rev. B}\ }\textbf {\bibinfo {volume} {83}},\ \bibinfo
		{pages} {205101} (\bibinfo {year} {2011})}\BibitemShut {NoStop}%
	\bibitem [{\citenamefont {Neumeier}\ \emph {et~al.}(1994)\citenamefont
		{Neumeier}, \citenamefont {Hundley}, \citenamefont {Smith}, \citenamefont
		{Thompson}, \citenamefont {Allgeier}, \citenamefont {Xie}, \citenamefont
		{Yelon},\ and\ \citenamefont {Kim}}]{Neumeier94}%
	\BibitemOpen
	\bibfield  {author} {\bibinfo {author} {\bibfnamefont {J.}~\bibnamefont
			{Neumeier}}, \bibinfo {author} {\bibfnamefont {M.}~\bibnamefont {Hundley}},
		\bibinfo {author} {\bibfnamefont {M.}~\bibnamefont {Smith}}, \bibinfo
		{author} {\bibfnamefont {J.}~\bibnamefont {Thompson}}, \bibinfo {author}
		{\bibfnamefont {C.}~\bibnamefont {Allgeier}}, \bibinfo {author}
		{\bibfnamefont {H.}~\bibnamefont {Xie}}, \bibinfo {author} {\bibfnamefont
			{W.}~\bibnamefont {Yelon}},\ and\ \bibinfo {author} {\bibfnamefont
			{J.}~\bibnamefont {Kim}},\ }\href@noop {} {\bibfield  {journal} {\bibinfo
			{journal} {Phys. Rev. B}\ }\textbf {\bibinfo {volume} {50}},\ \bibinfo
		{pages} {17910} (\bibinfo {year} {1994})}\BibitemShut {NoStop}%
	\bibitem [{\citenamefont {Puchkov}\ \emph {et~al.}(1998)\citenamefont
		{Puchkov}, \citenamefont {Shen}, \citenamefont {Kimura},\ and\ \citenamefont
		{Tokura}}]{Puchkov98}%
	\BibitemOpen
	\bibfield  {author} {\bibinfo {author} {\bibfnamefont {A.}~\bibnamefont
			{Puchkov}}, \bibinfo {author} {\bibfnamefont {Z.-X.}\ \bibnamefont {Shen}},
		\bibinfo {author} {\bibfnamefont {T.}~\bibnamefont {Kimura}},\ and\ \bibinfo
		{author} {\bibfnamefont {Y.}~\bibnamefont {Tokura}},\ }\href@noop {}
	{\bibfield  {journal} {\bibinfo  {journal} {Phys. Rev. B}\ }\textbf {\bibinfo
			{volume} {58}},\ \bibinfo {pages} {R13322} (\bibinfo {year}
		{1998})}\BibitemShut {NoStop}%
	\bibitem [{\citenamefont {Lu}\ \emph {et~al.}(1996)\citenamefont {Lu},
		\citenamefont {Schmidt}, \citenamefont {Cummins}, \citenamefont {Schuppler},
		\citenamefont {Lichtenberg},\ and\ \citenamefont {Bednorz}}]{Lu96}%
	\BibitemOpen
	\bibfield  {author} {\bibinfo {author} {\bibfnamefont {D.~H.}\ \bibnamefont
			{Lu}}, \bibinfo {author} {\bibfnamefont {M.}~\bibnamefont {Schmidt}},
		\bibinfo {author} {\bibfnamefont {T.~R.}\ \bibnamefont {Cummins}}, \bibinfo
		{author} {\bibfnamefont {S.}~\bibnamefont {Schuppler}}, \bibinfo {author}
		{\bibfnamefont {F.}~\bibnamefont {Lichtenberg}},\ and\ \bibinfo {author}
		{\bibfnamefont {J.~G.}\ \bibnamefont {Bednorz}},\ }\href
	{https://doi.org/10.1103/PhysRevLett.76.4845} {\bibfield  {journal} {\bibinfo
			{journal} {Phys. Rev. Lett.}\ }\textbf {\bibinfo {volume} {76}},\ \bibinfo
		{pages} {4845} (\bibinfo {year} {1996})}\BibitemShut {NoStop}%
	\bibitem [{\citenamefont {Damascelli}\ \emph {et~al.}(2000)\citenamefont
		{Damascelli}, \citenamefont {Lu}, \citenamefont {Shen}, \citenamefont
		{Armitage}, \citenamefont {Ronning}, \citenamefont {Feng}, \citenamefont
		{Kim}, \citenamefont {Shen}, \citenamefont {Kimura}, \citenamefont {Tokura},
		\citenamefont {Mao},\ and\ \citenamefont {Maeno}}]{Damascelli00}%
	\BibitemOpen
	\bibfield  {author} {\bibinfo {author} {\bibfnamefont {A.}~\bibnamefont
			{Damascelli}}, \bibinfo {author} {\bibfnamefont {D.~H.}\ \bibnamefont {Lu}},
		\bibinfo {author} {\bibfnamefont {K.~M.}\ \bibnamefont {Shen}}, \bibinfo
		{author} {\bibfnamefont {N.~P.}\ \bibnamefont {Armitage}}, \bibinfo {author}
		{\bibfnamefont {F.}~\bibnamefont {Ronning}}, \bibinfo {author} {\bibfnamefont
			{D.~L.}\ \bibnamefont {Feng}}, \bibinfo {author} {\bibfnamefont
			{C.}~\bibnamefont {Kim}}, \bibinfo {author} {\bibfnamefont {Z.-X.}\
			\bibnamefont {Shen}}, \bibinfo {author} {\bibfnamefont {T.}~\bibnamefont
			{Kimura}}, \bibinfo {author} {\bibfnamefont {Y.}~\bibnamefont {Tokura}},
		\bibinfo {author} {\bibfnamefont {Z.~Q.}\ \bibnamefont {Mao}},\ and\ \bibinfo
		{author} {\bibfnamefont {Y.}~\bibnamefont {Maeno}},\ }\href
	{https://doi.org/10.1103/PhysRevLett.85.5194} {\bibfield  {journal} {\bibinfo
			{journal} {Phys. Rev. Lett.}\ }\textbf {\bibinfo {volume} {85}},\ \bibinfo
		{pages} {5194} (\bibinfo {year} {2000})}\BibitemShut {NoStop}%
	\bibitem [{\citenamefont {Mackenzie}\ \emph {et~al.}(1996)\citenamefont
		{Mackenzie}, \citenamefont {Julian}, \citenamefont {Diver}, \citenamefont
		{McMullan}, \citenamefont {Ray}, \citenamefont {Lonzarich}, \citenamefont
		{Maeno}, \citenamefont {Nishizaki},\ and\ \citenamefont
		{Fujita}}]{Mackenzie96}%
	\BibitemOpen
	\bibfield  {author} {\bibinfo {author} {\bibfnamefont {A.~P.}\ \bibnamefont
			{Mackenzie}}, \bibinfo {author} {\bibfnamefont {S.~R.}\ \bibnamefont
			{Julian}}, \bibinfo {author} {\bibfnamefont {A.~J.}\ \bibnamefont {Diver}},
		\bibinfo {author} {\bibfnamefont {G.~J.}\ \bibnamefont {McMullan}}, \bibinfo
		{author} {\bibfnamefont {M.~P.}\ \bibnamefont {Ray}}, \bibinfo {author}
		{\bibfnamefont {G.~G.}\ \bibnamefont {Lonzarich}}, \bibinfo {author}
		{\bibfnamefont {Y.}~\bibnamefont {Maeno}}, \bibinfo {author} {\bibfnamefont
			{S.}~\bibnamefont {Nishizaki}},\ and\ \bibinfo {author} {\bibfnamefont
			{T.}~\bibnamefont {Fujita}},\ }\href
	{https://doi.org/10.1103/PhysRevLett.76.3786} {\bibfield  {journal} {\bibinfo
			{journal} {Phys. Rev. Lett.}\ }\textbf {\bibinfo {volume} {76}},\ \bibinfo
		{pages} {3786} (\bibinfo {year} {1996})}\BibitemShut {NoStop}%
	\bibitem [{\citenamefont {Mackenzie}\ \emph {et~al.}(1998)\citenamefont
		{Mackenzie}, \citenamefont {Ikeda}, \citenamefont {Maeno}, \citenamefont
		{Fujita}, \citenamefont {Julian},\ and\ \citenamefont
		{Lonzarich}}]{Mackenzie98}%
	\BibitemOpen
	\bibfield  {author} {\bibinfo {author} {\bibfnamefont {A.~P.}\ \bibnamefont
			{Mackenzie}}, \bibinfo {author} {\bibfnamefont {S.-i.}\ \bibnamefont
			{Ikeda}}, \bibinfo {author} {\bibfnamefont {Y.}~\bibnamefont {Maeno}},
		\bibinfo {author} {\bibfnamefont {T.}~\bibnamefont {Fujita}}, \bibinfo
		{author} {\bibfnamefont {S.~R.}\ \bibnamefont {Julian}},\ and\ \bibinfo
		{author} {\bibfnamefont {G.~G.}\ \bibnamefont {Lonzarich}},\ }\href@noop {}
	{\bibfield  {journal} {\bibinfo  {journal} {J. Phys. Soc. Jpn.}\ }\textbf
		{\bibinfo {volume} {67}},\ \bibinfo {pages} {385} (\bibinfo {year}
		{1998})}\BibitemShut {NoStop}%
	\bibitem [{\citenamefont {Chang}\ \emph {et~al.}(2011)\citenamefont {Chang},
		\citenamefont {Chang}, \citenamefont {Jang}, \citenamefont {Jeong},
		\citenamefont {Jung}, \citenamefont {Kim}, \citenamefont {Chung},\ and\
		\citenamefont {Noh}}]{Chang11}%
	\BibitemOpen
	\bibfield  {author} {\bibinfo {author} {\bibfnamefont {S.~H.}\ \bibnamefont
			{Chang}}, \bibinfo {author} {\bibfnamefont {Y.~J.}\ \bibnamefont {Chang}},
		\bibinfo {author} {\bibfnamefont {S.~Y.}\ \bibnamefont {Jang}}, \bibinfo
		{author} {\bibfnamefont {D.~W.}\ \bibnamefont {Jeong}}, \bibinfo {author}
		{\bibfnamefont {C.~U.}\ \bibnamefont {Jung}}, \bibinfo {author}
		{\bibfnamefont {Y.-J.}\ \bibnamefont {Kim}}, \bibinfo {author} {\bibfnamefont
			{J.-S.}\ \bibnamefont {Chung}},\ and\ \bibinfo {author} {\bibfnamefont
			{T.~W.}\ \bibnamefont {Noh}},\ }\href
	{https://doi.org/10.1103/PhysRevB.84.104101} {\bibfield  {journal} {\bibinfo
			{journal} {Phys. Rev. B}\ }\textbf {\bibinfo {volume} {84}},\ \bibinfo
		{pages} {104101} (\bibinfo {year} {2011})}\BibitemShut {NoStop}%
	\bibitem [{\citenamefont {Sohn}\ \emph {et~al.}(2018)\citenamefont {Sohn},
		\citenamefont {Kim}, \citenamefont {Park}, \citenamefont {Choi},
		\citenamefont {Moon}, \citenamefont {Choi}, \citenamefont {Choi},
		\citenamefont {Noh}, \citenamefont {Zhou}, \citenamefont {Chang} \emph
		{et~al.}}]{Sohn18}%
	\BibitemOpen
	\bibfield  {author} {\bibinfo {author} {\bibfnamefont {B.}~\bibnamefont
			{Sohn}}, \bibinfo {author} {\bibfnamefont {B.}~\bibnamefont {Kim}}, \bibinfo
		{author} {\bibfnamefont {S.~Y.}\ \bibnamefont {Park}}, \bibinfo {author}
		{\bibfnamefont {H.~Y.}\ \bibnamefont {Choi}}, \bibinfo {author}
		{\bibfnamefont {J.~Y.}\ \bibnamefont {Moon}}, \bibinfo {author}
		{\bibfnamefont {T.}~\bibnamefont {Choi}}, \bibinfo {author} {\bibfnamefont
			{Y.~J.}\ \bibnamefont {Choi}}, \bibinfo {author} {\bibfnamefont {T.~W.}\
			\bibnamefont {Noh}}, \bibinfo {author} {\bibfnamefont {H.}~\bibnamefont
			{Zhou}}, \bibinfo {author} {\bibfnamefont {S.~H.}\ \bibnamefont {Chang}},
		\emph {et~al.},\ }\href@noop {} {\bibfield  {journal} {\bibinfo  {journal}
			{arXiv preprint arXiv:1810.01615}\ } (\bibinfo {year} {2018})}\BibitemShut
	{NoStop}%
	\bibitem [{\citenamefont {Oh}\ \emph {et~al.}()\citenamefont {Oh} \emph
		{et~al.}}]{Oh20}%
	\BibitemOpen
	\bibfield  {author} {\bibinfo {author} {\bibfnamefont {J.~S.}\ \bibnamefont
			{Oh}} \emph {et~al.},\ }\href@noop {} {\bibinfo  {journal} {unpublished}\
	}\BibitemShut {NoStop}%
	\bibitem [{\citenamefont {Hase}\ and\ \citenamefont
		{Nishihara}(1997)}]{Hase97}%
	\BibitemOpen
	\bibfield  {journal} {  }\bibfield  {author} {\bibinfo {author} {\bibfnamefont
			{I.}~\bibnamefont {Hase}}\ and\ \bibinfo {author} {\bibfnamefont
			{Y.}~\bibnamefont {Nishihara}},\ }\href@noop {} {\bibfield  {journal}
		{\bibinfo  {journal} {J. Phys. Soc. Jpn.}\ }\textbf {\bibinfo {volume}
			{66}},\ \bibinfo {pages} {3517} (\bibinfo {year} {1997})}\BibitemShut
	{NoStop}%
	\bibitem [{\citenamefont {Singh}\ and\ \citenamefont {Mazin}(2001)}]{Singh01}%
	\BibitemOpen
	\bibfield  {author} {\bibinfo {author} {\bibfnamefont {D.}~\bibnamefont
			{Singh}}\ and\ \bibinfo {author} {\bibfnamefont {I.}~\bibnamefont {Mazin}},\
	}\href@noop {} {\bibfield  {journal} {\bibinfo  {journal} {Phys. Rev. B}\
		}\textbf {\bibinfo {volume} {63}},\ \bibinfo {pages} {165101} (\bibinfo
		{year} {2001})}\BibitemShut {NoStop}%
	\bibitem [{\citenamefont {Chang}\ \emph {et~al.}(2009)\citenamefont {Chang},
		\citenamefont {Kim}, \citenamefont {Phark}, \citenamefont {Kim},
		\citenamefont {Yu},\ and\ \citenamefont {Noh}}]{Chang09}%
	\BibitemOpen
	\bibfield  {author} {\bibinfo {author} {\bibfnamefont {Y.~J.}\ \bibnamefont
			{Chang}}, \bibinfo {author} {\bibfnamefont {C.~H.}\ \bibnamefont {Kim}},
		\bibinfo {author} {\bibfnamefont {S.-H.}\ \bibnamefont {Phark}}, \bibinfo
		{author} {\bibfnamefont {Y.~S.}\ \bibnamefont {Kim}}, \bibinfo {author}
		{\bibfnamefont {J.}~\bibnamefont {Yu}},\ and\ \bibinfo {author}
		{\bibfnamefont {T.~W.}\ \bibnamefont {Noh}},\ }\href
	{https://doi.org/10.1103/PhysRevLett.103.057201} {\bibfield  {journal}
		{\bibinfo  {journal} {Phys. Rev. Lett.}\ }\textbf {\bibinfo {volume} {103}},\
		\bibinfo {pages} {057201} (\bibinfo {year} {2009})}\BibitemShut {NoStop}%
	\bibitem [{\citenamefont {Jeong}\ \emph {et~al.}(2013)\citenamefont {Jeong},
		\citenamefont {Choi}, \citenamefont {Kim}, \citenamefont {Chang},
		\citenamefont {Sohn}, \citenamefont {Park}, \citenamefont {Kang},
		\citenamefont {Cho}, \citenamefont {Baek}, \citenamefont {Eom}, \citenamefont
		{Shim}, \citenamefont {Yu}, \citenamefont {Kim}, \citenamefont {Moon},\ and\
		\citenamefont {Noh}}]{Jeong13}%
	\BibitemOpen
	\bibfield  {author} {\bibinfo {author} {\bibfnamefont {D.~W.}\ \bibnamefont
			{Jeong}}, \bibinfo {author} {\bibfnamefont {H.~C.}\ \bibnamefont {Choi}},
		\bibinfo {author} {\bibfnamefont {C.~H.}\ \bibnamefont {Kim}}, \bibinfo
		{author} {\bibfnamefont {S.~H.}\ \bibnamefont {Chang}}, \bibinfo {author}
		{\bibfnamefont {C.~H.}\ \bibnamefont {Sohn}}, \bibinfo {author}
		{\bibfnamefont {H.~J.}\ \bibnamefont {Park}}, \bibinfo {author}
		{\bibfnamefont {T.~D.}\ \bibnamefont {Kang}}, \bibinfo {author}
		{\bibfnamefont {D.-Y.}\ \bibnamefont {Cho}}, \bibinfo {author} {\bibfnamefont
			{S.~H.}\ \bibnamefont {Baek}}, \bibinfo {author} {\bibfnamefont {C.~B.}\
			\bibnamefont {Eom}}, \bibinfo {author} {\bibfnamefont {J.~H.}\ \bibnamefont
			{Shim}}, \bibinfo {author} {\bibfnamefont {J.}~\bibnamefont {Yu}}, \bibinfo
		{author} {\bibfnamefont {K.~W.}\ \bibnamefont {Kim}}, \bibinfo {author}
		{\bibfnamefont {S.~J.}\ \bibnamefont {Moon}},\ and\ \bibinfo {author}
		{\bibfnamefont {T.~W.}\ \bibnamefont {Noh}},\ }\href
	{https://doi.org/10.1103/PhysRevLett.110.247202} {\bibfield  {journal}
		{\bibinfo  {journal} {Phys. Rev. Lett.}\ }\textbf {\bibinfo {volume} {110}},\
		\bibinfo {pages} {247202} (\bibinfo {year} {2013})}\BibitemShut {NoStop}%
	\bibitem [{\citenamefont {Zhang}\ \emph {et~al.}(2011)\citenamefont {Zhang},
		\citenamefont {Richard}, \citenamefont {Qian}, \citenamefont {Xu},
		\citenamefont {Dai},\ and\ \citenamefont {Ding}}]{Zhang11}%
	\BibitemOpen
	\bibfield  {author} {\bibinfo {author} {\bibfnamefont {P.}~\bibnamefont
			{Zhang}}, \bibinfo {author} {\bibfnamefont {P.}~\bibnamefont {Richard}},
		\bibinfo {author} {\bibfnamefont {T.}~\bibnamefont {Qian}}, \bibinfo {author}
		{\bibfnamefont {Y.-M.}\ \bibnamefont {Xu}}, \bibinfo {author} {\bibfnamefont
			{X.}~\bibnamefont {Dai}},\ and\ \bibinfo {author} {\bibfnamefont
			{H.}~\bibnamefont {Ding}},\ }\href@noop {} {\bibfield  {journal} {\bibinfo
			{journal} {Rev. Sci. Instrum.}\ }\textbf {\bibinfo {volume} {82}},\ \bibinfo
		{pages} {043712} (\bibinfo {year} {2011})}\BibitemShut {NoStop}%
	\bibitem [{\citenamefont {Matsuno}\ \emph {et~al.}(2016)\citenamefont
		{Matsuno}, \citenamefont {Ogawa}, \citenamefont {Yasuda}, \citenamefont
		{Kagawa}, \citenamefont {Koshibae}, \citenamefont {Nagaosa}, \citenamefont
		{Tokura},\ and\ \citenamefont {Kawasaki}}]{Matsuno16}%
	\BibitemOpen
	\bibfield  {author} {\bibinfo {author} {\bibfnamefont {J.}~\bibnamefont
			{Matsuno}}, \bibinfo {author} {\bibfnamefont {N.}~\bibnamefont {Ogawa}},
		\bibinfo {author} {\bibfnamefont {K.}~\bibnamefont {Yasuda}}, \bibinfo
		{author} {\bibfnamefont {F.}~\bibnamefont {Kagawa}}, \bibinfo {author}
		{\bibfnamefont {W.}~\bibnamefont {Koshibae}}, \bibinfo {author}
		{\bibfnamefont {N.}~\bibnamefont {Nagaosa}}, \bibinfo {author} {\bibfnamefont
			{Y.}~\bibnamefont {Tokura}},\ and\ \bibinfo {author} {\bibfnamefont
			{M.}~\bibnamefont {Kawasaki}},\ }\href@noop {} {\bibfield  {journal}
		{\bibinfo  {journal} {Sci. Adv.}\ }\textbf {\bibinfo {volume} {2}},\ \bibinfo
		{pages} {e1600304} (\bibinfo {year} {2016})}\BibitemShut {NoStop}%
	\bibitem [{\citenamefont {Sohn}\ \emph {et~al.}(2020)\citenamefont {Sohn},
		\citenamefont {Kim}, \citenamefont {Choi}, \citenamefont {Chang},
		\citenamefont {Han},\ and\ \citenamefont {Kim}}]{Sohn20}%
	\BibitemOpen
	\bibfield  {author} {\bibinfo {author} {\bibfnamefont {B.}~\bibnamefont
			{Sohn}}, \bibinfo {author} {\bibfnamefont {B.}~\bibnamefont {Kim}}, \bibinfo
		{author} {\bibfnamefont {J.~W.}\ \bibnamefont {Choi}}, \bibinfo {author}
		{\bibfnamefont {S.~H.}\ \bibnamefont {Chang}}, \bibinfo {author}
		{\bibfnamefont {J.~H.}\ \bibnamefont {Han}},\ and\ \bibinfo {author}
		{\bibfnamefont {C.}~\bibnamefont {Kim}},\ }\href
	{https://doi.org/https://doi.org/10.1016/j.cap.2019.10.021} {\bibfield
		{journal} {\bibinfo  {journal} {Curr. Appl. Phys.}\ }\textbf {\bibinfo
			{volume} {20}},\ \bibinfo {pages} {186 } (\bibinfo {year}
		{2020})}\BibitemShut {NoStop}%
	\bibitem [{\citenamefont {Jin}\ \emph {et~al.}(2020{\natexlab{a}})\citenamefont
		{Jin}, \citenamefont {Zhang}, \citenamefont {Liu}, \citenamefont {Dai},
		\citenamefont {Shen}, \citenamefont {Wang},\ and\ \citenamefont
		{Liu}}]{jin2020two}%
	\BibitemOpen
	\bibfield  {author} {\bibinfo {author} {\bibfnamefont {L.}~\bibnamefont
			{Jin}}, \bibinfo {author} {\bibfnamefont {X.}~\bibnamefont {Zhang}}, \bibinfo
		{author} {\bibfnamefont {Y.}~\bibnamefont {Liu}}, \bibinfo {author}
		{\bibfnamefont {X.}~\bibnamefont {Dai}}, \bibinfo {author} {\bibfnamefont
			{X.}~\bibnamefont {Shen}}, \bibinfo {author} {\bibfnamefont {L.}~\bibnamefont
			{Wang}},\ and\ \bibinfo {author} {\bibfnamefont {G.}~\bibnamefont {Liu}},\
	}\href@noop {} {\bibfield  {journal} {\bibinfo  {journal} {Phys. Rev. B}\
		}\textbf {\bibinfo {volume} {102}},\ \bibinfo {pages} {125118} (\bibinfo
		{year} {2020}{\natexlab{a}})}\BibitemShut {NoStop}%
	\bibitem [{\citenamefont {Zhou}\ \emph {et~al.}(2021)\citenamefont {Zhou},
		\citenamefont {Liu}, \citenamefont {Kuang}, \citenamefont {Wang},
		\citenamefont {Wang}, \citenamefont {Yang}, \citenamefont {Wang},
		\citenamefont {Cheng},\ and\ \citenamefont {Zhang}}]{zhou2021time}%
	\BibitemOpen
	\bibfield  {author} {\bibinfo {author} {\bibfnamefont {F.}~\bibnamefont
			{Zhou}}, \bibinfo {author} {\bibfnamefont {Y.}~\bibnamefont {Liu}}, \bibinfo
		{author} {\bibfnamefont {M.}~\bibnamefont {Kuang}}, \bibinfo {author}
		{\bibfnamefont {P.}~\bibnamefont {Wang}}, \bibinfo {author} {\bibfnamefont
			{J.}~\bibnamefont {Wang}}, \bibinfo {author} {\bibfnamefont {T.}~\bibnamefont
			{Yang}}, \bibinfo {author} {\bibfnamefont {X.}~\bibnamefont {Wang}}, \bibinfo
		{author} {\bibfnamefont {Z.}~\bibnamefont {Cheng}},\ and\ \bibinfo {author}
		{\bibfnamefont {G.}~\bibnamefont {Zhang}},\ }\href@noop {} {\bibfield
		{journal} {\bibinfo  {journal} {Nanoscale}\ }\textbf {\bibinfo {volume}
			{13}},\ \bibinfo {pages} {8235} (\bibinfo {year} {2021})}\BibitemShut
	{NoStop}%
	\bibitem [{\citenamefont {Jin}\ \emph {et~al.}(2020{\natexlab{b}})\citenamefont
		{Jin}, \citenamefont {Zhang}, \citenamefont {He}, \citenamefont {Meng},
		\citenamefont {Dai},\ and\ \citenamefont {Liu}}]{jin2020ferromagnetic}%
	\BibitemOpen
	\bibfield  {author} {\bibinfo {author} {\bibfnamefont {L.}~\bibnamefont
			{Jin}}, \bibinfo {author} {\bibfnamefont {X.}~\bibnamefont {Zhang}}, \bibinfo
		{author} {\bibfnamefont {T.}~\bibnamefont {He}}, \bibinfo {author}
		{\bibfnamefont {W.}~\bibnamefont {Meng}}, \bibinfo {author} {\bibfnamefont
			{X.}~\bibnamefont {Dai}},\ and\ \bibinfo {author} {\bibfnamefont
			{G.}~\bibnamefont {Liu}},\ }\href@noop {} {\bibfield  {journal} {\bibinfo
			{journal} {Appl. Surf. Sci.}\ }\textbf {\bibinfo {volume} {520}},\ \bibinfo
		{pages} {146376} (\bibinfo {year} {2020}{\natexlab{b}})}\BibitemShut
	{NoStop}%
	\bibitem [{\citenamefont {Sun}\ \emph {et~al.}(2009)\citenamefont {Sun},
		\citenamefont {Yao}, \citenamefont {Fradkin},\ and\ \citenamefont
		{Kivelson}}]{sun2009topological}%
	\BibitemOpen
	\bibfield  {author} {\bibinfo {author} {\bibfnamefont {K.}~\bibnamefont
			{Sun}}, \bibinfo {author} {\bibfnamefont {H.}~\bibnamefont {Yao}}, \bibinfo
		{author} {\bibfnamefont {E.}~\bibnamefont {Fradkin}},\ and\ \bibinfo {author}
		{\bibfnamefont {S.~A.}\ \bibnamefont {Kivelson}},\ }\href@noop {} {\bibfield
		{journal} {\bibinfo  {journal} {Phys. Rev. Lett.}\ }\textbf {\bibinfo
			{volume} {103}},\ \bibinfo {pages} {046811} (\bibinfo {year}
		{2009})}\BibitemShut {NoStop}%
	\bibitem [{\citenamefont {Chong}\ \emph {et~al.}(2008)\citenamefont {Chong},
		\citenamefont {Wen},\ and\ \citenamefont
		{Solja{\v{c}}i{\'c}}}]{chong2008effective}%
	\BibitemOpen
	\bibfield  {author} {\bibinfo {author} {\bibfnamefont {Y.~D.}\ \bibnamefont
			{Chong}}, \bibinfo {author} {\bibfnamefont {X.-G.}\ \bibnamefont {Wen}},\
		and\ \bibinfo {author} {\bibfnamefont {M.}~\bibnamefont
			{Solja{\v{c}}i{\'c}}},\ }\href@noop {} {\bibfield  {journal} {\bibinfo
			{journal} {Phys. Rev. B}\ }\textbf {\bibinfo {volume} {77}},\ \bibinfo
		{pages} {235125} (\bibinfo {year} {2008})}\BibitemShut {NoStop}%
	\bibitem [{\citenamefont {Xiao}\ \emph {et~al.}(2007)\citenamefont {Xiao},
		\citenamefont {Yao},\ and\ \citenamefont {Niu}}]{Xiao07}%
	\BibitemOpen
	\bibfield  {author} {\bibinfo {author} {\bibfnamefont {D.}~\bibnamefont
			{Xiao}}, \bibinfo {author} {\bibfnamefont {W.}~\bibnamefont {Yao}},\ and\
		\bibinfo {author} {\bibfnamefont {Q.}~\bibnamefont {Niu}},\ }\href@noop {}
	{\bibfield  {journal} {\bibinfo  {journal} {Phys. Rev. Lett.}\ }\textbf
		{\bibinfo {volume} {99}},\ \bibinfo {pages} {236809} (\bibinfo {year}
		{2007})}\BibitemShut {NoStop}%
	\bibitem [{\citenamefont {Go}\ \emph {et~al.}(2018)\citenamefont {Go},
		\citenamefont {Jo}, \citenamefont {Kim},\ and\ \citenamefont {Lee}}]{Go18}%
	\BibitemOpen
	\bibfield  {author} {\bibinfo {author} {\bibfnamefont {D.}~\bibnamefont
			{Go}}, \bibinfo {author} {\bibfnamefont {D.}~\bibnamefont {Jo}}, \bibinfo
		{author} {\bibfnamefont {C.}~\bibnamefont {Kim}},\ and\ \bibinfo {author}
		{\bibfnamefont {H.-W.}\ \bibnamefont {Lee}},\ }\href@noop {} {\bibfield
		{journal} {\bibinfo  {journal} {Phys. Rev. Lett.}\ }\textbf {\bibinfo
			{volume} {121}},\ \bibinfo {pages} {086602} (\bibinfo {year}
		{2018})}\BibitemShut {NoStop}%
	\bibitem [{\citenamefont {Liu}\ \emph {et~al.}(2011)\citenamefont {Liu},
		\citenamefont {Bian}, \citenamefont {Miller},\ and\ \citenamefont
		{Chiang}}]{Liu11}%
	\BibitemOpen
	\bibfield  {author} {\bibinfo {author} {\bibfnamefont {Y.}~\bibnamefont
			{Liu}}, \bibinfo {author} {\bibfnamefont {G.}~\bibnamefont {Bian}}, \bibinfo
		{author} {\bibfnamefont {T.}~\bibnamefont {Miller}},\ and\ \bibinfo {author}
		{\bibfnamefont {T.-C.}\ \bibnamefont {Chiang}},\ }\href@noop {} {\bibfield
		{journal} {\bibinfo  {journal} {Phys. Rev. Lett.}\ }\textbf {\bibinfo
			{volume} {107}},\ \bibinfo {pages} {166803} (\bibinfo {year}
		{2011})}\BibitemShut {NoStop}%
	\bibitem [{\citenamefont {Park}\ \emph
		{et~al.}(2012{\natexlab{a}})\citenamefont {Park}, \citenamefont {Kim},
		\citenamefont {Rhim},\ and\ \citenamefont {Han}}]{Park12_2}%
	\BibitemOpen
	\bibfield  {author} {\bibinfo {author} {\bibfnamefont {J.-H.}\ \bibnamefont
			{Park}}, \bibinfo {author} {\bibfnamefont {C.~H.}\ \bibnamefont {Kim}},
		\bibinfo {author} {\bibfnamefont {J.-W.}\ \bibnamefont {Rhim}},\ and\
		\bibinfo {author} {\bibfnamefont {J.~H.}\ \bibnamefont {Han}},\ }\href@noop
	{} {\bibfield  {journal} {\bibinfo  {journal} {Phys. Rev. B}\ }\textbf
		{\bibinfo {volume} {85}},\ \bibinfo {pages} {195401} (\bibinfo {year}
		{2012}{\natexlab{a}})}\BibitemShut {NoStop}%
	\bibitem [{\citenamefont {Cho}\ \emph {et~al.}(2018)\citenamefont {Cho},
		\citenamefont {Park}, \citenamefont {Hong}, \citenamefont {Jung},
		\citenamefont {Kim}, \citenamefont {Han}, \citenamefont {Kyung},
		\citenamefont {Kim}, \citenamefont {Mo}, \citenamefont {Denlinger} \emph
		{et~al.}}]{Cho18}%
	\BibitemOpen
	\bibfield  {author} {\bibinfo {author} {\bibfnamefont {S.}~\bibnamefont
			{Cho}}, \bibinfo {author} {\bibfnamefont {J.-H.}\ \bibnamefont {Park}},
		\bibinfo {author} {\bibfnamefont {J.}~\bibnamefont {Hong}}, \bibinfo {author}
		{\bibfnamefont {J.}~\bibnamefont {Jung}}, \bibinfo {author} {\bibfnamefont
			{B.~S.}\ \bibnamefont {Kim}}, \bibinfo {author} {\bibfnamefont
			{G.}~\bibnamefont {Han}}, \bibinfo {author} {\bibfnamefont {W.}~\bibnamefont
			{Kyung}}, \bibinfo {author} {\bibfnamefont {Y.}~\bibnamefont {Kim}}, \bibinfo
		{author} {\bibfnamefont {S.-K.}\ \bibnamefont {Mo}}, \bibinfo {author}
		{\bibfnamefont {J.}~\bibnamefont {Denlinger}}, \emph {et~al.},\ }\href@noop
	{} {\bibfield  {journal} {\bibinfo  {journal} {Phys. Rev. Lett.}\ }\textbf
		{\bibinfo {volume} {121}},\ \bibinfo {pages} {186401} (\bibinfo {year}
		{2018})}\BibitemShut {NoStop}%
	\bibitem [{\citenamefont {{\"U}nzelmann}\ \emph {et~al.}(2020)\citenamefont
		{{\"U}nzelmann}, \citenamefont {Bentmann}, \citenamefont {Figgemeier},
		\citenamefont {Eck}, \citenamefont {Neu}, \citenamefont {Geldiyev},
		\citenamefont {Diekmann}, \citenamefont {Rohlf}, \citenamefont {Buck},
		\citenamefont {Hoesch}, \citenamefont {Kalläne}, \citenamefont {Rossnagel},
		\citenamefont {Thomale}, \citenamefont {Siegrist}, \citenamefont
		{Sangiovanni}, \citenamefont {Sante},\ and\ \citenamefont
		{Reinert}}]{2020momentumspace}%
	\BibitemOpen
	\bibfield  {author} {\bibinfo {author} {\bibfnamefont {M.}~\bibnamefont
			{{\"U}nzelmann}}, \bibinfo {author} {\bibfnamefont {H.}~\bibnamefont
			{Bentmann}}, \bibinfo {author} {\bibfnamefont {T.}~\bibnamefont
			{Figgemeier}}, \bibinfo {author} {\bibfnamefont {P.}~\bibnamefont {Eck}},
		\bibinfo {author} {\bibfnamefont {J.~N.}\ \bibnamefont {Neu}}, \bibinfo
		{author} {\bibfnamefont {B.}~\bibnamefont {Geldiyev}}, \bibinfo {author}
		{\bibfnamefont {F.}~\bibnamefont {Diekmann}}, \bibinfo {author}
		{\bibfnamefont {S.}~\bibnamefont {Rohlf}}, \bibinfo {author} {\bibfnamefont
			{J.}~\bibnamefont {Buck}}, \bibinfo {author} {\bibfnamefont {M.}~\bibnamefont
			{Hoesch}}, \bibinfo {author} {\bibfnamefont {M.}~\bibnamefont {Kalläne}},
		\bibinfo {author} {\bibfnamefont {K.}~\bibnamefont {Rossnagel}}, \bibinfo
		{author} {\bibfnamefont {R.}~\bibnamefont {Thomale}}, \bibinfo {author}
		{\bibfnamefont {T.}~\bibnamefont {Siegrist}}, \bibinfo {author}
		{\bibfnamefont {G.}~\bibnamefont {Sangiovanni}}, \bibinfo {author}
		{\bibfnamefont {D.~D.}\ \bibnamefont {Sante}},\ and\ \bibinfo {author}
		{\bibfnamefont {F.}~\bibnamefont {Reinert}},\ }\href@noop {} {\bibfield
		{journal} {\bibinfo  {journal} {arXiv preprint arXiv:2012.06996}\ } (\bibinfo
		{year} {2020})}\BibitemShut {NoStop}%
	\bibitem [{\citenamefont {Park}\ \emph
		{et~al.}(2012{\natexlab{b}})\citenamefont {Park}, \citenamefont {Han},
		\citenamefont {Kim}, \citenamefont {Koh}, \citenamefont {Kim}, \citenamefont
		{Lee}, \citenamefont {Choi}, \citenamefont {Han}, \citenamefont {Lee},
		\citenamefont {Hur} \emph {et~al.}}]{Park12}%
	\BibitemOpen
	\bibfield  {author} {\bibinfo {author} {\bibfnamefont {S.~R.}\ \bibnamefont
			{Park}}, \bibinfo {author} {\bibfnamefont {J.}~\bibnamefont {Han}}, \bibinfo
		{author} {\bibfnamefont {C.}~\bibnamefont {Kim}}, \bibinfo {author}
		{\bibfnamefont {Y.~Y.}\ \bibnamefont {Koh}}, \bibinfo {author} {\bibfnamefont
			{C.}~\bibnamefont {Kim}}, \bibinfo {author} {\bibfnamefont {H.}~\bibnamefont
			{Lee}}, \bibinfo {author} {\bibfnamefont {H.~J.}\ \bibnamefont {Choi}},
		\bibinfo {author} {\bibfnamefont {J.~H.}\ \bibnamefont {Han}}, \bibinfo
		{author} {\bibfnamefont {K.~D.}\ \bibnamefont {Lee}}, \bibinfo {author}
		{\bibfnamefont {N.~J.}\ \bibnamefont {Hur}}, \emph {et~al.},\ }\href@noop {}
	{\bibfield  {journal} {\bibinfo  {journal} {Phys. Rev. Lett.}\ }\textbf
		{\bibinfo {volume} {108}},\ \bibinfo {pages} {046805} (\bibinfo {year}
		{2012}{\natexlab{b}})}\BibitemShut {NoStop}%
	\bibitem [{\citenamefont {Sch{\"u}ler}\ \emph {et~al.}(2020)\citenamefont
		{Sch{\"u}ler}, \citenamefont {De~Giovannini}, \citenamefont {H{\"u}bener},
		\citenamefont {Rubio}, \citenamefont {Sentef},\ and\ \citenamefont
		{Werner}}]{Schuler20}%
	\BibitemOpen
	\bibfield  {author} {\bibinfo {author} {\bibfnamefont {M.}~\bibnamefont
			{Sch{\"u}ler}}, \bibinfo {author} {\bibfnamefont {U.}~\bibnamefont
			{De~Giovannini}}, \bibinfo {author} {\bibfnamefont {H.}~\bibnamefont
			{H{\"u}bener}}, \bibinfo {author} {\bibfnamefont {A.}~\bibnamefont {Rubio}},
		\bibinfo {author} {\bibfnamefont {M.~A.}\ \bibnamefont {Sentef}},\ and\
		\bibinfo {author} {\bibfnamefont {P.}~\bibnamefont {Werner}},\ }\href@noop {}
	{\bibfield  {journal} {\bibinfo  {journal} {Sci. Adv.}\ }\textbf {\bibinfo
			{volume} {6}},\ \bibinfo {pages} {eaay2730} (\bibinfo {year}
		{2020})}\BibitemShut {NoStop}%
	\bibitem [{\citenamefont {Wang}\ and\ \citenamefont {Gedik}(2013)}]{Wang13}%
	\BibitemOpen
	\bibfield  {author} {\bibinfo {author} {\bibfnamefont {Y.}~\bibnamefont
			{Wang}}\ and\ \bibinfo {author} {\bibfnamefont {N.}~\bibnamefont {Gedik}},\
	}\href@noop {} {\bibfield  {journal} {\bibinfo  {journal} {Phys. Status
				Solidi RRL}\ }\textbf {\bibinfo {volume} {7}},\ \bibinfo {pages} {64}
		(\bibinfo {year} {2013})}\BibitemShut {NoStop}%
	\bibitem [{\citenamefont {Wang}\ \emph {et~al.}(2011)\citenamefont {Wang},
		\citenamefont {Hsieh}, \citenamefont {Pilon}, \citenamefont {Fu},
		\citenamefont {Gardner}, \citenamefont {Lee},\ and\ \citenamefont
		{Gedik}}]{Wang11}%
	\BibitemOpen
	\bibfield  {author} {\bibinfo {author} {\bibfnamefont {Y.}~\bibnamefont
			{Wang}}, \bibinfo {author} {\bibfnamefont {D.}~\bibnamefont {Hsieh}},
		\bibinfo {author} {\bibfnamefont {D.}~\bibnamefont {Pilon}}, \bibinfo
		{author} {\bibfnamefont {L.}~\bibnamefont {Fu}}, \bibinfo {author}
		{\bibfnamefont {D.}~\bibnamefont {Gardner}}, \bibinfo {author} {\bibfnamefont
			{Y.}~\bibnamefont {Lee}},\ and\ \bibinfo {author} {\bibfnamefont
			{N.}~\bibnamefont {Gedik}},\ }\href@noop {} {\bibfield  {journal} {\bibinfo
			{journal} {Phys. Rev. Lett.}\ }\textbf {\bibinfo {volume} {107}},\ \bibinfo
		{pages} {207602} (\bibinfo {year} {2011})}\BibitemShut {NoStop}%
	\bibitem [{\citenamefont {Schultz}\ \emph {et~al.}(2009)\citenamefont
		{Schultz}, \citenamefont {Levy}, \citenamefont {Reiner},\ and\ \citenamefont
		{Klein}}]{Schultz09}%
	\BibitemOpen
	\bibfield  {author} {\bibinfo {author} {\bibfnamefont {M.}~\bibnamefont
			{Schultz}}, \bibinfo {author} {\bibfnamefont {S.}~\bibnamefont {Levy}},
		\bibinfo {author} {\bibfnamefont {J.~W.}\ \bibnamefont {Reiner}},\ and\
		\bibinfo {author} {\bibfnamefont {L.}~\bibnamefont {Klein}},\ }\href@noop {}
	{\bibfield  {journal} {\bibinfo  {journal} {Phys. Rev. B}\ }\textbf {\bibinfo
			{volume} {79}},\ \bibinfo {pages} {125444} (\bibinfo {year}
		{2009})}\BibitemShut {NoStop}%
	\bibitem [{\citenamefont {Mathieu}\ \emph {et~al.}(2004)\citenamefont
		{Mathieu}, \citenamefont {Asamitsu}, \citenamefont {Yamada}, \citenamefont
		{Takahashi}, \citenamefont {Kawasaki}, \citenamefont {Fang}, \citenamefont
		{Nagaosa},\ and\ \citenamefont {Tokura}}]{mathieu2004scaling}%
	\BibitemOpen
	\bibfield  {author} {\bibinfo {author} {\bibfnamefont {R.}~\bibnamefont
			{Mathieu}}, \bibinfo {author} {\bibfnamefont {A.}~\bibnamefont {Asamitsu}},
		\bibinfo {author} {\bibfnamefont {H.}~\bibnamefont {Yamada}}, \bibinfo
		{author} {\bibfnamefont {K.}~\bibnamefont {Takahashi}}, \bibinfo {author}
		{\bibfnamefont {M.}~\bibnamefont {Kawasaki}}, \bibinfo {author}
		{\bibfnamefont {Z.}~\bibnamefont {Fang}}, \bibinfo {author} {\bibfnamefont
			{N.}~\bibnamefont {Nagaosa}},\ and\ \bibinfo {author} {\bibfnamefont
			{Y.}~\bibnamefont {Tokura}},\ }\href@noop {} {\bibfield  {journal} {\bibinfo
			{journal} {Phys. Rev. Lett.}\ }\textbf {\bibinfo {volume} {93}},\ \bibinfo
		{pages} {016602} (\bibinfo {year} {2004})}\BibitemShut {NoStop}%
	\bibitem [{\citenamefont {Zhang}\ \emph {et~al.}(2019)\citenamefont {Zhang},
		\citenamefont {Wang}, \citenamefont {Lu}, \citenamefont {Sui}, \citenamefont
		{Xu}, \citenamefont {Yu},\ and\ \citenamefont {Xue}}]{Zhang19}%
	\BibitemOpen
	\bibfield  {author} {\bibinfo {author} {\bibfnamefont {D.}~\bibnamefont
			{Zhang}}, \bibinfo {author} {\bibfnamefont {Y.}~\bibnamefont {Wang}},
		\bibinfo {author} {\bibfnamefont {N.}~\bibnamefont {Lu}}, \bibinfo {author}
		{\bibfnamefont {X.}~\bibnamefont {Sui}}, \bibinfo {author} {\bibfnamefont
			{Y.}~\bibnamefont {Xu}}, \bibinfo {author} {\bibfnamefont {P.}~\bibnamefont
			{Yu}},\ and\ \bibinfo {author} {\bibfnamefont {Q.-K.}\ \bibnamefont {Xue}},\
	}\href@noop {} {\bibfield  {journal} {\bibinfo  {journal} {Phys. Rev. B}\
		}\textbf {\bibinfo {volume} {100}},\ \bibinfo {pages} {060403} (\bibinfo
		{year} {2019})}\BibitemShut {NoStop}%
	\bibitem [{\citenamefont {Koster}\ \emph {et~al.}(1998)\citenamefont {Koster},
		\citenamefont {Kropman}, \citenamefont {Rijnders}, \citenamefont {Blank},\
		and\ \citenamefont {Rogalla}}]{Koster98}%
	\BibitemOpen
	\bibfield  {author} {\bibinfo {author} {\bibfnamefont {G.}~\bibnamefont
			{Koster}}, \bibinfo {author} {\bibfnamefont {B.~L.}\ \bibnamefont {Kropman}},
		\bibinfo {author} {\bibfnamefont {G.~J.}\ \bibnamefont {Rijnders}}, \bibinfo
		{author} {\bibfnamefont {D.~H.}\ \bibnamefont {Blank}},\ and\ \bibinfo
		{author} {\bibfnamefont {H.}~\bibnamefont {Rogalla}},\ }\href@noop {}
	{\bibfield  {journal} {\bibinfo  {journal} {Appl. Phys. Lett.}\ }\textbf
		{\bibinfo {volume} {73}},\ \bibinfo {pages} {2920} (\bibinfo {year}
		{1998})}\BibitemShut {NoStop}%
	\bibitem [{\citenamefont {Klein}\ \emph {et~al.}(1996)\citenamefont {Klein},
		\citenamefont {Dodge}, \citenamefont {Ahn}, \citenamefont {Reiner},
		\citenamefont {Mieville}, \citenamefont {Geballe}, \citenamefont {Beasley},\
		and\ \citenamefont {Kapitulnik}}]{klein96}%
	\BibitemOpen
	\bibfield  {author} {\bibinfo {author} {\bibfnamefont {L.}~\bibnamefont
			{Klein}}, \bibinfo {author} {\bibfnamefont {J.}~\bibnamefont {Dodge}},
		\bibinfo {author} {\bibfnamefont {C.}~\bibnamefont {Ahn}}, \bibinfo {author}
		{\bibfnamefont {J.}~\bibnamefont {Reiner}}, \bibinfo {author} {\bibfnamefont
			{L.}~\bibnamefont {Mieville}}, \bibinfo {author} {\bibfnamefont
			{T.}~\bibnamefont {Geballe}}, \bibinfo {author} {\bibfnamefont
			{M.}~\bibnamefont {Beasley}},\ and\ \bibinfo {author} {\bibfnamefont
			{A.}~\bibnamefont {Kapitulnik}},\ }\href@noop {} {\bibfield  {journal}
		{\bibinfo  {journal} {J. Phys. Condens. Matter}\ }\textbf {\bibinfo {volume}
			{8}},\ \bibinfo {pages} {10111} (\bibinfo {year} {1996})}\BibitemShut
	{NoStop}%
	\bibitem [{\citenamefont {Kresse}\ and\ \citenamefont
		{Furthm{\"u}ller}(1996)}]{Kresse96}%
	\BibitemOpen
	\bibfield  {author} {\bibinfo {author} {\bibfnamefont {G.}~\bibnamefont
			{Kresse}}\ and\ \bibinfo {author} {\bibfnamefont {J.}~\bibnamefont
			{Furthm{\"u}ller}},\ }\href@noop {} {\bibfield  {journal} {\bibinfo
			{journal} {Phys. Rev. B}\ }\textbf {\bibinfo {volume} {54}},\ \bibinfo
		{pages} {11169} (\bibinfo {year} {1996})}\BibitemShut {NoStop}%
	\bibitem [{\citenamefont {Kresse}\ and\ \citenamefont
		{Joubert}(1999)}]{Kresse99}%
	\BibitemOpen
	\bibfield  {author} {\bibinfo {author} {\bibfnamefont {G.}~\bibnamefont
			{Kresse}}\ and\ \bibinfo {author} {\bibfnamefont {D.}~\bibnamefont
			{Joubert}},\ }\href@noop {} {\bibfield  {journal} {\bibinfo  {journal} {Phys.
				Rev. B}\ }\textbf {\bibinfo {volume} {59}},\ \bibinfo {pages} {1758}
		(\bibinfo {year} {1999})}\BibitemShut {NoStop}%
	\bibitem [{\citenamefont {Perdew}\ \emph {et~al.}(1996)\citenamefont {Perdew},
		\citenamefont {Burke},\ and\ \citenamefont {Ernzerhof}}]{Perdew96}%
	\BibitemOpen
	\bibfield  {author} {\bibinfo {author} {\bibfnamefont {J.~P.}\ \bibnamefont
			{Perdew}}, \bibinfo {author} {\bibfnamefont {K.}~\bibnamefont {Burke}},\ and\
		\bibinfo {author} {\bibfnamefont {M.}~\bibnamefont {Ernzerhof}},\ }\href@noop
	{} {\bibfield  {journal} {\bibinfo  {journal} {Phys. Rev. Lett.}\ }\textbf
		{\bibinfo {volume} {77}},\ \bibinfo {pages} {3865} (\bibinfo {year}
		{1996})}\BibitemShut {NoStop}%
	\bibitem [{\citenamefont {Bl{\"o}chl}(1994)}]{Blochl94}%
	\BibitemOpen
	\bibfield  {author} {\bibinfo {author} {\bibfnamefont {P.~E.}\ \bibnamefont
			{Bl{\"o}chl}},\ }\href@noop {} {\bibfield  {journal} {\bibinfo  {journal}
			{Phys. Rev. B}\ }\textbf {\bibinfo {volume} {50}},\ \bibinfo {pages} {17953}
		(\bibinfo {year} {1994})}\BibitemShut {NoStop}%
	\bibitem [{\citenamefont {Pizzi}\ \emph {et~al.}(2020)\citenamefont {Pizzi},
		\citenamefont {Vitale}, \citenamefont {Arita}, \citenamefont {Bl{\"u}gel},
		\citenamefont {Freimuth}, \citenamefont {G{\'e}ranton}, \citenamefont
		{Gibertini}, \citenamefont {Gresch}, \citenamefont {Johnson}, \citenamefont
		{Koretsune} \emph {et~al.}}]{Pizzi20}%
	\BibitemOpen
	\bibfield  {author} {\bibinfo {author} {\bibfnamefont {G.}~\bibnamefont
			{Pizzi}}, \bibinfo {author} {\bibfnamefont {V.}~\bibnamefont {Vitale}},
		\bibinfo {author} {\bibfnamefont {R.}~\bibnamefont {Arita}}, \bibinfo
		{author} {\bibfnamefont {S.}~\bibnamefont {Bl{\"u}gel}}, \bibinfo {author}
		{\bibfnamefont {F.}~\bibnamefont {Freimuth}}, \bibinfo {author}
		{\bibfnamefont {G.}~\bibnamefont {G{\'e}ranton}}, \bibinfo {author}
		{\bibfnamefont {M.}~\bibnamefont {Gibertini}}, \bibinfo {author}
		{\bibfnamefont {D.}~\bibnamefont {Gresch}}, \bibinfo {author} {\bibfnamefont
			{C.}~\bibnamefont {Johnson}}, \bibinfo {author} {\bibfnamefont
			{T.}~\bibnamefont {Koretsune}}, \emph {et~al.},\ }\href@noop {} {\bibfield
		{journal} {\bibinfo  {journal} {J. Phys. Condens. Matter.}\ }\textbf
		{\bibinfo {volume} {32}},\ \bibinfo {pages} {165902} (\bibinfo {year}
		{2020})}\BibitemShut {NoStop}%
	\bibitem [{\citenamefont {Liechtenstein}\ \emph {et~al.}(1995)\citenamefont
		{Liechtenstein}, \citenamefont {Anisimov},\ and\ \citenamefont
		{Zaanen}}]{Liechtenstein95}%
	\BibitemOpen
	\bibfield  {author} {\bibinfo {author} {\bibfnamefont {A.~I.}\ \bibnamefont
			{Liechtenstein}}, \bibinfo {author} {\bibfnamefont {V.~I.}\ \bibnamefont
			{Anisimov}},\ and\ \bibinfo {author} {\bibfnamefont {J.}~\bibnamefont
			{Zaanen}},\ }\href {https://doi.org/10.1103/PhysRevB.52.R5467} {\bibfield
		{journal} {\bibinfo  {journal} {Phys. Rev. B}\ }\textbf {\bibinfo {volume}
			{52}},\ \bibinfo {pages} {R5467} (\bibinfo {year} {1995})}\BibitemShut
	{NoStop}%
	\bibitem [{\citenamefont {Vaugier}\ \emph {et~al.}(2012)\citenamefont
		{Vaugier}, \citenamefont {Jiang},\ and\ \citenamefont
		{Biermann}}]{vaugier2012hubbard}%
	\BibitemOpen
	\bibfield  {author} {\bibinfo {author} {\bibfnamefont {L.}~\bibnamefont
			{Vaugier}}, \bibinfo {author} {\bibfnamefont {H.}~\bibnamefont {Jiang}},\
		and\ \bibinfo {author} {\bibfnamefont {S.}~\bibnamefont {Biermann}},\
	}\href@noop {} {\bibfield  {journal} {\bibinfo  {journal} {Phys. Rev. B}\
		}\textbf {\bibinfo {volume} {86}},\ \bibinfo {pages} {165105} (\bibinfo
		{year} {2012})}\BibitemShut {NoStop}%
	\bibitem [{\citenamefont {Bezdicka}\ \emph {et~al.}(1993)\citenamefont
		{Bezdicka}, \citenamefont {Wattiaux}, \citenamefont {Grenier}, \citenamefont
		{Pouchard},\ and\ \citenamefont {Hagenmuller}}]{Bezdicka93}%
	\BibitemOpen
	\bibfield  {author} {\bibinfo {author} {\bibfnamefont {P.}~\bibnamefont
			{Bezdicka}}, \bibinfo {author} {\bibfnamefont {A.}~\bibnamefont {Wattiaux}},
		\bibinfo {author} {\bibfnamefont {J.}~\bibnamefont {Grenier}}, \bibinfo
		{author} {\bibfnamefont {M.}~\bibnamefont {Pouchard}},\ and\ \bibinfo
		{author} {\bibfnamefont {P.}~\bibnamefont {Hagenmuller}},\ }\href@noop {}
	{\bibfield  {journal} {\bibinfo  {journal} {Z. Anorg. Allg. Chem.}\ }\textbf
		{\bibinfo {volume} {619}},\ \bibinfo {pages} {7} (\bibinfo {year}
		{1993})}\BibitemShut {NoStop}%
\end{thebibliography}

%

\end{document}


\title{\bf{Supplementary Materials: Sign-tunable anomalous Hall effect induced by symmetry-protected nodal structures in ferromagnetic perovskite oxide thin films}}

\author[1,2]{Byungmin Sohn\thanks{These authors contributed equally to this work.}}
\author[1,2,3]{Eunwoo Lee\thanks{These authors contributed equally to this work.}}
\author[4,5]{Se Young Park}
\author[1,2]{Wonshik Kyung}
\author[6]{Jinwoong Hwang}
\author[6]{Jonathan. D. Denlinger}
\author[1,2]{Minsoo Kim}
\author[1,2]{Donghan Kim}
\author[1,2]{Bongju Kim}
\author[1,2]{Hanyoung Ryu}
\author[1,2]{Soonsang Huh}
\author[1,2]{Ji Seop Oh}
\author[1,2]{Jong Keun Jung}
\author[1,2]{Dongjin Oh}
\author[1,2]{Younsik Kim}
\author[7]{Moonsup Han}
\author[1,2]{Tae Won Noh}
\author[1,2,3]{Bohm-Jung Yang\thanks{Electronic address:$~~${bjyang@snu.ac.kr}}}
\author[1,2]{Changyoung Kim\thanks{Electronic address:$~~${changyoung@snu.ac.kr}}}

\affil[1]{Department of Physics and Astronomy, Seoul National University, Seoul 08826, Korea}
\affil[2]{Center for Correlated Electron Systems, Institute for Basic Science, Seoul 08826, Korea}
\affil[3]{Center for Theoretical Physics (CTP), Seoul National University, Seoul 08826, Korea}
\affil[4]{Department of Physics, Soongsil University, Seoul, 06978 Korea}
\affil[5]{Integrative Institute of Basic Sciences, Soongsil University, Seoul, 06978 Korea}
\affil[6]{Advanced Light Source, Lawrence Berkeley National Laboratory, Berkeley, CA 94720, USA}
\affil[7]{Department of Physics, University of Seoul, Seoul, 02504, Korea}

\date{}

\maketitle
\clearpage
\tableofcontents

\clearpage

\begin{center}
\section{Characterization of SrRuO$_3$ ultrathin films}
\end{center}

\begin{figure*}[h!]
\centering
\includegraphics[width=17cm]{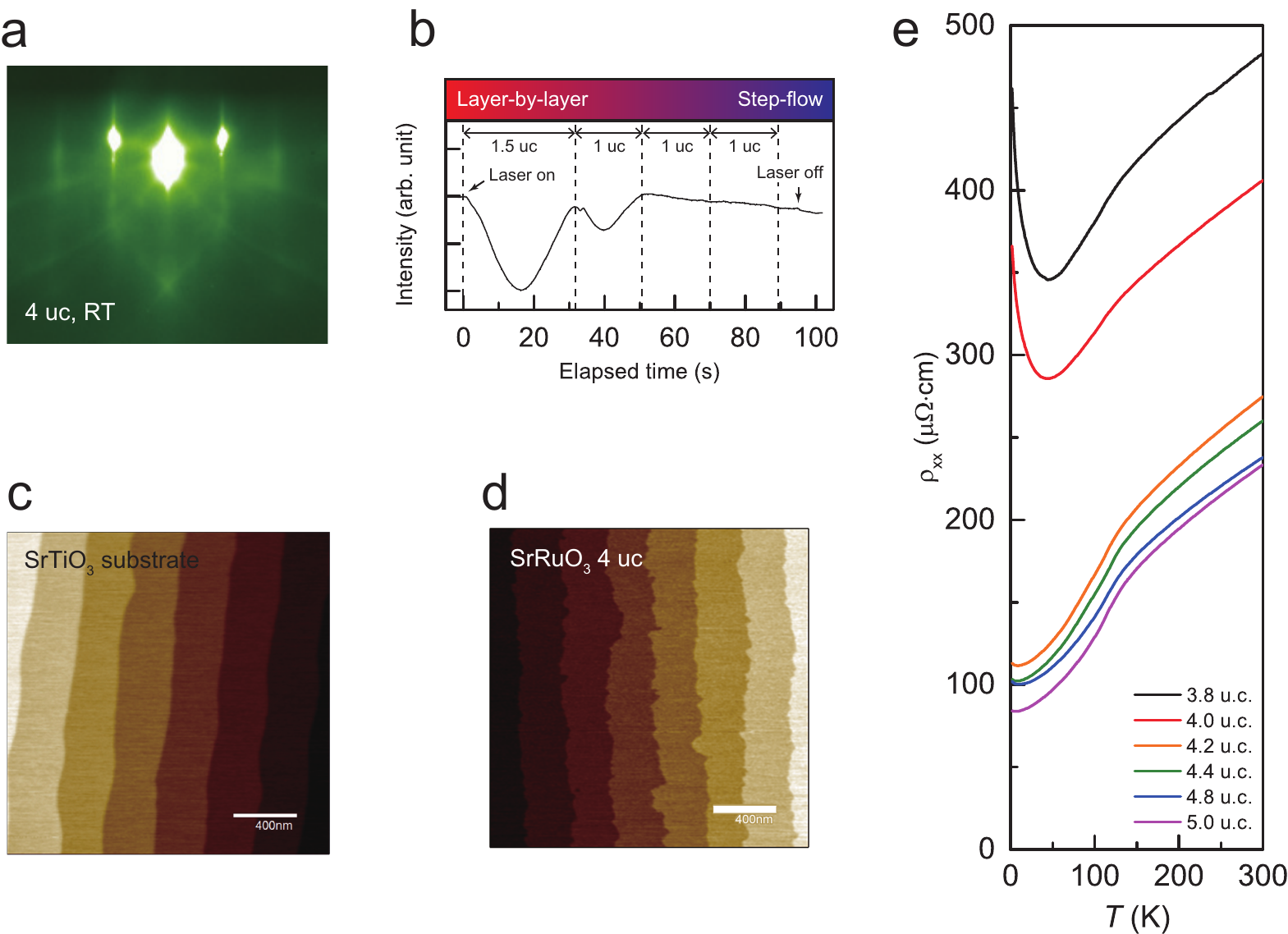}
\caption{{\bf Characterization of SrRuO$_3$ (SRO) ultrathin films.} (a) An {\it in-situ} reflection high-energy electron diffraction (RHEED) pattern of 4 unit cell (uc) SRO thin film taken at room temperature. (b) {\it In-situ} RHEED intensity $vs.$ elapsed time plot during 4.4~uc SRO growth. Growth mode changes from the initial layer-by-layer to step-flow mode after 2 uc growth. 40~laser pulses are necessary to grow 1~uc. (c) Atomic force microscopy (AFM) image of TiO$_2$-terminated SrTiO$_3$ (001) substrate. (d) AFM image of 4~uc SRO thin film. The dimension for both AFM images is $2$~$\mu$m~$\times$~$2$~$\mu$m. (e) Temperature dependent resistivity for various thickness SRO ultrathin films.}
\label{Ch1}
\end{figure*}

In order to monitor the sample growth process, {\it in-situ} reflection high-energy electron diffraction (RHEED) was performed in the chamber during the growth. Figure \ref{Ch1}a shows a RHEED pattern from 4~unit-cell (uc) SrRuO$_3$ (SRO) thin film. Reciprocal rods have spot-shape rather than streak-shape, indicating that the SRO thin film has flat and single-crystalline surface~\cite{Hasegawa12}. Figure \ref{Ch1}b shows RHEED intensity as a function of the growth time. It is known that an annealed SrTiO$_3$ (STO) (001) substrate has TiO$_2$-terminated (B-site terminated) surface~\cite{Ohnishi04}. Since SRO thin films grow to have SrO-terminated (A-site terminated) surface, it takes about $1.5$~times longer to grow the first unit cell of the film~\cite{Choi01}. 40~laser pulses (20~seconds with 2~Hz growth rate) are required for 1~uc SRO growth.

During the film growth, a variety of surface diffusion processes are expected depending on the temperature, pressure, and deposition rate in the pulsed laser deposition (PLD) method~\cite{Choi01,Lippmaa00}. In the case of SRO (001), the growth mode changes from the layer-by-layer mode to the step-flow mode under appropriate conditions~\cite{Choi01}. We also have observed that there is a change in the growth mode after 2~uc growth (Fig. \ref{Ch1}b). Assuming that the growth rate does not change with growth mode, 40~laser pulses are equivalent to 1~uc SRO in the step-flow growth mode. We estimate the thickness of the SRO ultrathin film based on these values, even to a fractional thickness.

Figures \ref{Ch1}c and \ref{Ch1}d are atomic force microscopy (AFM) images of an STO substrate and a 4~uc SRO thin film, showing the surface morphology. Clear step terraces are seen with a width of $\sim300$~nm. Clear terraces seen on SRO film imply that the film was grown in the step-flow growth mode. Figure \ref{Ch1}e shows temperature and thickness dependent resistivity ($\rho_{xx}$) for a thickness range of 3.8 - 5.0~uc. The low-temperature up-turn of ultrathin samples disappears as the film thickness increases.

\begin{figure*}[h!]
\centering
\includegraphics[width=17cm]{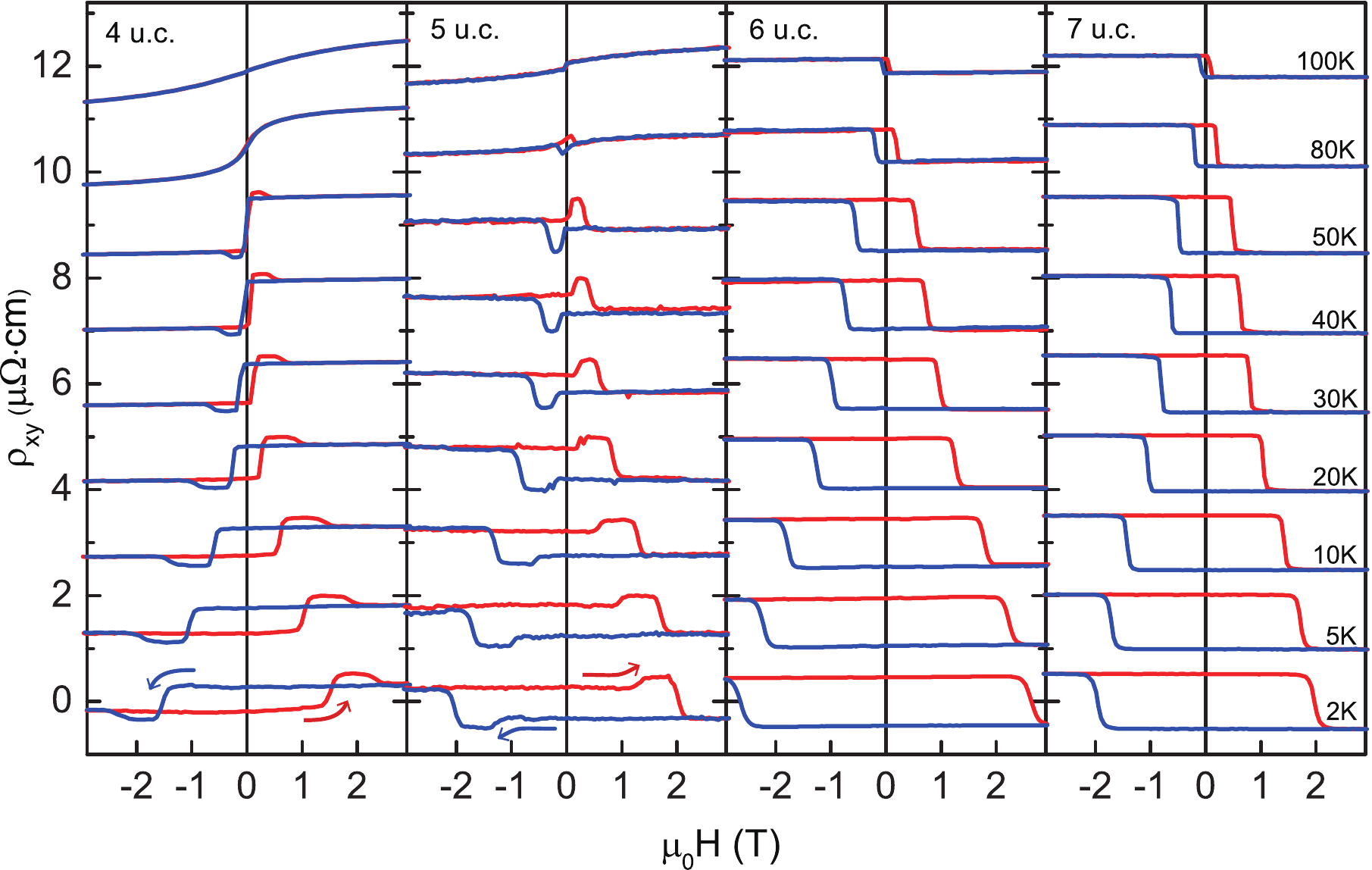}
\caption{{\bf Hall resistivity of $4$-$7$~uc SRO thin films.} (Ref. B. Sohn {\it et al.}, arXiv:1810.01615 [cond-mat.str-el])}
\label{Ch2}
\end{figure*}

Plotted in Fig. \ref{Ch2} are Hall resistivity ($\rho_{xy}$) of 4, 5, 6, and 7 uc SRO thin films taken at various temperatures~\cite{sohn18}. Anomalous Hall resistivity varies with both film thickness and temperature. In particular, we note that the sign of anomalous Hall effect (AHE) is inverted between 4 and 5~uc SRO thin films. We calculate the Hall conductivity by using the equation $\sigma_{xy}=\rho_{xy}/(\rho_{xx}^2+\rho_{xy}^2)$ and plot it in Figs. 4d and 4e.

\begin{figure*}[h!]
\centering
\includegraphics[width=17cm]{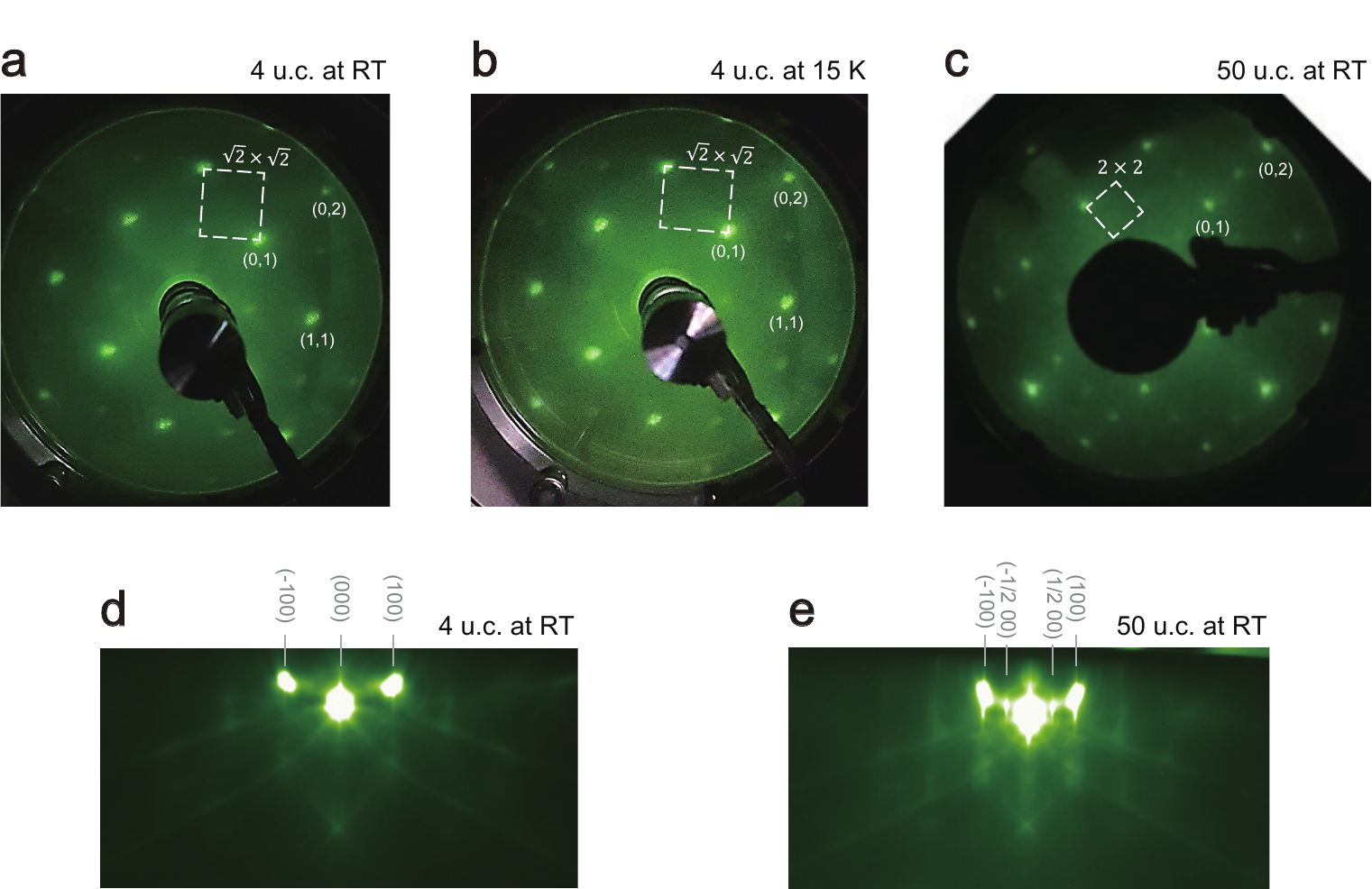}
\caption{{\bf \textit{In-situ} low-energy electron diffraction (LEED) and RHEED patterns of SRO thin film.} (a), (b) LEED patterns of 4~uc SRO at room temperature and 15~K with the electron beam energy of 142~eV. (c) LEED pattern of 50~uc SRO at room temperature with the electron beam energy of 150~eV. (d) RHEED image of 4~uc SRO at room temperature. (e) RHEED image of 50~uc SRO at room temperature.}
\label{Ch3}
\end{figure*}

After the SRO thin films were synthesized in the PLD chamber, films were characterized \textit{in-situ} by low-energy electron diffraction (LEED). LEED patterns of 4 and 50~uc SRO thin films are shown in Figs. \ref{Ch3}a, b, and c. Figures \ref{Ch3}a and b are LEED patterns of the 4 uc thin film at room temperature and 16~K. Both LEED patterns are taken with the electron beam energy of 142~eV. The pseudo-cubic diffraction peaks are indicated in figures. White dotted rectangles represent primitive cells of SRO. The 4~uc SRO thin film shows $\sqrt{2}\times\sqrt{2}$ spots. In contrast, the 50~uc SRO film data in Fig. \ref{Ch3}c shows $2 \times 2$ spots, indicating that the in-plane real space unit cell is doubled with respect to that of the 4 uc SRO thin film.

RHEED patterns were also obtained after the growth. Figures \ref{Ch3}d and e show RHEED patterns of 4 and 50~uc SRO thin films, respectively, taken at room temperature with the electron beam direction along the [100]$_c$ azimuth. Bragg spots at the 0\textsuperscript{th} and 1\textsuperscript{st} Laue circles and Kikuchi lines are clearly observed, implying that our SRO thin films have high crystalline order at the surface. In Fig. \ref{Ch3}d, only (-100), (000) and (100) Bragg peaks are observed whereas (${1\over2}$00) and (-${1\over2}$00) Bragg peaks are observed additionally in Fig. \ref{Ch3}e, also representing that the in-plane unit cell of 50~uc SRO thin film is doubled with respect to the 4~uc SRO thin film. A previous study showed that thick SRO films grown on the STO substrate at room temperature or below have the orthorhombic structure (Pbnm, $a^{-}a^{-}c^{+}$), while SRO thin films thinner than 17~uc are stabilized with the tetragonal structure (I4/mcm, $a^{0}a^{0}c^{-}$)~\cite{Chang11}. Our 4~uc SRO thin film also maintains the tetragonal structure (I4/mcm, $a^{0}a^{0}c^{-}$) without any octahedral tilting at room temperature and 15~K.

\clearpage

\vspace{\baselineskip}
\begin{center}
\section{Angle-resolved photoemission spectroscopy (ARPES) results of SrRuO$_3$ ultrathin film with He-I$\alpha$ light}
\end{center}

\begin{figure*}[h]
\centering
\includegraphics[width=1\textwidth]{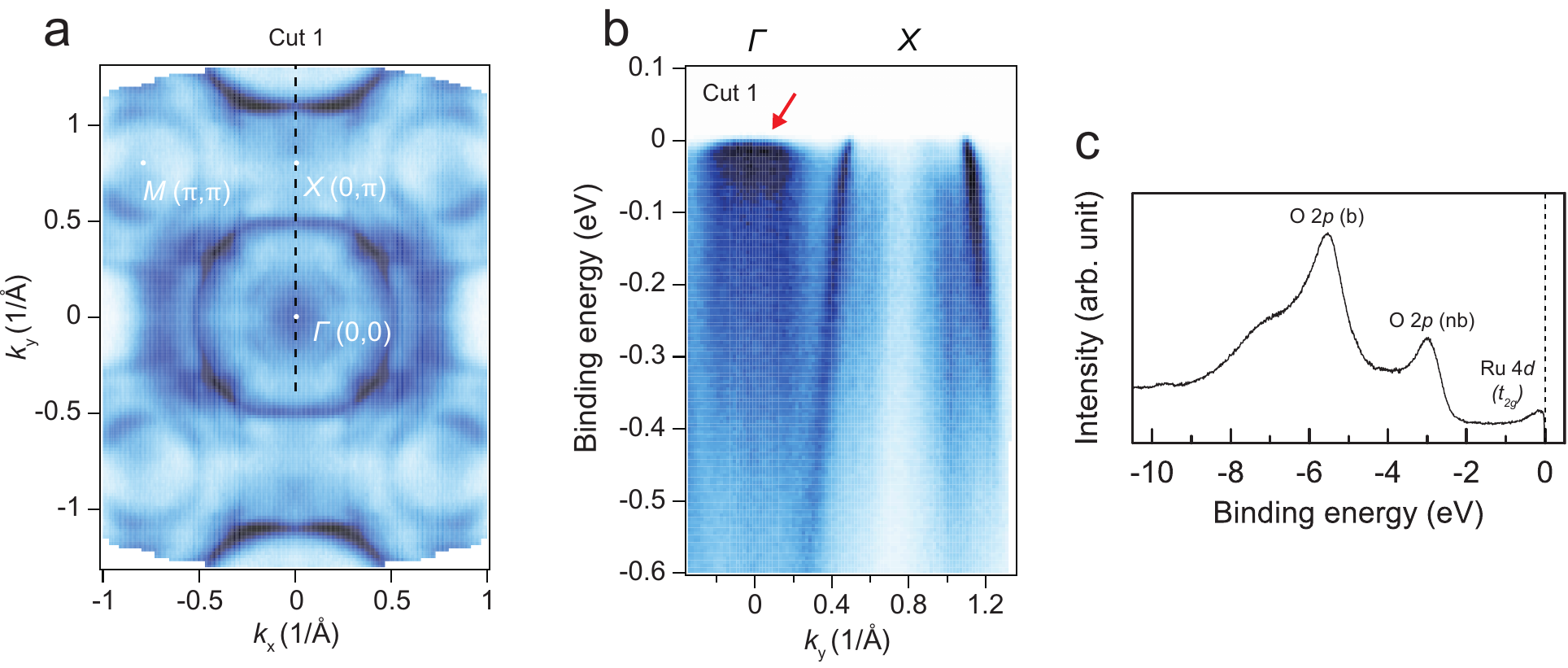}
\caption{{\bf Home-lab based {\it in-situ} ARPES data from 4~uc SRO thin film with He I$\alpha$ photon.} (a) Fermi surface map at 10~K obtained by integrating the spectral function within an energy window of E$_F\pm 10$~meV. (b) $\Gamma$-$X$ high symmetry cut (Cut 1 in (a)). Clear $\beta$ and folded $A$ bands are observed. The folded $A$ band is indicated with the red arrow. (c) Angle-integrated photoemission data taken at 10~K.}
\label{HL}
\end{figure*}

{\it In-situ} angle-resolved photoemission spectroscopy (ARPES) measurements were performed on 4~uc SRO thin films with He I$\alpha$ at 10~K using a home-lab based system. Clear Fermi surfaces were observed as shown in Fig. \ref{HL}a (symmetrized). Figure \ref{HL}b shows the $\Gamma$-$X$ high symmetry cut (Cut 1 in panel a). The folded $A$ band (see Fg. 2f) is located near the $\Gamma$ point. A Fermi level crossing of the $\beta$ band is observed at $k_y=0.514$~${\AA^{-1}}$. We obtained angle-integrated photoemission data as shown in Fig. \ref{HL}c. An O~2$p$ bonding (non-bonding) state is located near $-5.5$ ($-3$)~eV as reported in the previous literature~\cite{Fujioka97, Shai13}. Near the Fermi level, the spectral weight has a main contribution from Ru 4$d$ $t_{2g}$ states~\cite{Kim05}.

\begin{figure*}[h]
\centering
\includegraphics[width=0.7\textwidth]{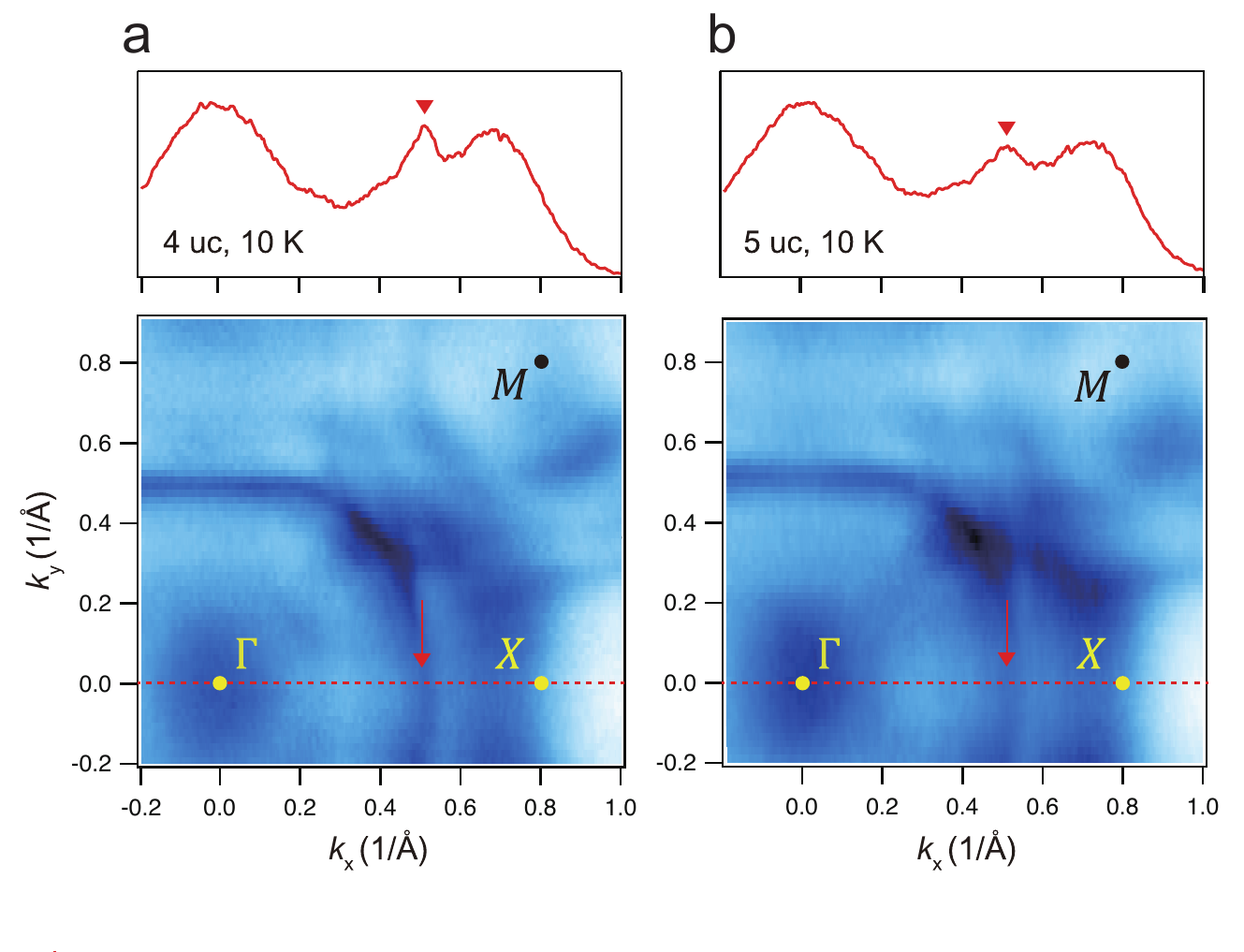}
\vspace{-0.5cm}
\caption{{\bf Fermi surface of 4 and 5~uc SRO thin films measured by {\it in-situ} ARPES.} (a, b) Fermi surface maps of 4 and 5~uc SRO thin films in ($k_x$, $k_y$) plane measured at 10~K. ARPES data are integrated within E$_F\pm$ 10~meV to produce the Fermi surface. Overall shapes of the Fermi surfaces are almost the same in two cases. (upper) Momentum distribution curves at Fermi level along the $\Gamma$-$X$ line, the red dashed line in the lower figures. Red inverted triangles and red arrows represent the peaks from $\beta$ band. The positions of red inverted triangles are at $k_x$ = 0.515 and 0.514~$\AA^{-1}$, respectively, which shows that $\beta$ electron pockets have almost the same size (the difference is less than 1~$\%$).}
\label{Thickness}
\end{figure*}

We also performed {\it in-situ} ARPES measurements on SRO thin films of different thicknesses (4 and 5 uc) to investigate thickness-dependent properties. Figure \ref{Thickness} shows momentum distribution curves along $k_y=0$ at the Fermi level (upper panels) and ($k_x,k_y$) Fermi surfaces (lower panels). The ARPES measurements were performed at 10~K with He I$\alpha$ using a home-lab based system.
	
Overall, the ARPES data show a similar fermiology in two cases. We extracted the $\beta$ band position from the Fermi surface maps. Inverted triangles in the upper panels of Fig. \ref{Thickness} mark peaks of the $\beta$ band along the $\Gamma$-$X$ line. The position of the band is found to be $k_x=0.515$ and $0.514$~${\AA^{-1}}$ for 4 and 5~uc thin films, respectively, which shows that the $\beta$ band pocket has almost the same size (difference less than 1~\%).
	
A study of thickness-dependent Fermi surfaces in SrVO$_3$ ultrathin films has been reported by Yoshimatsu {\it et al.}~\cite{Yoshimatsu11}. The results show that shapes of the Fermi surfaces in SrVO$_3$ ultrathin films remain similar regardless of the thickness. In case of SrRuO$_3$ ultrathin films, the overall band structure at the Fermi level for metallic ultrathin films remains similar regardless of the thickness as in the case of SrVO$_3$. We speculate that the overall similarity in the Fermi surface topology among films with different thicknesses comes from combination of the photon energy we used (21.2~eV) and the characteristic of t$_{2g}$ bands, which have semi-straight sections.

\clearpage

\vspace{\baselineskip}
\begin{center}
\section{Spin-resolved ARPES (SARPES) results on 4~uc SrRuO$_3$ thin film}
\end{center}

\begin{figure*}[h]
\centering
\includegraphics[width=1\textwidth]{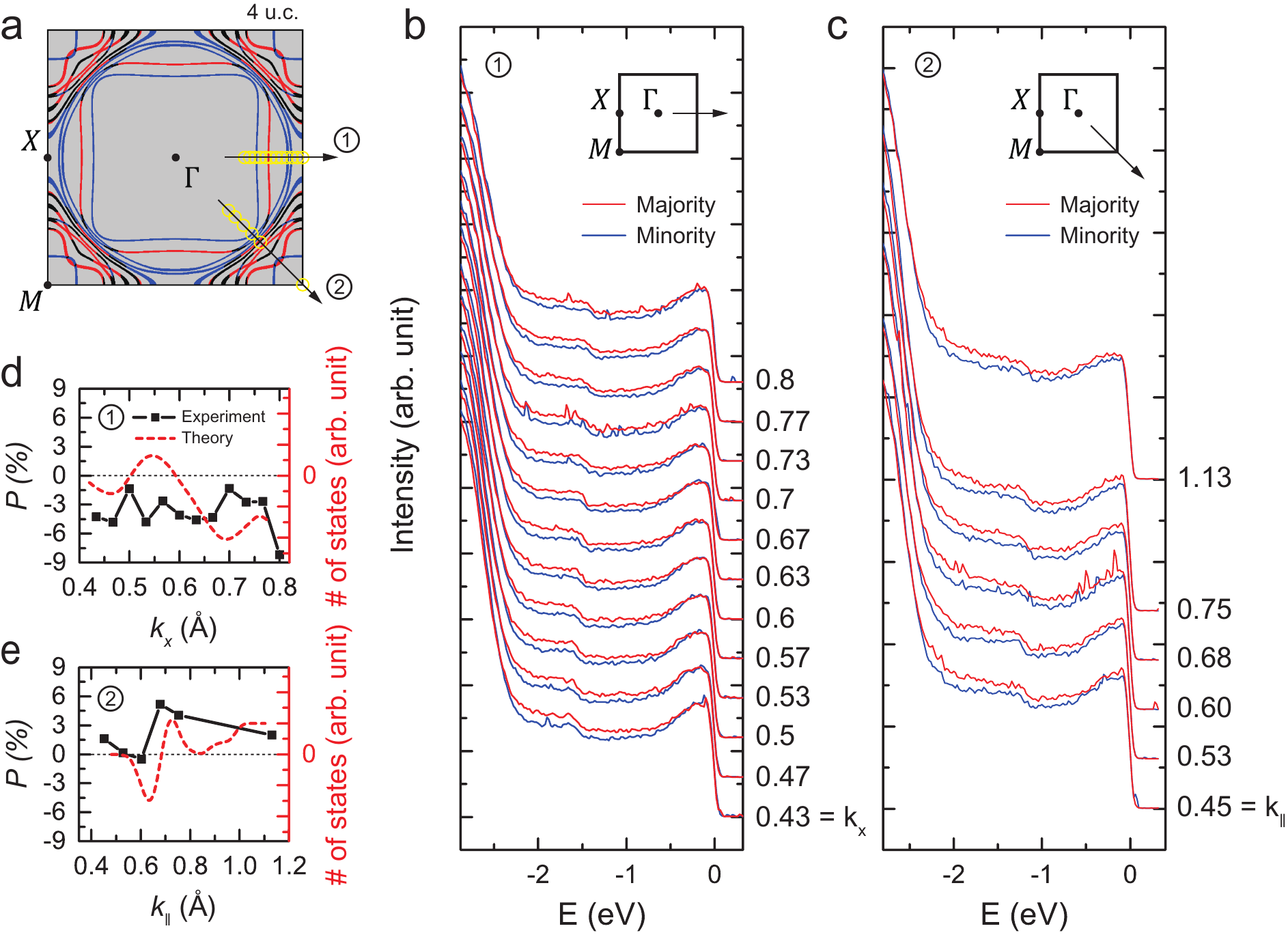}
\vspace{0cm}
\caption{{\bf Spin-resolved {\it in-situ} ARPES data of 4~uc SRO thin film along high-symmetry lines.}
(a) Calculated band structures of 4 uc SRO. Yellow open circles represent the points where spin ARPES is measured. The spin-resolved ARPES was performed along the high symmetry lines indicated by black arrows. (b, c) Spin-resolved ARPES data of the valence band along the $\Gamma$-$X$ and $\Gamma$-$M$ lines. (d, e) Spin polarizations at Fermi surface (FS) along the $\Gamma$-$X$ and $\Gamma$-$M$ lines. Here the spin-resolved ARPES data is integrated from E$_F$ to E$_F$ - 60~meV. For comparison, the number of spin-polarized states at each momentum, calculated by using tight binding model, is also plotted.}
\label{SpinARPES}
\end{figure*}

\begin{figure*}[h]
\centering
\includegraphics[width=0.5\textwidth]{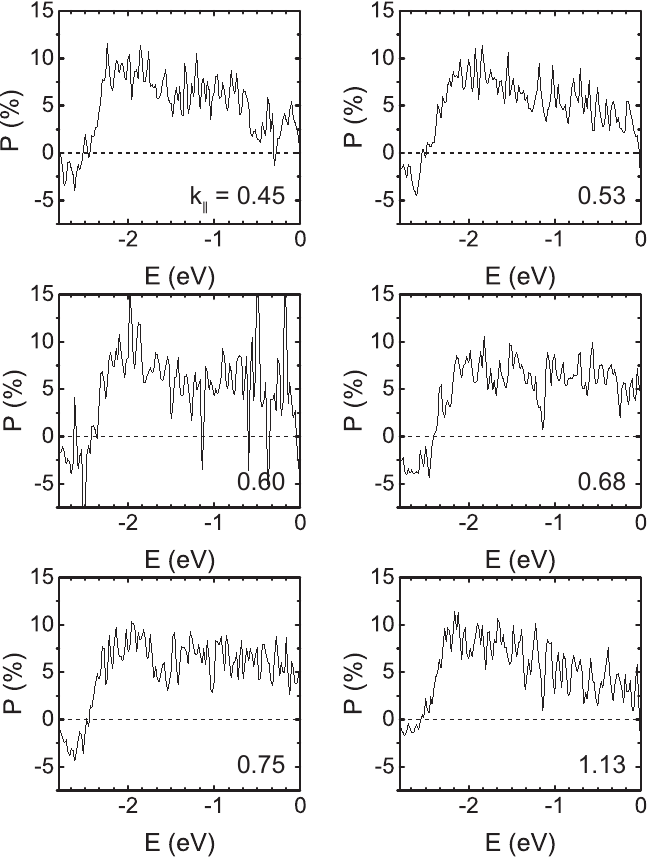}
\vspace{0cm}
\caption{{\bf Spin polarization data of 4 uc SRO at 10~K along the $\Gamma$-$M$ line.} Each spin polarization is presented as a function of the electron momentum, $k$.}
\label{GtoM}
\end{figure*}

\begin{figure*}[h]
\centering
\includegraphics[width=0.5\textwidth]{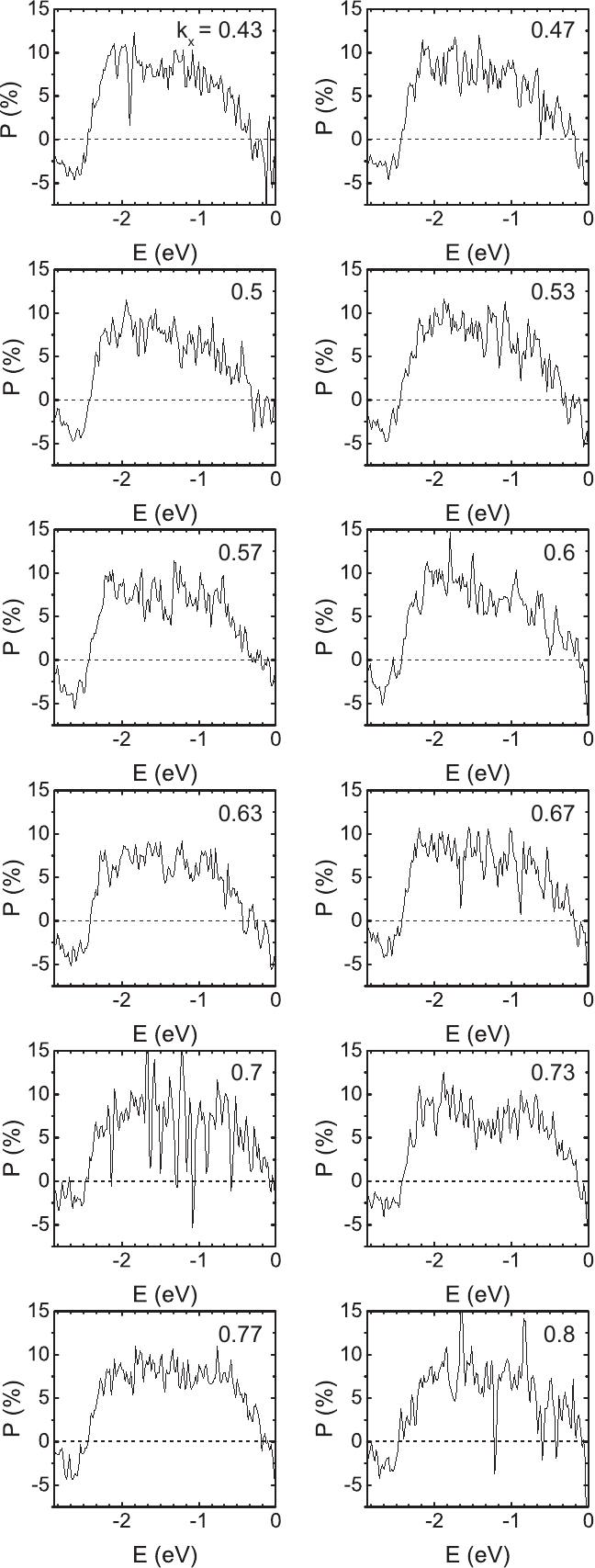}
\vspace{0cm}
\caption{{\bf Spin polarization data of 4 uc SRO at 10~K along the $\Gamma$-$X$ line.} Each spin polarization is presented as a function of the electron momentum, $k$.}
\label{GtoX}
\end{figure*}

We performed spin-resolved ARPES (SARPES) on 4~uc SRO thin films at 10~K, and compare the experimentally obtained spin polarization at the Fermi level with that from tight-binding model calculation. Figure \ref{SpinARPES}a shows calculated band structures of 4~uc SRO thin film. SARPES were obtained along high-symmetry cuts of $\Gamma$-$X$ and $\Gamma$-$M$ as indicated by black arrows and yellow open circles. Figures \ref{SpinARPES}b and c show the spin-resolved ARPES data along the two high-symmetry cuts. Spin polarizations, $P=(I_{\uparrow}-I_{\downarrow})/(I_{\uparrow}+I_{\downarrow})$, along $\Gamma$-$X$ and $\Gamma$-$M$ cuts are plotted in Figs. \ref{GtoM} and \ref{GtoX}, respectively. Based on these results, we obtain the spin polarizations near the Fermi level and plot them in Figs. \ref{SpinARPES}d and e along with calculated spin polarizations. We see that behaviors of the experimental and calculated spin polarization show similar trend (within a factor).


There are only a handful of studies on the spin polarization of SRO thin films. A negative spin polarization of -9.5~\% was initially reported for 50~nm SRO film by using tunneling measurements~\cite{Worledge00}. Subsequently. A positive value of 50~\% spin polarization was obtained for 120 nm thick SRO films by using Andreev reflection spectroscopy~\cite{Nadgorny03}. These experiments only show local spin polarization value of SRO thick films. On the other hand, our SARPES measurements provide momentum-dependent spin polarization values of SRO ultrathin films. Interestingly, the polarization difference is observed up to 2.5 eV below the Fermi level, which shows that the magnetization is not confined to Ru t$_{2g}$ orbital but has a contribution from Ru-O hybridized orbitals. In addition, we observe that the sign and magnitude of the spin polarization varies with the momentum. Our SARPES work shows momentum-dependent spin polarization in SRO ultrathin film, which in turn suggests that the material possesses an itinerant ferromagnetism character.

\clearpage

\vspace{\baselineskip}
\begin{center}
\section{Synchrotron ARPES measurements on 4~uc SrRuO$_3$ thin film}
\end{center}

\begin{figure*}[h]
\centering
\includegraphics[width=0.5\textwidth]{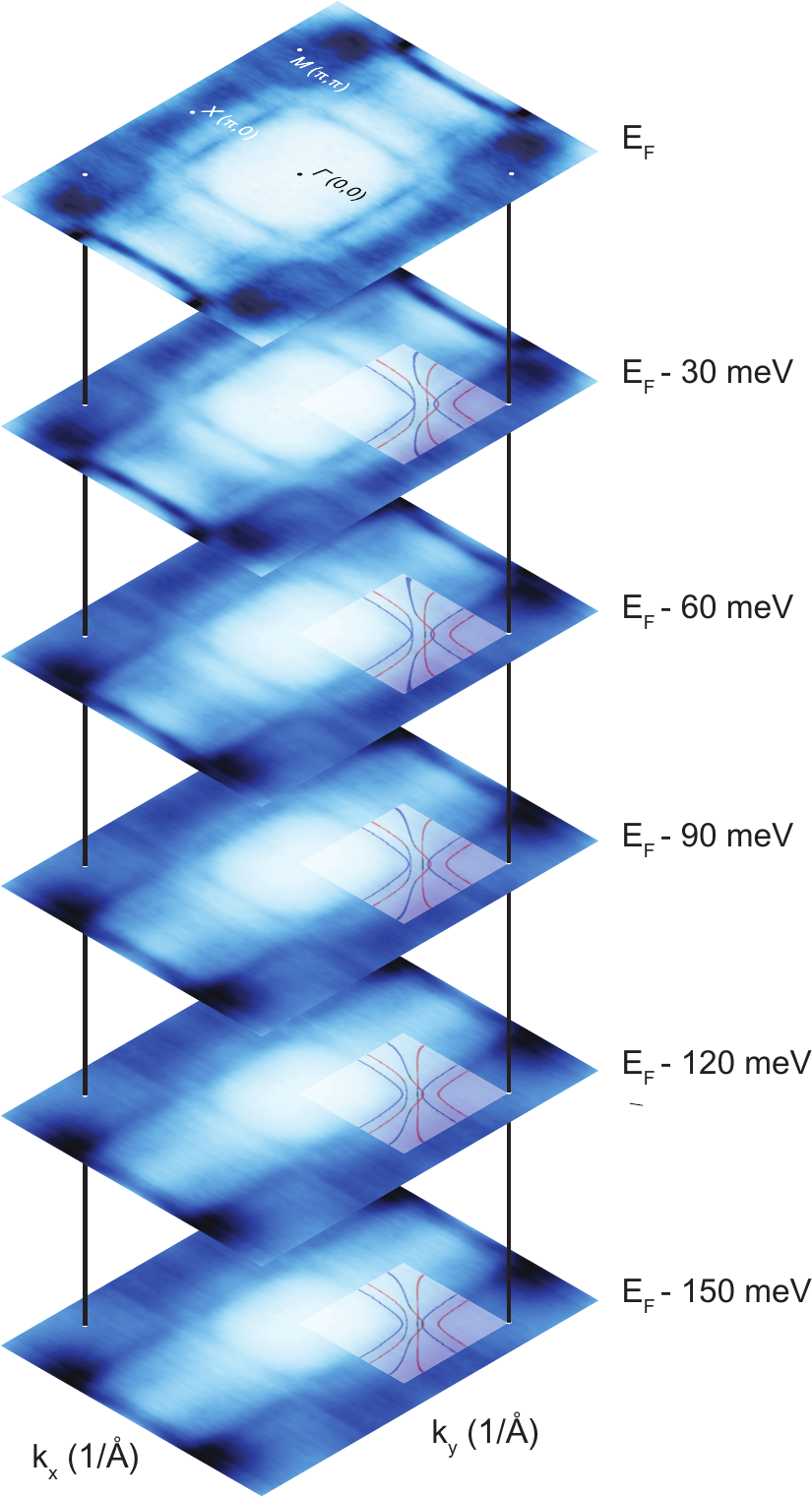}
\caption{{\bf Constant-energy maps at the binding energy of 0, 30, 60, 90, 120, and 150~meV along with tight binding model fits.}}
\label{ALS2}
\end{figure*}

Synchrotron ARPES measurements were performed on 4~uc films. Figure \ref{ALS2} shows constant energy maps in the $k_x$-$k_y$ plane at various binding energies for data taken with 80~eV photon. Note that folded bands are weaker in the synchrotron data due to matrix element effects. Overlaid on the maps are theoretical fits based on the tight-binding calculation. Our calculation and experimental data are quite consistent with each other not only near the Fermi energy but also at high binding energies.

Note that the spin majority bands are not clearly seen in the data. Within the itinerant magnetism, the bands should split into majority and minority spin bands when magnetization is nonzero. As one can see in Figs. 2 and \ref{ALS2}, the sharp bands are minority spin bands, whereas the corresponding spin majority bands are the broad spectral weight. According to the DMFT calculation results by Kim {\it et al.}~\cite{Kim15}, majority spin bands are mostly incoherent in the case of SRO, like in the case of Fe, Co, and Ni~\cite{Lichtenstein01, Monastra02, Grechnev07}. Thus, it is not abnormal to see that majority bands are more incoherent than minority bands in SRO.

\begin{figure*}[h]
\centering
\includegraphics[width=1.0\textwidth]{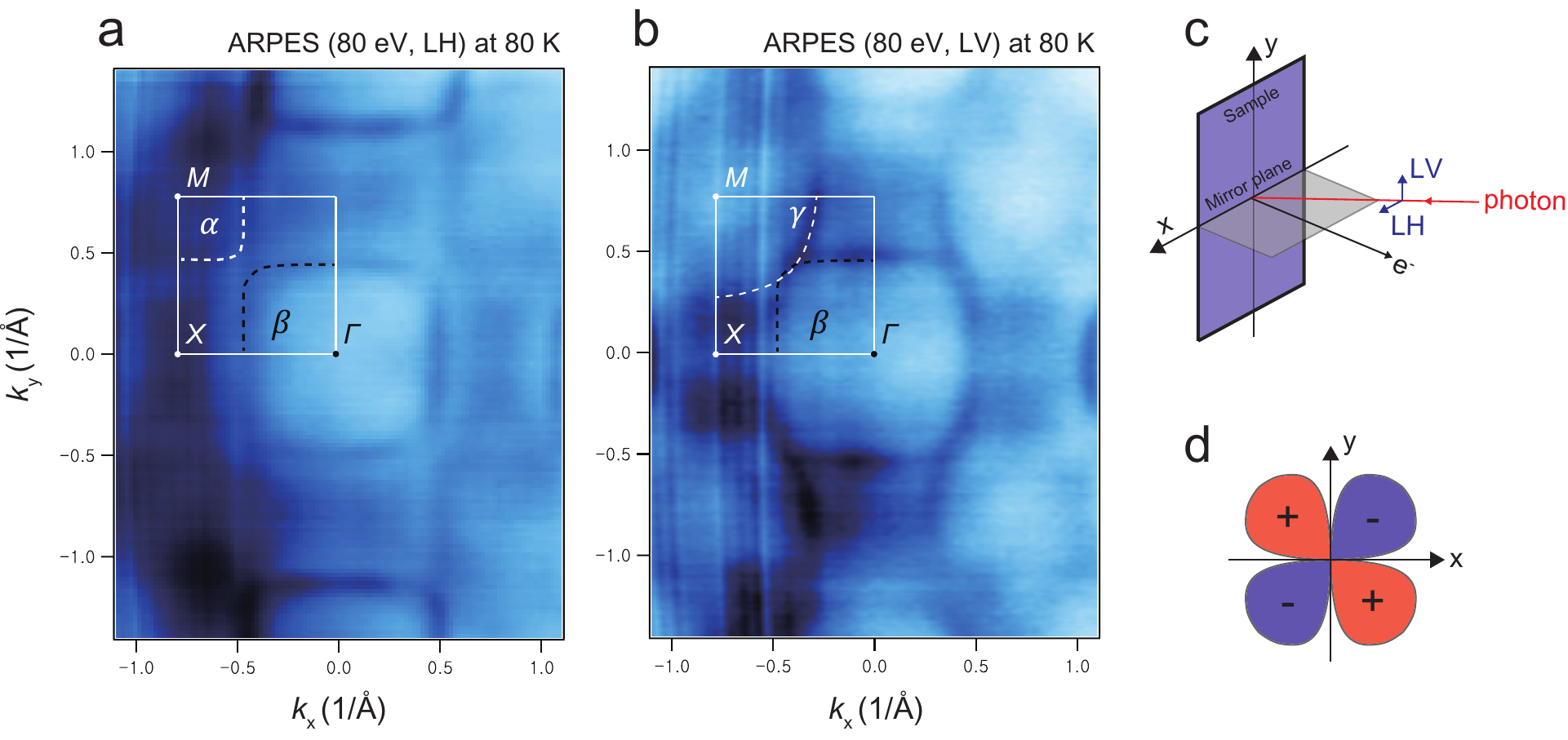}
\caption{{\bf Polarization-dependent ARPES data of 4~uc SRO thin film at 80~K.} (a, b) Fermi surface map in the ($k_x$, $k_y$) plane taken with linear horizontal (LH) and linear vertical (LV) light with a photon energy of 80~eV. (c) Schematic diagram for the ARPES experimental geometry. LV and LH show the two polarizations used in the experiment. A mirror plane coincides with the $xz$-plane. Note that even- (odd-) symmetic initial states are only observable with p- (s-) polarization following the dipole selection rule. (d) A schematic for $d_{xy}$ orbital. Notice that $d_{xy}$ orbital is odd-symmetric to the mirror plane.}
\label{LHLV}
\end{figure*}

In order to precisely determine the band configuration in the Fermi surfaces, we performed polarization-dependent ARPES measurements at 80~K with the photon energy of 80~eV, as shown in Fig. \ref{LHLV}. Orbital characters of the bands can be determined via polarization-dependent ARPES measurements; one can obtain the wave function parity of the initial state with respect to the experimental mirror plane. Following the dipole selection rule~\cite{Eberhardt80}, the even- (odd-)symmetric initial states are only observable with p (s) polarization~\cite{Iwasawa10}. As Fig. \ref{LHLV}a shows, one can see that $\alpha$ and $\beta$ bands are clearly observed with linear-horizontal (LH) beam. On the other hand, in Fig. \ref{LHLV}b, the $\gamma$ band is clearly observed with linear-vertical (LV) beam. Since the LH (LV) beam is p- (s-)polarized in our ARPES experimental geometry, we can identify that the $\gamma$ band is odd-symmetric to the mirror plane (Fig. \ref{LHLV}c). As expected, the $\gamma$ band consists of Ru $d_{xy}$ orbital since $d_{xy}$ orbital is odd-symmetric to the mirror plane as Fig. \ref{LHLV}d shows. This experimental result is also consistent with our theoretical result.


Since the topological structures generate multi-sign Berry curvature on the Fermi surfaces, it would be desirable to directly show the Berry curvature in the momentum space. During the past decade, measurements of the orbital angular momentum (OAM) and Berry curvature with circular-dichroism (CD) ARPES have been well established~\cite{Liu11, Park12, Park12_2, Cho18, Schuler20, 2020momentumspace}. We can utilize CD ARPES on SRO films to study OAM and Berry curvature. Such a study can then verify whether the multiple sign of Berry curvature exists or not.

We performed CD ARPES on a 4 uc SRO film to investigate the Berry curvature. ARPES data were taken with left- and right-circularly polarized (LCP and RCP, respectively) 80~eV light at 20~K. To magnetize the sample, the 4 uc SRO film was field-cooled with a permanent magnet. The light incident angle was $45 \pm 10^{\circ}$ to the sample surface normal, so the out-of-plane OAM could be observed~\cite{Park12}.

\begin{figure*}[h]
\centering
\includegraphics[width=0.65\textwidth]{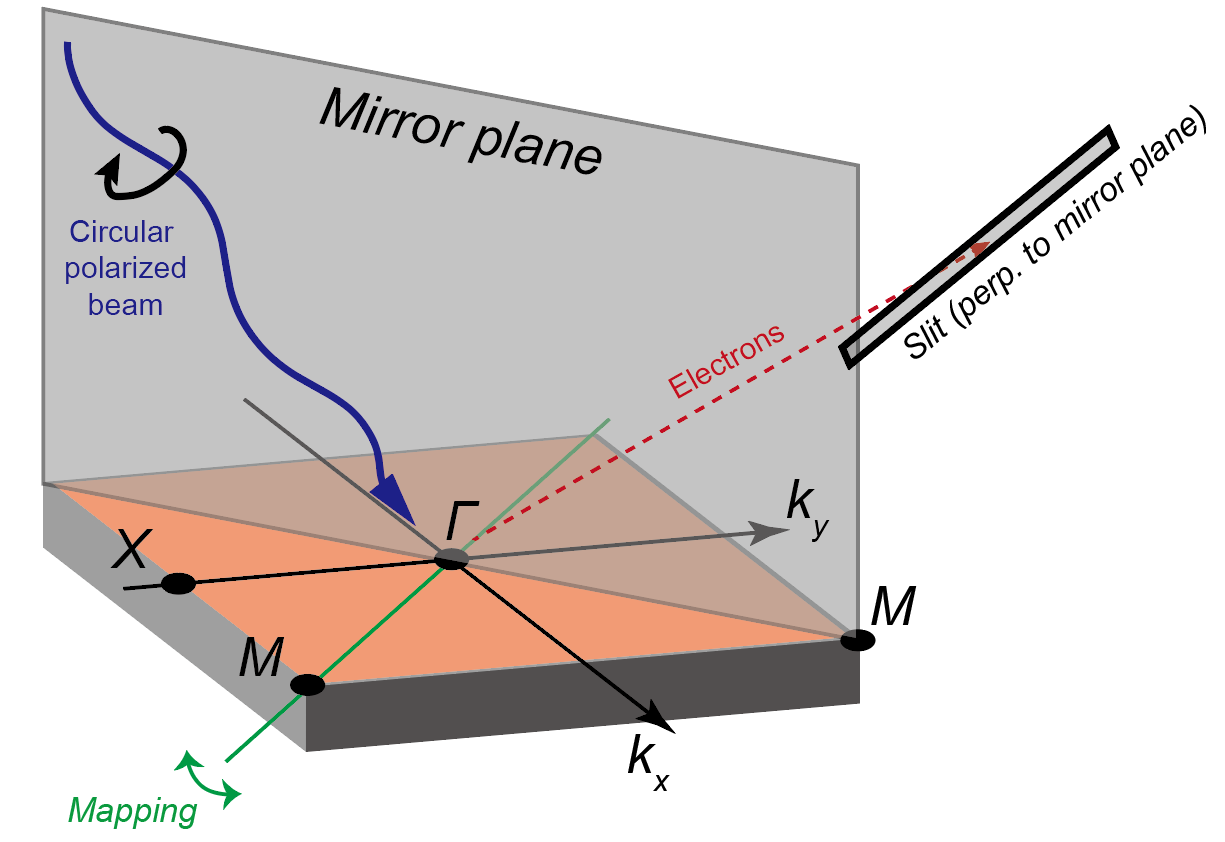}
\caption{{\bf Schematic illustration of the geometry for the circular-dichroism (CD) ARPES experiment.}}
\label{CD}
\end{figure*}

The experimental mirror plane, which was normal to the sample surface and contained the incident light wave vector, was precisely aligned to cross both the $\Gamma$ and $M$ points (Fig.~\ref{CD}). Then, the CD ARPES was taken along the momentum perpendicular to the mirror plane (still parallel to the $\Gamma$-$M$ direction due to the four-fold symmetry), $i.e.$ the slit was perpendicular to the mirror plane. The geometrical effect which is caused by a mirror symmetry breaking in the experimental geometry can be removed using the fact that it is an odd function along the direction perpendicular to the mirror plane. Since electronic structures (and OAM) are an even function about the mirror plane with the experimental geometry, we can take the OAM contribution by extracting only the even part of the CD spectra~\cite{Cho18}.

\begin{figure*}[h]
\centering
\includegraphics[width=0.5\textwidth]{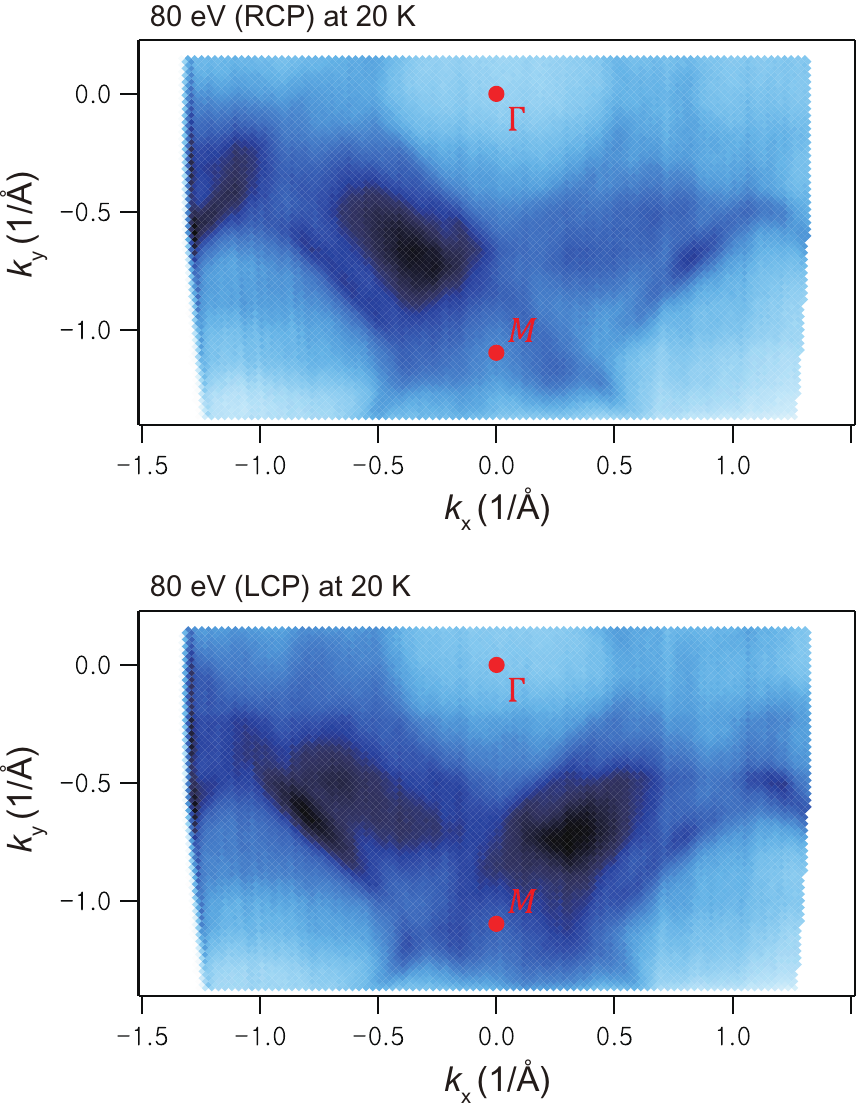}
\caption{{\bf Fermi surface of 4~uc SRO measured with (upper) right and (lower) left circularly-polarized (RCP and LCP, respectively) light at 20~K.} The photon energy in each measurement was 80~eV.}
\label{RLCP}
\end{figure*}

\begin{figure*}[h]
\centering
\includegraphics[width=0.5\textwidth]{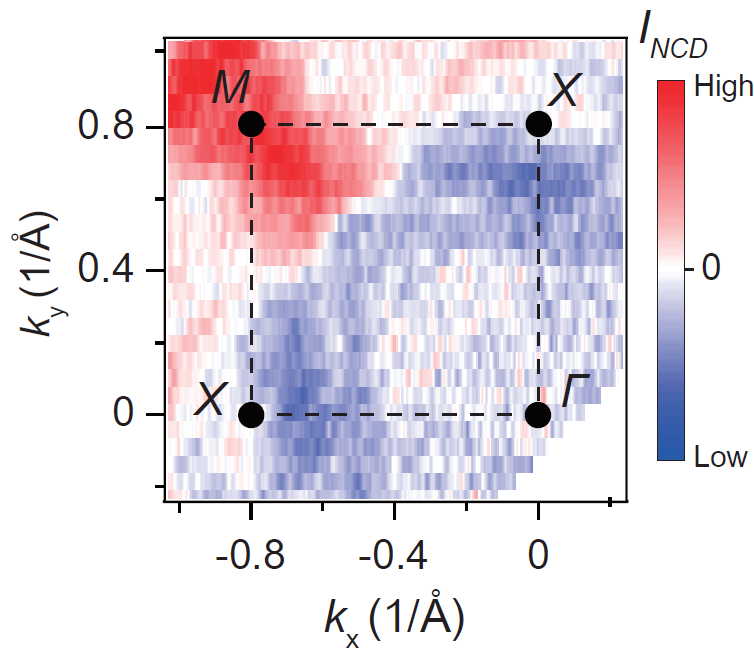}
\caption{\bf Even part of the normalized-CD (NCD) ARPES intensity near the Fermi energy.}
\label{CDFS}
\end{figure*}

The Fermi surfaces (FSs) ($E_F \pm 27$~meV) measured with RCP and LCP light are shown in Fig.~\ref{RLCP}. The normalized CD (NCD) intensity is defined as $I_{NCD}=\frac{I_{RCP}-I_{LCP}}{I_{RCP}+I_{LCP}}$, where $I_{RCP}$ and $I_{LCP}$ are the intensity of spectra measured with RCP and LCP light, respectively. Using the data taken with RCP and LCP shown in Fig.~\ref{RLCP}, we can build an NCD map. What is shown in Fig.~\ref{CDFS} is the even part of such NCD ARPES data. As one can see in the figure, the NCD intensity varies from zero to negative and then to positive as we move from $\Gamma$ to $M$. A clear sign-changing behavior is observed as a function of the momentum.


\clearpage

\vspace{\baselineskip}

\begin{center}
\section{Results of ionic liquid gating on 4 and 5~uc SrRuO$_3$ thin films}
\end{center}

\begin{figure*}[h]
\centering
\includegraphics[width=1\textwidth]{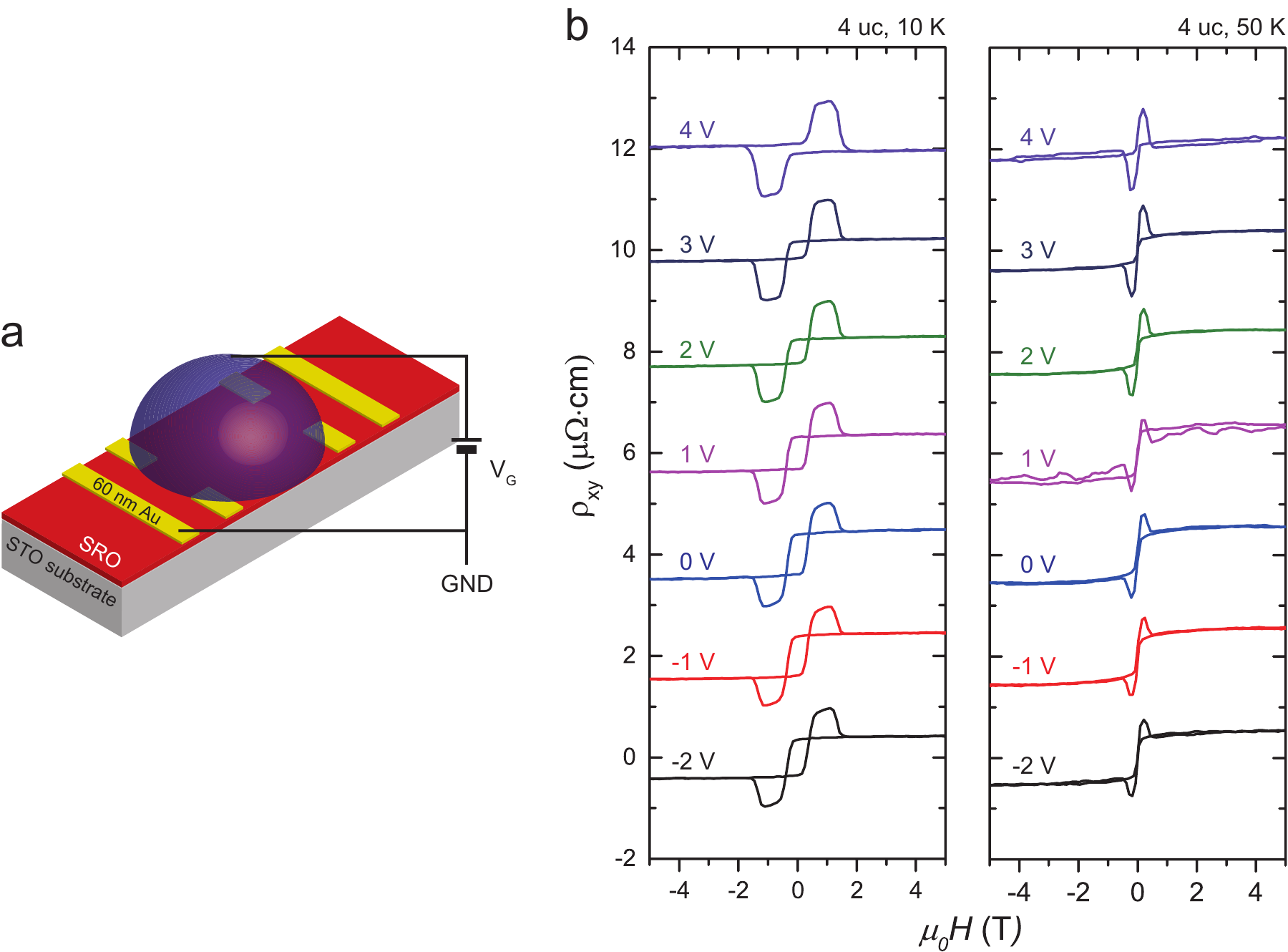}
\caption{{\bf Hall effect measurement of 4~uc SRO thin film with ionic liquid gating.} (a) A schematic for the device used in ionic liquid gating experiment. (b) Hall resistivity at 10 and 50~K for various ionic liquid gating voltages. Note that the anomalous Hall resistivity varies as the gate voltage changes.}
\label{ga1}
\end{figure*}

\begin{figure*}[h]
\centering
\includegraphics[width=0.5\textwidth]{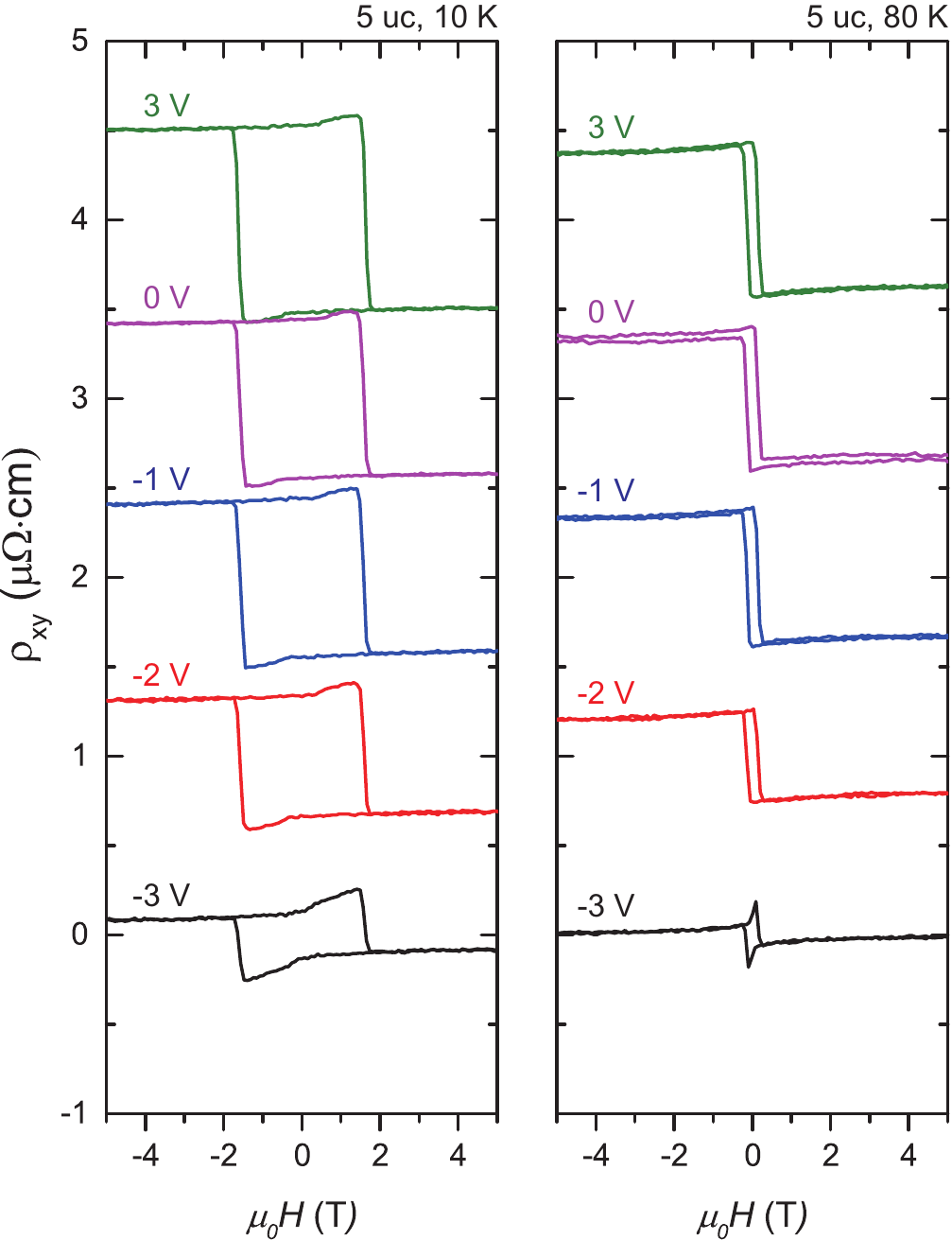}
\caption{{\bf Hall effect measurement of 5~uc SRO thin film with ionic liquid gating at 10 and 80~K.} Anomalous Hall resistivity varies as the gate voltage changes.}
\label{ga2}
\end{figure*}

We measured Hall resistivity of 4 and 5~uc SRO thin films with ionic liquid gating (Figs. \ref{ga1} and \ref{ga2}). Figure \ref{ga1}a shows schematics for the Hall bar and the ionic liquid gating device. By tuning the gate voltage, chemical potential of the SRO thin films is expected to change. The change in the chemical potential in turn changes the anomalous Hall resistivity (that is, AHE is controlled) as shown in Fig. \ref{ga1}b and Fig. \ref{ga2}.

\clearpage
\begin{center}
\section{Band structures obtained by {\it ab-initio} and tight-binding method}
\end{center}

\begin{figure*}[h]
\centering
\includegraphics[width=0.8\textwidth]{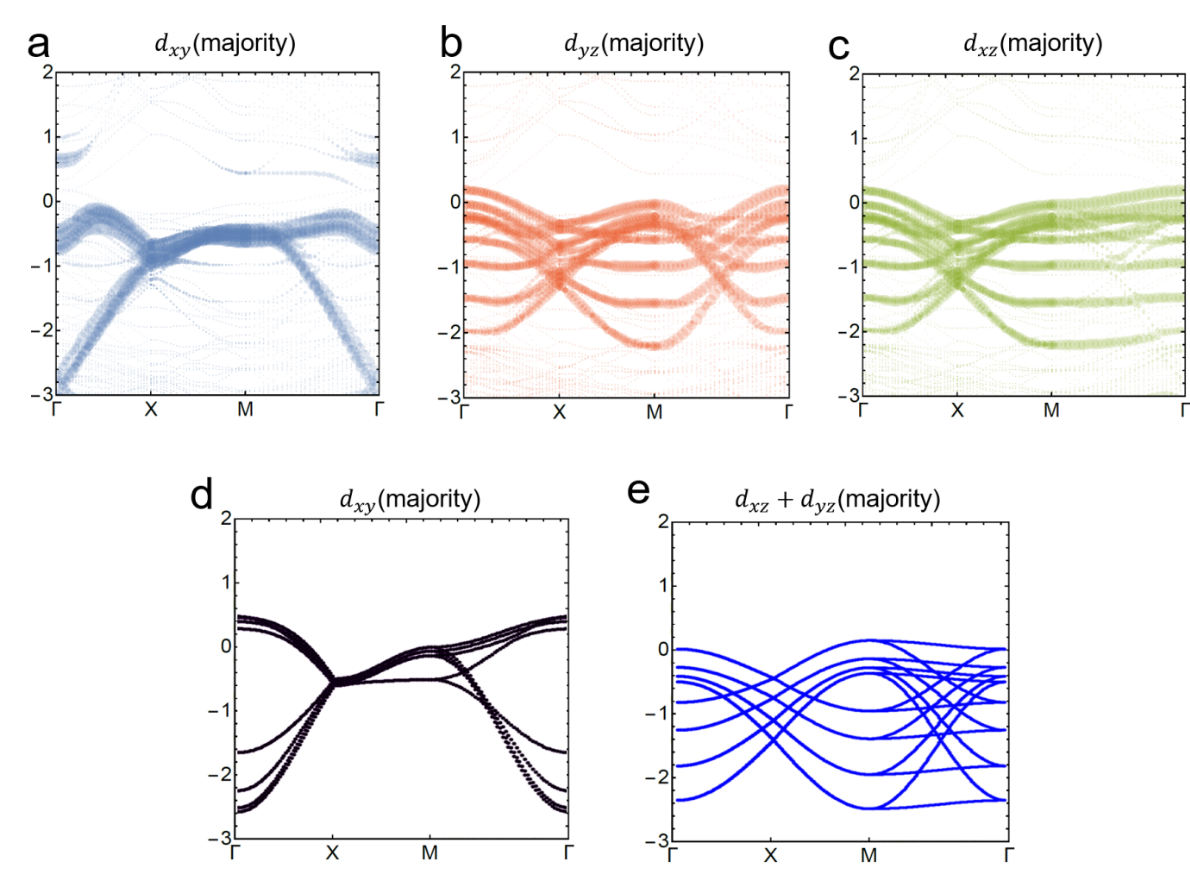}
\caption{{\bf Projected $t_{2g}$ majority bands along the high-symmetry lines calculated by DFT and tight-binding model calculations.} We neglected spin-orbit coupling (SOC) for both of the cases. (a, b, c) The band structures from first-principles calculations, which are performed based on the atomic structure obtained by coherent Bragg rod analysis (COBRA). The calculation takes into account lattice rotation and other orbitals in addition to $t_{2g}$ orbitals. (d, e) The band structures from tight-binding calculations, which are performed on 4~uc SRO where the parameters are extracted from Wannierized Hamiltonian obtained by DFT calculations. For simplicity, only $t_{2g}$ orbitals are taken into account and lattice rotation is neglected. One can see that band structures calculated with $ab-initio$ and tight-binding methods match well.}
\label{fig-bands}
\end{figure*}

{\it ab-initio} band structure calculations were performed using the atomic structure information of 4~uc SRO ultrathin films. The electronic structure of 4~uc SRO film is obtained by calculating a slab of SRO/STO heterostructure consisting of 4 layers of SRO and 4.5 layers STO (including an additional SrO layer at the surface), and a vacuum of 19 $\AA$ . We used the experimental atomic positions from COBRA results~\cite{sohn18}, having a $\sqrt{2}\times\sqrt{2}$ in-plane unit cell to include octahedral rotation. Fig.~\ref{fig-bands} (a), (b), and (c) show the band structure of 4 uc SRO film from our {\it ab-initio} calculation.

Based on the {\it ab-initio} calculation result, we construct 4~uc SRO tight-binding model where the tight-binding parameters for Ru-$d$ derived bands of SRO/STO heterostructure are obtained using Wannier90 package~\cite{Pizzi20}. In the tight-binding Hamiltonian, only $t_{2g}$ orbitals are considered since they account for the most of the density of states near the Fermi level. Fig.~\ref{fig-bands} (d) and (e) show the band structures of 4 uc SRO film from tight-binding calculation. One can clearly see that band structures calculated by ab-initio and tight-binding methods are very similar.

\clearpage

\begin{center}
\section{Band structures and Fermi surfaces of freestanding single-layer and four-layer SrRuO$_3$ slabs}
\end{center}

\begin{figure*}[h]
\centering
\includegraphics[width=1\textwidth]{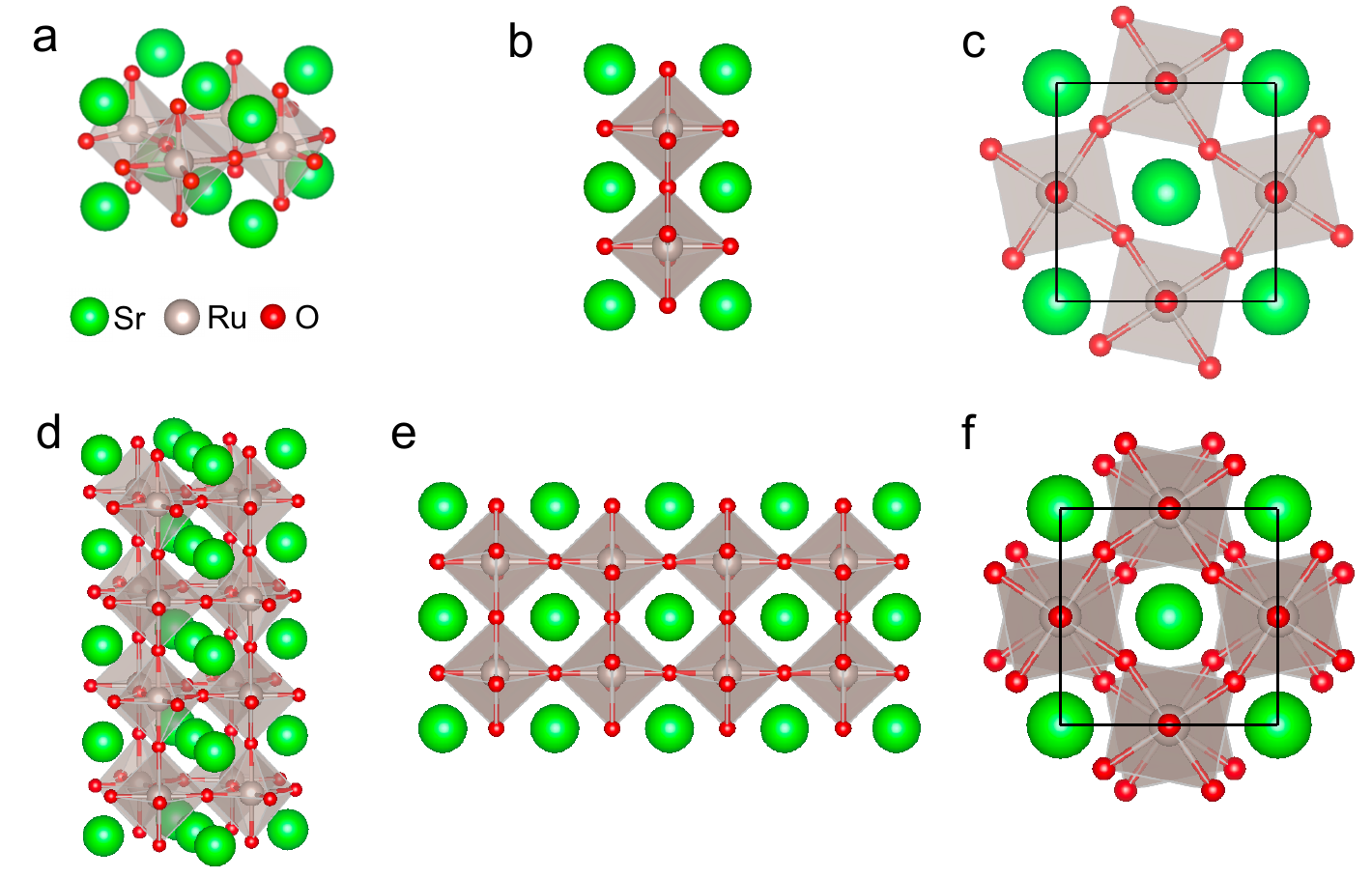}
\caption{{\bf Atomic arrangements of the calculated SRO slabs.} (a-c) Perspective, side, and top view of the single-layer SRO slab. (d-f) Perspective, side, and top view of the four-layer SRO slab. A vacuum of 15 Å is used for both calculations. The block box in panels (c) and (h) represent the in-plane unit cell.}
\label{fig-sroslabs}
\end{figure*}

In order to investigate the effect of the dimensional confinement, we have calculated the electronic structures of freestanding single- and four-layer SRO slabs (Fig.~\ref{fig-sroslabs}), fully relaxed with fixed in-plane lattice constant of experimental substrate ($a$= 3.89 {\AA} of SrTiO$_3$ substrate). We find that there is no significant change in the magnetic moment of Ru ions with respect to the thickness; the calculated average moments of the single- and four-layer SRO are 1.40 and 1.35 $\mu_B$, respectively. We also find that inclusion of on-site Coulomb interaction does not change the magnetic moment significantly as including on-site Coulomb interaction by using GGA+$U$ scheme ($U$=3.23 eV, $J$=0.74 eV from constrained random-phase approximation calculation~\cite{vaugier2012hubbard}) only slightly increases the magnetic moments of single- and four-layer SRO to 1.49 and 1.42 $\mu_B$, respectively. 

\begin{figure*}[h]
\centering
\includegraphics[width=1\textwidth]{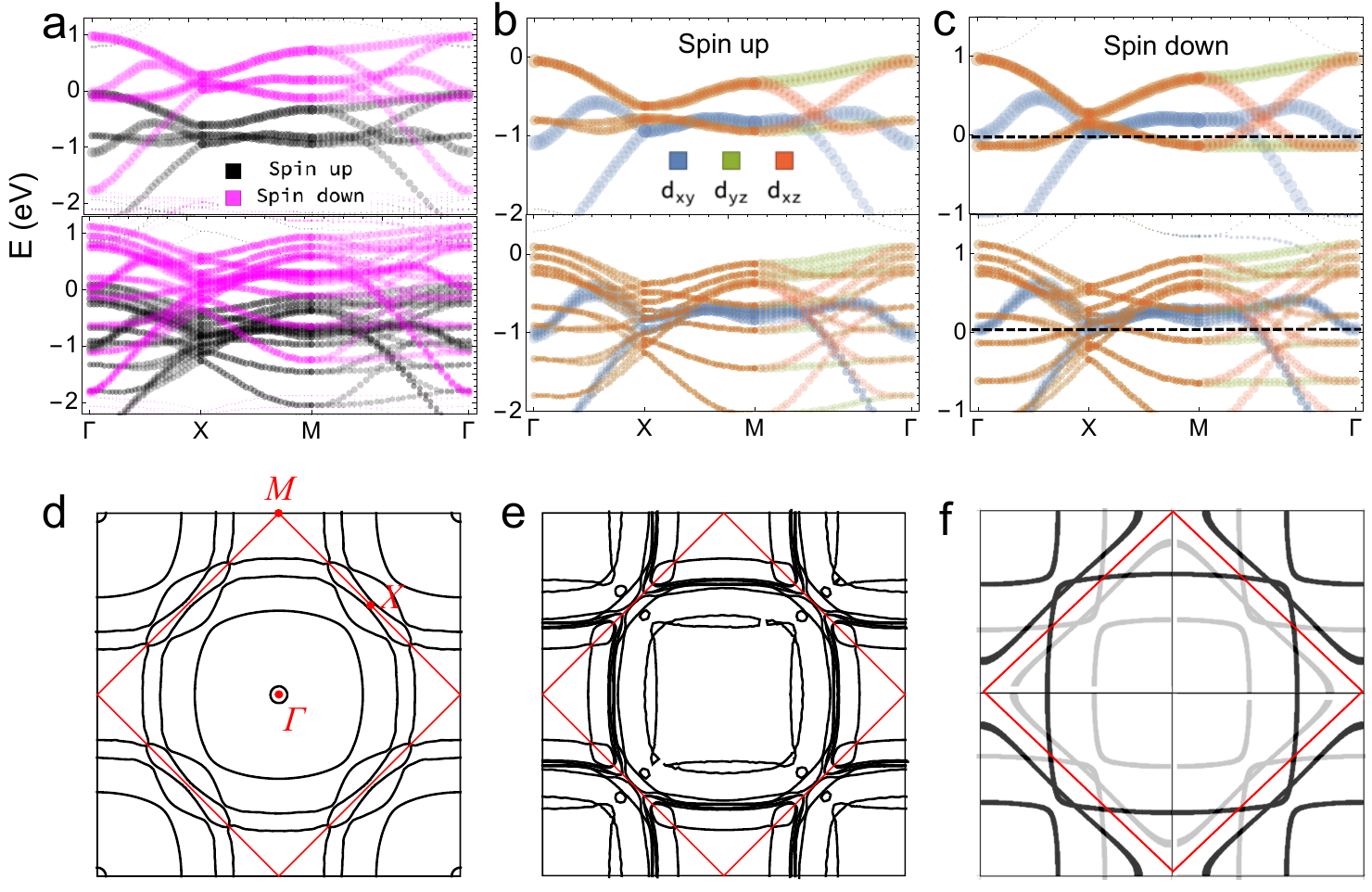}
\caption{{\bf Band structures and Fermi surface maps of single- and four-layer SRO slabs.} (a) Comparison of exchange splittings in single- (top panel) and four-layer (bottom panel) SRO slab. Ru-$d$ orbital projected band structures of (b) spin-up and (c) spin down electrons (up: single-layer SRO, down: four-layer SRO). The dotted horizontal lines are the energy that the Fermi surfaces are evaluated. Fermi surface of the spin down bands of (d) single-layer and (e) four-layer SRO slab. The red lines and letters are the Brillouin zone and high-symmetry points of $\sqrt{2}\times\sqrt{2}$ in-plane unit cell, respectively. (f) Fermi surface of spin down bands obtained by our tight-binding model (Fig. 1(b)). The gray lines are the bands folded into the Brillouin zone of the $\sqrt{2}\times\sqrt{2}$ in-plane unit cell, shown for better comparison. }
\label{fig-sroslabbd}
\end{figure*}

We note that, as the thickness of the film gets smaller, in general, a long-range magnetic order is expected to be suppressed in Heisenberg magnets~\cite{Mermin66}, although Ising magnetocrystralline anisotropy in thin-film SRO can stabilize the long-range magnetism~\cite{bander1988ferromagnetism, Koster98}. Therefore, for proper description of the magnetic moments in SRO thin films, the dynamical magnetic fluctuation, which is not considered in GGA and GGA+$U$ calculations, can be important. Namely, the overestimated magnetic moments in DFT calculations might be due to the missing magnetic fluctuation in static mean-field treatment of the electronic correlations. This is further evidenced by the decrease in the magnetic moments in dynamical mean-field theory calculations of SRO~\cite{Kim15}, showing reduced moments. Thus, capturing the correct value of the magnetic moment requires methods beyond DFT that consider, for example, magnetic fluctuations. This requires further investigation in the future. As an alternative approach, we treat the magnetic moment (or exchange splitting) as a parameter in our tight-binding model with which we obtain band structures consistent with the ARPES data. 

For the comparison between the band structures from tight-binding parameters and of our first-principles calculations, we calculate band structures of single- and four-layer SRO slabs, presented in Fig.~\ref{fig-sroslabbd}. We find that the band structures of the single- and four-layer SRO slabs have similar exchange splitting (panel a), consistent with the small difference in Ru magnetic moment between the two systems. Moreover, both the band dispersion and orbital character share the common characteristic features, but we find sub-band splitting in $d_{yz}/d_{xz}$ bands from the confinement effect. Although there are a larger number of bands that cross the Fermi energy for four-layer SRO slab, we find that our tight-binding model constructed from a single layer of SRO slab successfully reproduces the main features of the four-layer slab band structures as presented in Fig.~\ref{fig-sroslabbd} (d-f). Since spin up and down bands are approximately related by a rigid shift, we compare only spin down Fermi surfaces. We find that the overall structure of the Fermi surface is similar between single- and four-layer SRO slabs (panels d and e), which is consistent with the similar AHE versus magnetization curve calculated from our tight-binding model (Fig. 4(d)). (We note that the rectangular bands around the $\Gamma$ point is from the band folding effect from octahedral rotation, expected to have reduced spectral weight.) More importantly, our tight-binding model successfully reproduces the four-layer-SRO Fermi surfaces (panels e and f), which clearly shows the validity of our tight-binding band structures to describe the electronic properties of the SRO film.

\clearpage

\begin{center}
	\section{Anomalous Hall Effect in three-dimensional SrRuO$_3$}
\end{center}

The AHE $vs.$ magnetization scaling behavior in 3D SRO shows the similar sign changing behavior to that in 2D SRO~\cite{Fang03}. However, in general, there is no particular reason to expect similar AHE between 2D and 3D SRO. This is because the spatial dimensionality and the relevant symmetry strongly constrain the band crossing condition and the topological characteristics of the resulting band crossing points, so that topological nodes in 2D band structure are fundamentally different from the 3D case. 
For example, 3D SRO band structure can be obtained from the band structure of vertically stacked 2D SRO layers by increasing the interlayer coupling.
When the interlayer coupling is sufficiently large, the band structure is deformed along the k$_z$ direction in a way that a number of Weyl points are created. 
Because of such additional band crossings induced by interlayer coupling, the 2D and 3D topological band structures are not adiabatically connected.

To understand the evolution of the topological nodal structures between 2D and 3D limits, we consider a Hamiltonian H$_{3D}$ which describes the band structure of 3D SRO. By replacing the interlayer hopping t$_z$ of $H_{3D}$ by $r$~$\times$~t$_z$, where 0~$\leq$~$r$~$\leq$~1, we define the Hamiltonian $H(r)$, which continuously interpolates the band structure from the 3D limit with $H(r=1)$~=~H$_{3D}$ to the 2D limit with $H(r=0)$. Here $r$~=~0 indicates the vanishing interlayer coupling so that the relevant system is composed of decoupled 2D SRO layers. 

\begin{figure*}[h]
	\centering
	\includegraphics[width=1\textwidth]{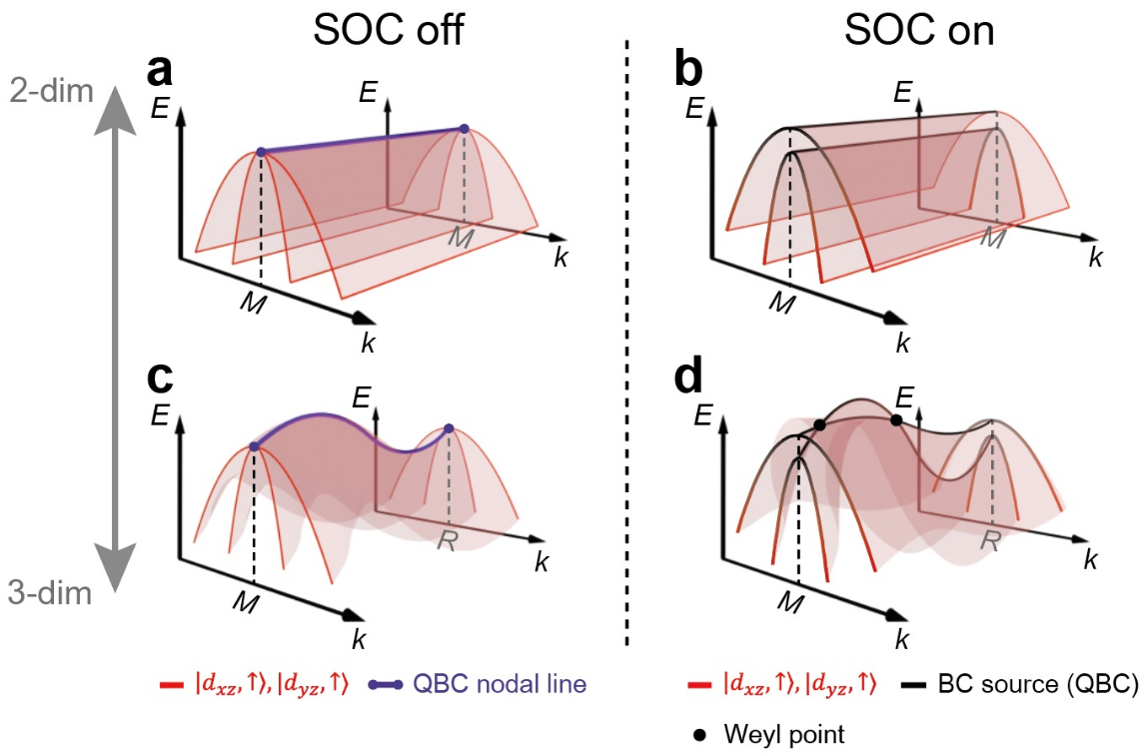}
	\caption{{\bf The nodal structure evolution of $H(r)$, especially focusing on the nodal line arising from the QBC of 2D SRO.} (a) A flat nodal line when $r$~=~0 and spin-orbit coupling (SOC) is absent. (b) The gapped flat nodal line when SOC is turned on. (c) The dispersing gapless nodal line when $r$~$>$~0 and SOC is zero. (d) The dispersing gapped nodal line with extra Weyl points from additional band crossing when $r$~$>$~0 and SOC is finite.}
	\label{3DSRO}
\end{figure*}

When $r$ is vanishingly small, the 3D band structure of $H(r)$ is given by simple extension of the 2D band structure along the k$_z$ direction without any dispersion. This means that the QBC (NL) of the 2D SRO becomes a flat nodal line (a flat nodal surface) in $H(r=0)$ without dispersion along the kz direction, when spin orbit coupling (SOC) is negligible [see Fig.~\ref{3DSRO}(a)]. In the presence of finite SOC, both the flat nodal line and the flat nodal surface become gapped, generating large Berry curvature [see Fig.~\ref{3DSRO}(b)]. The topological band structure and the relevant AHE of $H(r)$ with small $r$ would be almost the same as those of the 2D SRO, hence we expect the same AHE $vs.$ magnetization scaling.

On the other hand, as $r$ gets larger, one can expect the deviation of AHE $vs.$ magnetization scaling of $H(r)$ from that of $H(r=0)$. The main effects of finite $r$ are as follows: $i$) the flat nodal line and the flat nodal surface, arising from the QBC and the NL of 2D SRO, develop finite dispersion along the k$_z$ direction, which in turn induces non-uniform distribution of their Berry curvature along the k$_z$ direction. [see Fig.~\ref{3DSRO}(c)]. $ii$) the band inversion that occurs as $r$ increases can generate additional Weyl points, which make extra contribution to the total AHE. [see Fig.~\ref{3DSRO}(d)]. The combined effect of $i$) and $ii$) makes the AHE $vs.$ magnetization scaling of $H(r)$ to deviate from that of $H(r=0)$ as $r$ becomes larger. Because of this reason, one cannot expect a simple relationship between the AHE of 2D and 3D SRO in general.

To confirm the idea explained above, we performed additional numerical calculations of the lattice Hamiltonian
\begin{align}
	H=\sum_\textbf{k}[(\epsilon_{\textbf{k}\sigma}^a )\delta_{ab} \delta_{\sigma\sigma'}+f_{\textbf{k}}^{ab} \delta_{\sigma\sigma'}+i\lambda\epsilon^{abc} \tau_{\sigma\sigma'}^c] d_{\textbf{k}a\sigma}^{\dagger} d_{\textbf{k}b\sigma},
\end{align}
where
\begin{align}
	\epsilon_{k\sigma}^{1=yz}&=-2t_1(\cos{k_y}+r\cos{k_z})-2t_2\cos{k_x}-m\tau_{\sigma\sigma}^z, \nonumber\\
	\epsilon_{k\sigma}^{2=xz}&=-2t_1(\cos{k_x}+r\cos{k_z})-2t_2\cos{k_y}-m\tau_{\sigma\sigma}^z, \nonumber\\
	\epsilon_{k\sigma}^{3=xy}&=-2t_1(\cos{k_x}+\cos{k_y})-2rt_2\cos{k_z}-m\tau_{\sigma\sigma}^z,
\end{align}
and
\begin{align}
	f_{\textbf{k}}^{ab}&=-4f\sin{k_a}\sin{k_b}
\end{align}
for a,b $\neq$ z, and
\begin{align}
	f_{\textbf{k}}^{ab}&=-4rf\sin{k_a}\sin{k_b}
\end{align}
for one of (a,b)~=~z.

\begin{figure*}[h]
	\centering
	\includegraphics[width=0.75\textwidth]{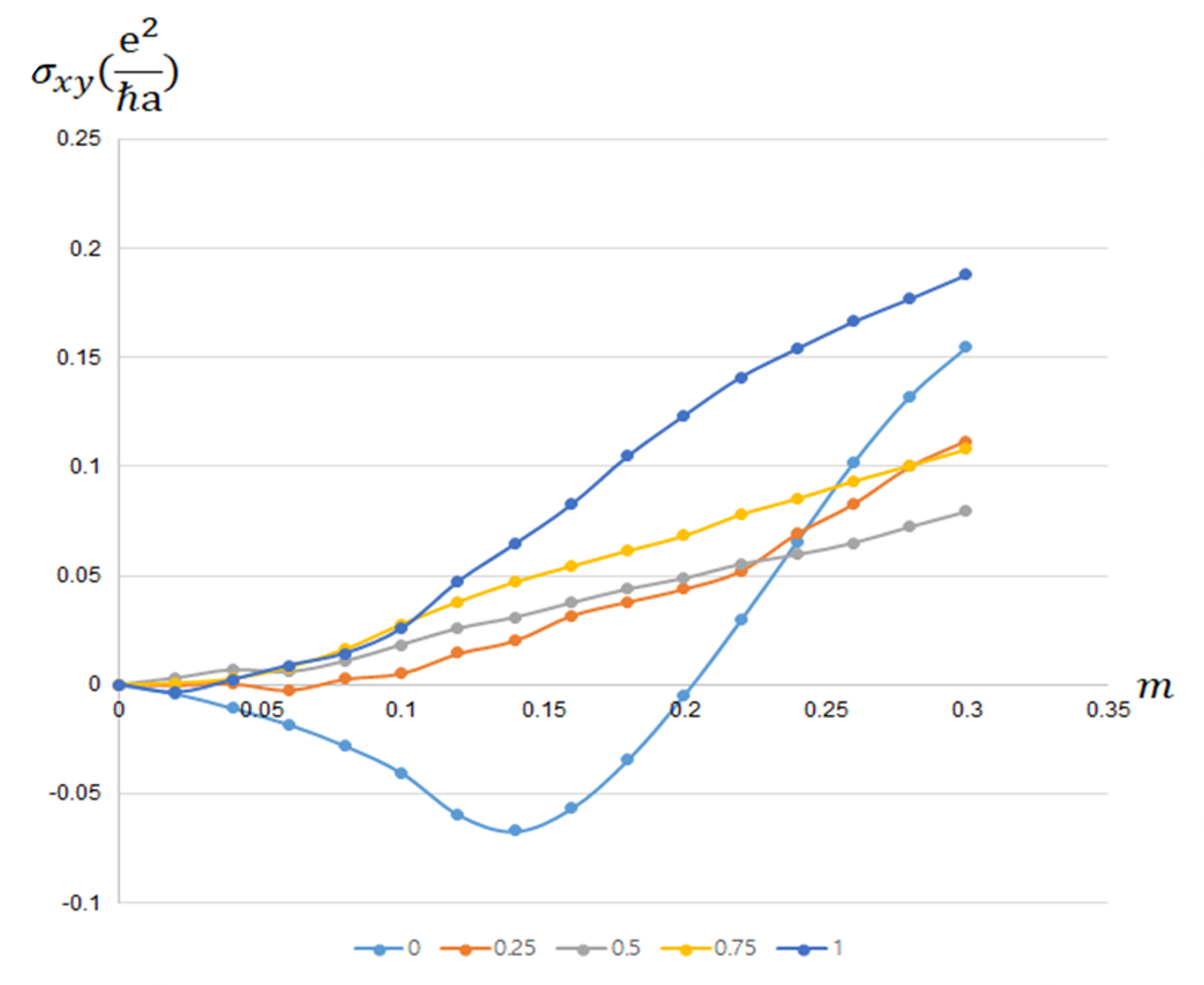}
	\caption{Anomalous Hall conductivity ($\sigma_{xy}$) as a function of the magnetization ($m$) computed from $H(r)$ for various $r$ values.}
	\label{3DSRO_2}
\end{figure*}

This Hamiltonian often has been used to describe the band structure of 3D SRO~\cite{carter2013theory, chen2013weyl}. We note that the relevant 2D Hamiltonian (with $r=0$) is the same as the Hamiltonian used in our manuscript, except that here we only consider nearest-neighbor hopping terms for simplicity. The behavior of AHE $vs.$ magnetization is computed using the above Hamiltonian for various $r$ values, as shown in Fig.~\ref{3DSRO_2} .

One can clearly observe that AHE $vs.$ magnetization behavior strongly deviates from the 2D case as $r$ grows. The deviation from the 2D behavior becomes more prominent as the two factors $i$) and $ii$) mentioned above become more important. Namely, $i$) the finite k$_z$ dispersion of the nodal line (nodal surface) arising from the QBC and NL of 2D SRO, $ii$) Weyl point generation arising from extra band inversion, make the behavior of the 2D and 3D systems different. 

Let us note that the AHE of 3D SRO in Fig.~\ref{3DSRO_2} ($r=1$ case) does not show sign reversal, because, in the calculations, we fixed the in-plane hopping parameters irrespective of the value $r$. However, adjusting the in-plane hopping parameters that can properly reproduce the Fermi surface of 3D SRO, the sign changing AHE can be correctly described as expected. The main purpose of displaying Fig.~\ref{3DSRO_2} is to understand how the topological band structure of SRO generally evolves from 2D to 3D.

Nonetheless, the observed similarity of AHE $vs.$ magnetization behavior in 2D and 3D, is indeed very interesting. To fully understand this interesting observation, careful examination of the layer-number dependent variation of the SRO topological band structure, both in theory and in experiment, is required. We leave this important question as a future research topic.

\clearpage
\begin{center}
\section{Symmetry-protected nodal structure in monolayer SrCoO$_{3}$}
\end{center}

\begin{figure*}[h]
\centering
\includegraphics[width=1\textwidth]{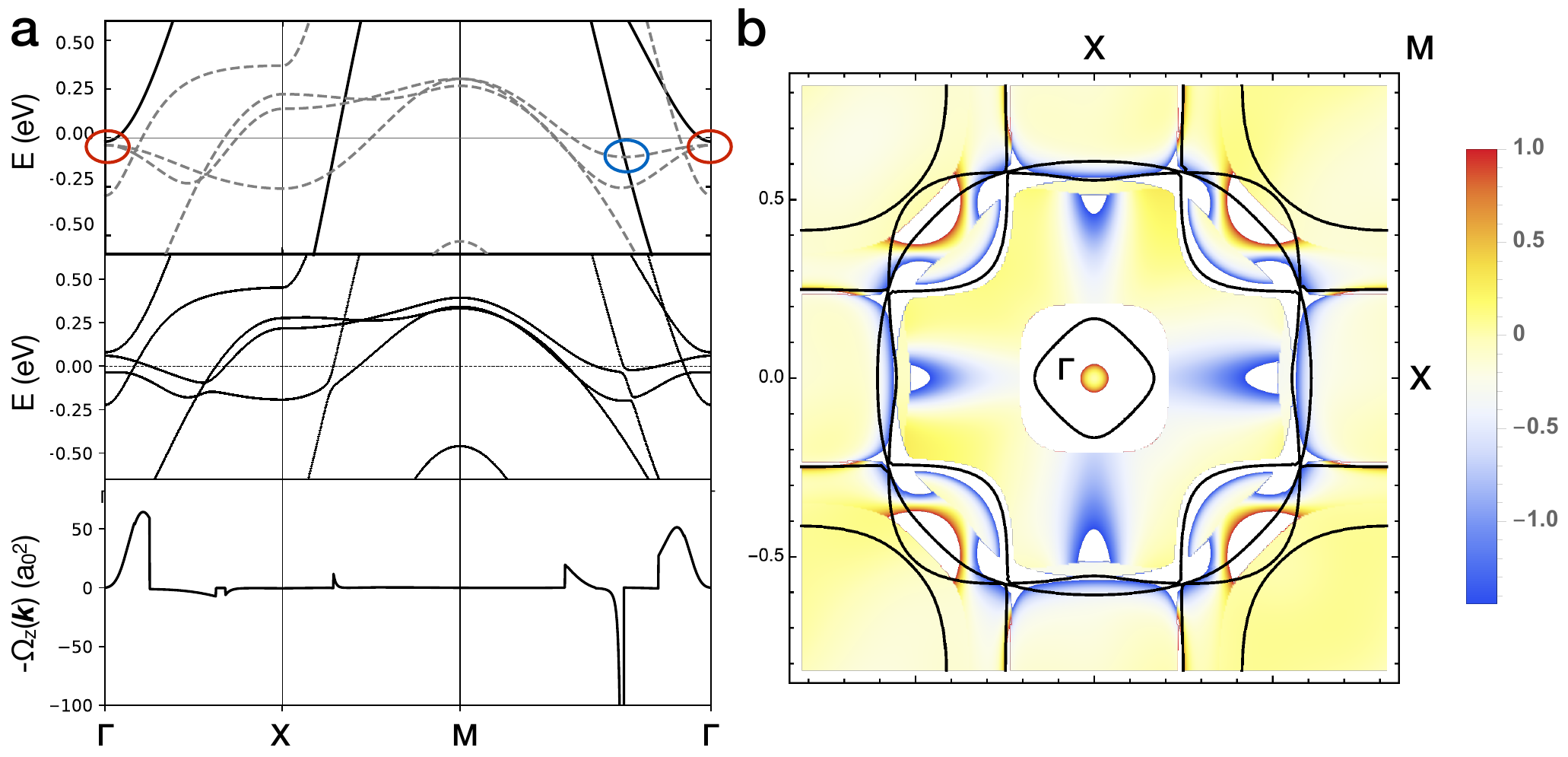}
\caption{{\bf Band structures and Berry curvature of SrCoO$_3$ monolayer.} (a) Top panel: band dispersions without spin-orbit coupling where solid and dashed lines represent the spin up and down bands, respectively. Middle panel: band dispersion with spin-orbit coupling. Bottom panel: Berry curvature integrated up to the chemical potential. ($a_{0}$ : Bohr radius). The zero energy is defined as 62~meV above the Fermi energy to emphasize the large Berry curvatures around the band crossings. A quadratic band crossing at $\Gamma$ point and the nodal line in the M$\Gamma$ direction, marked by red and blue circles, respectively, are major sources of enhanced Berry curvature as in SrRuO$_{3}$. (b) Integrated Berry curvature (color plot) and constant energy contour (black solid lines) at the zero energy, showing that the Berry curvature is dominantly contributed around the band crossings.}
\label{fig-sco}
\end{figure*}

We demonstrate the generality of the symmetry-protected nodal structures that we identified in two dimensional magnets. As an example, we perform first-principles calculations on SrCoO$_3$, for which the sign-changing AHE upon thickness change is reported \cite{Zhang19}. To emphasize the characteristic band crossings in two dimensional limit, we calculate the band structures of a 1.5 layer of SrCoO$_{3}$ with vacuum in which additional SrO layer is added to maintain inversion symmetry (see the method section for detail.). We find the ferromagnetic ground state with out-of-plane magnetization direction. The calculated band structures in the absence of spin-orbit coupling show two types of band crossings, quadratic band touching and nodal line (Fig.~\ref{fig-sco} (a)). With inclusion of spin-orbit coupling, we find that the Berry curvature is dominantly contributed from the $k$-points around the spin-orbit split quadratic band touching points and nodal lines as presented in Fig.~\ref{fig-sco} (b-c).

%